\documentclass[numberedappendix,twocolappendix,twocolumn]{aastex631}

\def\rev{}

\usepackage{bm}

\shorttitle{MRI and outflows in Binary Star Formation}
\shortauthors{Tomoaki Matsumoto}

\begin{document}

\title{Angular Momentum Transport in Binary Star Formation: The Enhancement of Magneto-Rotational Instability and Role of Outflows}

\correspondingauthor{Tomoaki Matsumoto}
\email{matsu@hosei.ac.jp}

\author[0000-0002-8125-4509]{Tomoaki Matsumoto}
\affiliation{Faculty of Sustainability Studies, Hosei University, Fujimi, Chiyoda-ku, Tokyo 102-8160, Japan}

\begin{abstract}
The formation of binary stars is highly influenced by magnetic fields, which play a crucial role in transporting angular momentum. We conducted three-dimensional numerical simulations of binary star accretion via a circumbinary disk, taking into account a magnetic field perpendicular to the disk and an infalling envelope. Our simulations reproduce the following phenomena: (1) spiral arms associated with circumstellar disks, (2) turbulence in the circumbinary disk, induced by magneto-rotational instability (MRI), (3) a fast outflow launched from each circumstellar disk, and (4) a slow outflow from the circumbinary disk. The binary models exhibit a higher $\alpha$-parameter than the corresponding single star models, indicating that the binary stars enhance MRI turbulence. Moreover, an infalling envelope also enhance the turbulence, leading to a high $\alpha$-parameter. 
While the spiral arms promotes radial flow, causing transfer of mass and angular momentum within the circumbinary disk, the MRI turbulence and outflows are main drivers of angular momentum transfer to reduce the specific angular momentum of the system. 
\end{abstract}

\keywords{
Star formation(1569) ---
Binary stars(154) --- 
Circumstellar disks(235) --- 
Stellar jets(1607) --- 
Magnetohydrodynamical simulations(1966)
}

\section{Introduction} \label{sec:intro}

About half of solar-type stars are formed as members of multiple systems, and the multiplicity increases for higher stellar masses \citep{Duchene13}. Therefore, binary star formation is an important process, and several formation scenarios have been proposed so far \citep{Reipurth14,Offner22}. Recent high-resolution observations with the Atacama Large Millimeter/Submillimeter Array (ALMA) have revealed the early phases of low-mass binary and multiple star formation \citep{Hioki07,Fukagawa13,Dutrey14,Takakuwa14,Alves19,Takakuwa20}. The binary protostars surrounded by circumbinary disks imply a contribution of formation mechanism based on the disk fragmentation scenario \citep{Matsumoto03,Kratter10}.

It is known that magnetic fields play a crucial role in star formation. Magnetic fields give rise to various phenomena in the star formation process, such as launching outflows, exerting magnetic braking on disks, \citep[e.g.,][]{Machida11,Tsukamoto22} and generating turbulence via magneto-rotational instability (MRI) \citep[e.g.,][]{Balbus91,Bai13}. However, researches on the role of magnetic fields have primarily focused on single star formation rather than binary star formation. These magnetic processes induce angular momentum transport within the system, and in the context of binary star formation, the magnetic fields are expected to influence fundamental binary parameters such as binary separation.

Binary accretion from a circumbinary disk has been intensively investigated in many studies, not only for binary black holes but also for binary star formation \citep[see recent review][]{Lai22}. 
These studies usually assume $\alpha$ viscosity model \citep{Shakura73}, where angular momentum transport occurs in the circumbinary disk. The assumed viscosity affects the orbital evolution of binary stars \citep[e.g.,][]{Dittmann22}. The origin of the $\alpha$ viscosity is thought to be MRI turbulence. Once a magnetic field is assumed in a disk model, it causes not only MRI \citep{Noble21} but also disk winds \citep{Suzuki14} and outflows \citep{Machida11}, which lead to the redistribution of angular momentum.

Several researchers have considered magnetic fields in binary accretion models; \citet{Noble12,Shi12,Shi15,Lopez-Armengol21,Noble21} incorporated magnetic fields in their simulations, but the initial magnetic field configuration was confined within the circumbinary disk, where the closed magnetic field lines followed the isodensity surfaces of the disk. 
\citet{Bowen18, Avara23} also took into account magnetic fields, using the initial conditions derived from snapshots of the simulations by \citet{Noble12}. 
Although the magnetic fields change during the evolution of the circumbinary disks, the initial configuration of the magnetic fields may affect the disk's evolution.
In the case of binary star formation with typical field strength, the natal molecular cloud core collapses along the magnetic fields, and a disk perpendicular to the open magnetic field naturally forms \citep[e.g.,][]{Matsumoto17,Tsukamoto18}. Moreover, the perpendicular component (vertical component) of the magnetic field is crucial for launching outflows \citep{Machida11,Gerrard19}, which play an important role in angular momentum transport.

Typical binary accretion models employ an infinitely extended circumbinary disk in a steady state as an initial condition \citep[e.g.,][]{Moody19}. However, in the context of binary star formation, recent ALMA observations reveal that circumbinary disks actually have a finite size \citep[e.g.,][]{Takakuwa20}. Furthermore, in the early stages of star formation, a young star is embedded in a cloud core and accretes gas from an infalling envelope \citep[e.g.,][]{Hayashi93}. Therefore, a circumbinary disk model with a finite size, which accretes gas from an infalling envelope, is more suitable for the context of binary star formation, as adopted by \citet{Bate97}. The assumed infalling envelope has the potential to influence the evolution of the binary system \citep{Bate00}.

In this paper, we examine the accretion of binary stars from a finite-sized circumbinary disk, considering a magnetic field that perpendicularly threads the disk. Additionally, we incorporate an infalling envelope, which significantly contributes to the mass and angular momentum supply to the system, as well as to angular momentum transport through magnetic braking. Despite its importance, this factor has often been overlooked in many previous studies. Consequently, our models not only reproduce the MRI in the circumbinary disk but also simulate the outflows from binary stars and accretion onto the circumbinary disk from an infalling envelope. By employing this model, we conduct a quantitative investigation into angular momentum transport in a binary system.
The present model reproduces two types of disks: a circumbinary disk and circumstellar disks. Our focus is primarily on the circumbinary disk, as the circumstellar disks are likely influenced by magnetic diffusion processes, such as Ohmic dissipation, ambipolar diffusion, and the Hall effect. We assume the ideal-MHD, and these processes are not included in our model for simplicity. The outflows are also investigated as a case of the ideal-MHD limit.

This paper is organized as follows. In Sections~\ref{sec:models} and \ref{sec:methods}, the model and methods are shown. The results of the simulations are presented in Section~\ref{sec:results}, and they are discussed in Section~\ref{sec:discussion}. Finally, the conclusions of this paper are given in Section~\ref{sec:summary}. The details of the analyses are shown in Appendix.

\section{Models}
\label{sec:models}

The numerical models presented in this paper are based on \citet{Matsumoto19}, which is extended to include magnetic fields. The primary and secondary stars have constant masses of $M_1$ and $M_2$, respectively. It is assumed that the stars rotate around the origin in fixed circular orbits at an angular velocity of $\Omega_\star = (G M_\mathrm{tot}/a_b^3)^{1/2}$, where $G$ is the gravitational constant, $M_\mathrm{tot}$ ($= M_1 + M_2$) is the total mass of the stars, and $a_b$ is binary separation. The rotation period of the binary stars is $T_\star = 2 \pi (a_b^3/GM_\mathrm{tot})^{1/2}$.

\begin{figure*}[t]
\plotone{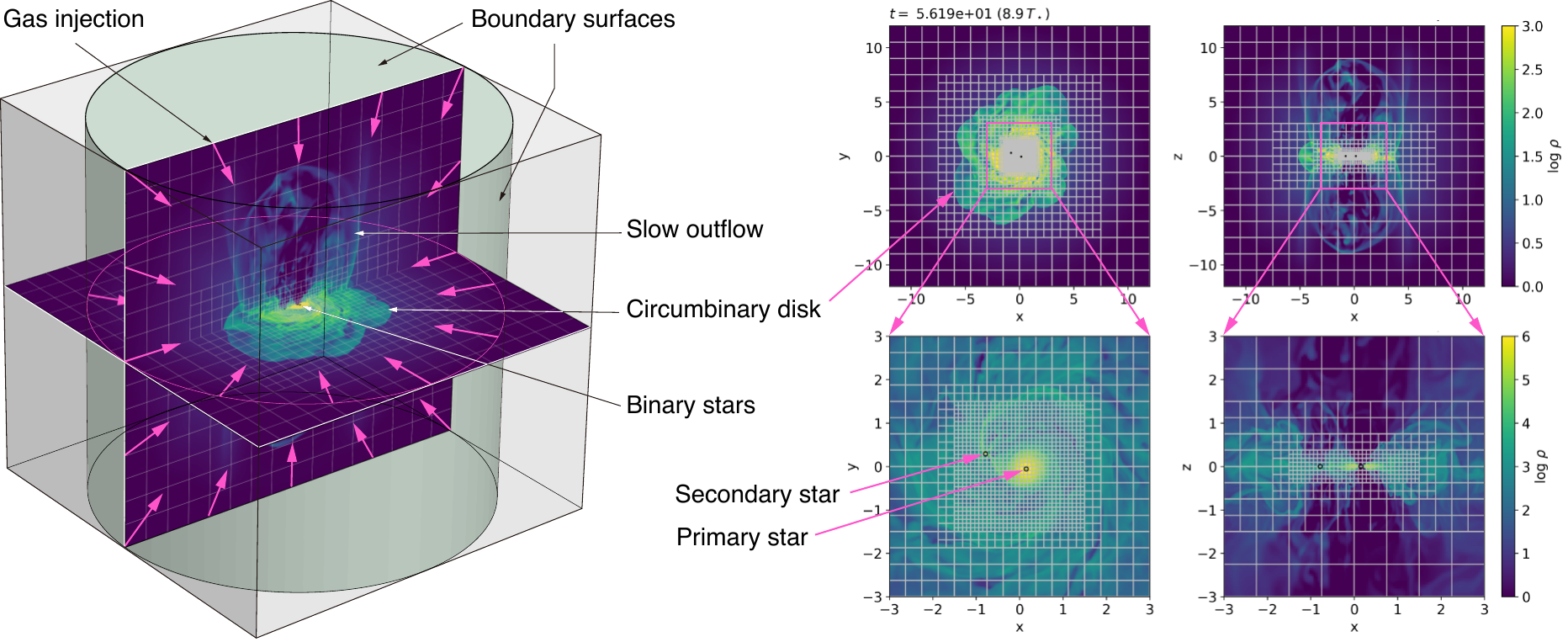}
\caption{The schematic diagram illustrates the computational domain. The left panel depicts a 3D view of the computational domain and the imposition of the boundary condition. The four right panels display a static hierarchical grid configuration that overlays the density distribution in cross sections. The gray grid represents the FMR blocks, with each block consisting of $16^3$ cells. Black circles indicate the positions of sink particles, which model binary stars. The radii of the circles are two times larger than the actual sink radius for ease of visual confirmation.
 \label{f1.pdf}}
\end{figure*}

A cylindrical computational domain, as shown in Figure \ref{f1.pdf}, is used in the simulations. The radius of the cylinder is set at $L$ ($=12 a_b$), and the height of the cylinder is $[-L, L]$. The gas is injected at the boundary surfaces, i.e., the side, top, and bottom of the cylinder. The gas injection mimics a rigidly rotating, spherical infalling envelope around binary protostars. The gas has a density distribution of $\rho(r) = \rho_0 (r/L)^{-1.5}$ at the boundary surfaces, where $r$ denotes the radius in spherical coordinates. The power index of $-1.5$ is for the infalling envelope of the mass accretion phase. The injected gas has a constant angular velocity of $\Omega_\mathrm{inf}$ and a radial velocity of $v_r = (2 G M_\mathrm{tot}/r - (\Omega_\mathrm{inf} R)^2)^{1/2}$ at the boundary surfaces, assuming freefall from a distance of infinity. Here, $R$ denotes the cylindrical radius. The injected gas has a range of specific angular momenta $0 \le j \le j_\mathrm{inf}$, where $j_\mathrm{inf} = \Omega_\mathrm{inf} L^2$ is the maximum specific angular momentum of the infalling gas.

The barotropic equation of state is assumed, in which the gas pressure is given by $p = c_s^2 \rho [1 + (\rho/\rho_\mathrm{cr})^{2/5}]$. This equation of state is an approximation obtained from radiation transfer calculations \citep{Masunaga98} and is commonly used in numerical simulations for protostellar collapse \citep[e.g.,][]{Matsumoto17}. The gas is approximated as isothermal if the density is less than $\rho_\mathrm{cr}$ and polytropic \rev{with an adiabatic index $\gamma = 7/5$} if the density is larger than $\rho_\mathrm{cr}$. In this paper, the critical density of $\rho_\mathrm{cr}$ is set at $10^4 \rho_0$. Consequently, the gas is isothermal in an infalling envelope and a circumbinary disk, while it is polytropic in circumstellar disks.
According to \citet{Masunaga98}, the critical density is given by $\rho_\mathrm{cr} \sim 10^{-13}\,\mathrm{g\,cm}^{-3}$ (with the corresponding number density $n_\mathrm{cr} = 10^{10-11}\,\mathrm{cm}^{-3}$). The envelope gas is therefore assumed to have a density of $\rho_0 \sim 10^{-17}\,\mathrm{g\,cm}^{-3}$ (with the corresponding number density $n_0 = 10^{6-7}\,\mathrm{cm}^{-3}$), which is typical for a scale of $1000\,\mathrm{au}$ around protostars \citep{Onishi02,Tokuda16,Tokuda20}.
The barotropic equation of state considers the energy balance between compression heating and radiation cooling during protostellar collapse. In the isothermal region, this equation assumes that the timescale for radiation cooling is significantly shorter than that for heating processes, such as shock heating and compression heating due to gravity. While the barotropic equation of state accurately replicates the gas temperature in the central region of a collapsing gas cloud, it underestimates the temperature by a factor of $2-3$ on a scale of approximately $10\, \mathrm{au}$ when compared with results from radiation hydrodynamics \citep{Tomida10}.

The self-gravity of the gas is ignored, implying that the models here can be applied to a binary system where the total mass of the gas is much less than those of the stars. Models that take into account the gas-to-star interaction will be reported in a future paper.

Before starting an MHD simulation, a hydrodynamical calculation is performed for 10 rotation periods of the binary stars without magnetic fields to construct the initial condition. This hydrodynamical calculation provides a quasi-steady state of the circumbinary and circumstellar disks, as shown in \citet{Matsumoto19}, and the magnetic field is imposed on this state. The simulation clock is set at $t=0$ when the MHD calculation begins. The initial magnetic field is uniform and parallel to the $z$-direction, and its strength is denoted by $B_{z,0}$. We assume ideal-MHD in the evolution, and magnetic diffusion processes such as Ohmic dissipation and ambipolar diffusion are not considered. 
We also assume an inviscid gas, excluding any artificial viscosity and $\alpha$-viscosity, both in constructing the initial condition and during the evolution after the magnetic field imposition.

We adopt the units of $a_b=1$, $GM_\mathrm{tot}=1$, and $\rho_0=1$. The model parameters are the mass ratio of the binary stars $q=M_2/M_1$, the isothermal gas sound speed $c_s$, the maximum specific angular momentum of the infalling gas $j_\mathrm{inf}$, and the initial magnetic field strength $B_{z,0}$. In this paper, we examine models with $q=0.2$, $c_s=0.1$, $j_\mathrm{inf}=1.2$, and varying $B_{z,0}$ to investigate the effects of the magnetic field (Table~\ref{table:model-parameters}). 

The sound speed of $c_s=0.1$ corresponds to $0.1(GM_\mathrm{tot}/a_b)^{1/2}=0.30\,\mathrm{km\,s}^{-1}$ when assuming $M_\mathrm{tot}=1 M_\sun$ and $a_b=100\,\mathrm{au}$. 
The disk thickness is related to the sound speed, given by $H/R = c_s (R/GM_\mathrm{tot})^{1/2}$.
In the case of $c_s=0.1$, the disk thickness is $H/R = 0.1 (R/a_b)^{1/2}$. Note that the simulated circumbinary disk has a thickness $H \sim a_b$ (Figure~\ref{f3.pdf}), which is much thicker than that estimation, because of disturbance by the spiral arms.
The circumstellar disk around the primary star exhibits a density of $\rho \sim 10^5\rho_0$ at its thickest radius (approximately $0.3a_b$). Given the barotropic equation of state assumed here, the corresponding sound speed is $0.21 (GM_\mathrm{tot}/a_b)^{1/2} $. This leads to an estimated disk thickness of $H/R =  \rev{0.21} [(1+q) R/a_b]^{1/2} \sim \rev{0.13}$ \rev{when $q=0.2$}, which is \rev{consistent with, but smaller than,} the thickness of the simulated disk, approximately $1/3$. Note that this paper mainly focus on the circumbinary disk rather than the circumstellar disks.

The initial magnetic field of $B_{z,0}=0.1$ corresponds to $94.2\,\mu\mathrm{Gauss}$, under the same assumptions and $\rho_0=10^{-17}\,\mathrm{g\,cm}^3$, which corresponds to a number density of $n_0=2.61\times10^6\,\mathrm{cm}^{-3}$ for a mean molecular weight of 2.3. The Alfv\'en speed in the infalling envelope is therefore $v_A=B_{z,0}/(4\pi\rho_0)^{1/2}=0.028(GM_\mathrm{tot}/a_b)^{1/2}$, corresponding to $v_A=0.084\,\mathrm{km\,s}^{-1}$. 
The most unstable mode of MRI is given by $k_z v_A/\Omega \simeq 1$ for a wave number $k_z$ \citep{Balbus91}. In the case of $B_{0,z}=0.1$, the corresponding wavelength is $\lambda_\mathrm{max} = 0.1 \pi^{1/2} (\rho_0/\rho)^{1/2} (R^3/a_b)^{1/2}$ when assuming the Keplerian rotation. At the initial stage, the density in the circumbinary disk is $\rho \sim 10^2 \rho_0 - 10^3 \rho_0$, corresponding to $\lambda_\mathrm{max} \sim 0.01 a_b - 0.05a_b$ at $R = 2 a_b$. The wavelength is less than the disk thickness ($H \sim a_b$). Therefore, the circumbinary disk is unstable against the MRI. A typical cell width covering the circumbinary disk is $\Delta x = 0.023a_b - 0.0058a_b$ for $\ell = 2-4$ (see section \ref{sec:methods}), indicating that the most unstable MRI mode is marginally resolved at the initial stage. In Appendix~\ref{eq:numerical_resolution}, we investigate the numerical resolution for resolving the MRI through the evolution.

To investigate the effects of the infalling envelope, we also computed models without the infalling envelope, where gas injection from the boundary surfaces is halted at $t = 0$. These models are denoted as ``N'' in the ``Infall'' column of Table~\ref{table:model-parameters}. Additionally, we examined models of a single star ($q = 0$) to draw comparisons with the binary.


\begin{deluxetable}{llll}
\tablecaption{Model parameters \label{table:model-parameters}}
\tablehead{
\colhead{$q$} &
\colhead{$B_{z,0}$} &
\colhead{Infall} &
\colhead{Comments}
}
\startdata
0.2 & 0.4   & Y & Strong field model\\
0.2 & 0.1   & Y & Fiducial model\\
0.2 & 0.025 & Y & Weak field model\\
0.2 & 0.01  & Y & Very weak field model\\
0.2 & 0     & Y & HD model\\
0.2 & 0.1   & N & Fiducial model w/o infall envelope\\
0.2 & 0     & N & HD model w/o infalling envelope\\
0   & 0.1   & Y & Single star model\\
0   & 0     & Y & Single star HD model\\
0   & 0.1   & N & Single star w/o infall envelope\\
0   & 0     & N & Single star HD model w/o infall envelope\\
\enddata
\end{deluxetable}

\section{Methods}
\label{sec:methods}

The numerical simulations were performed using the adaptive mesh refinement code, \texttt{SFUMATO} \citep{Matsumoto07}, which employs fixed mesh refinement (FMR) with a static hierarchical grid configuration shown in Figure~\ref{f1.pdf}. The computation domain is covered by $16^3$ blocks for a grid level of $\ell = 0$ (the base grid), and each block has $16^3$ cubic cells. The base grid has therefore a resolution of $256^3$ cells, and the cell width is $\Delta x_\mathrm{max} = 0.0938$.  The maximum grid level is set at $\ell_\mathrm{max} = 4$, and the minimum cell width is $\Delta x_\mathrm{min} = 0.00586$.

The simulations use a MHD equation solver with a third-order accuracy in space using the MUSCL method and second-order accuracy in time with the predictor-corrector method. During the course of the simulations, cells with extremely low density can appear, leading to an extremely high Alfv\'en speed and consequently, a short timestep. To solve this issue, a barotropic version of the Boris-HLLD scheme \citep{Matsumoto19HLLD} is adopted to calculate the numerical flux. This scheme is based on the HLLD Riemann solver \citep{Miyoshi05} and modified to solve the MHD equation with Boris correction \citep{Gombosi02}. The modified scheme suppresses the Alfv\'en speed below the reduced speed of light, which should be several times larger than the maximum velocity in the computation domain, $|\mathbf{v}|_\mathrm{max}$, for a stable calculation \citep{Matsumoto19HLLD}. The reduced speed of light $c_\mathrm{red}$ is varied according to the gas velocity as $c_\mathrm{red} = \max(4 |\mathbf{v}|_\mathrm{max}, 50)$ to avoid a very short timestep when the density becomes very low, allowing for long-term evolution of the simulations.

The sink particles are used to represent the binary stars in the simulations. These sink particles accrete gas within a sphere of radius $r_\mathrm{sink}$ and density higher than $\rho_\mathrm{sink}$ \citep{Matsumoto15}. We set $r_\mathrm{sink} = 4 \Delta x_\mathrm{min} = 0.0234 a_b$ for the sink radius, which is much smaller than the binary star separation, so the sink particles have little hydrodynamical impact on the circumstellar disk. Despite the $r_\mathrm{sink}$ assumed here being only four times larger than the cell size, influences from finite cell size, such as quadrupole structures, were not observed around the sink regions in the models. The density threshold for accretion onto the sink particles is set to $\rho_\mathrm{sink} = 10^6 \rho_0$, which is a typical density at the center of the circumstellar disks. Although the sink particles accrete mass and angular momentum, their masses and orbits remain fixed during the evolution.

Even though the sink particles accrete gas within the sink regions, gas with a density below the threshold $\rho_\mathrm{sink}$ remains. If this gas rotates and is coupled with the magnetic fields, outflows would be artificially launched from the sink regions. To avoid this issue, we impose Ohmic dissipation only in the sink regions with a high resistivity, $\eta = v_K r_\mathrm{sink}$, where $v_K = (G M_\mathrm{sink} / r_\mathrm{sink})^{1/2}$ is the Keplerian velocity for the sink particle with a mass of $M_\mathrm{sink}$. With this measure, the artificial launching of the outflows was not observed inside the sink regions.

During the accretion process of a sink particle, we extract the gas with a density exceeding $\rho_\mathrm{sink}$ within the sink region, while the magnetic field in that area is retained. This process results in an accumulation of magnetic flux in the sink region when the magnetic field lines have a open configuration, as it is in our present models. A more realistic treatment of magnetic fields related to a sink particle will be studied in a future paper.

\begin{figure*}[t]
\plotone{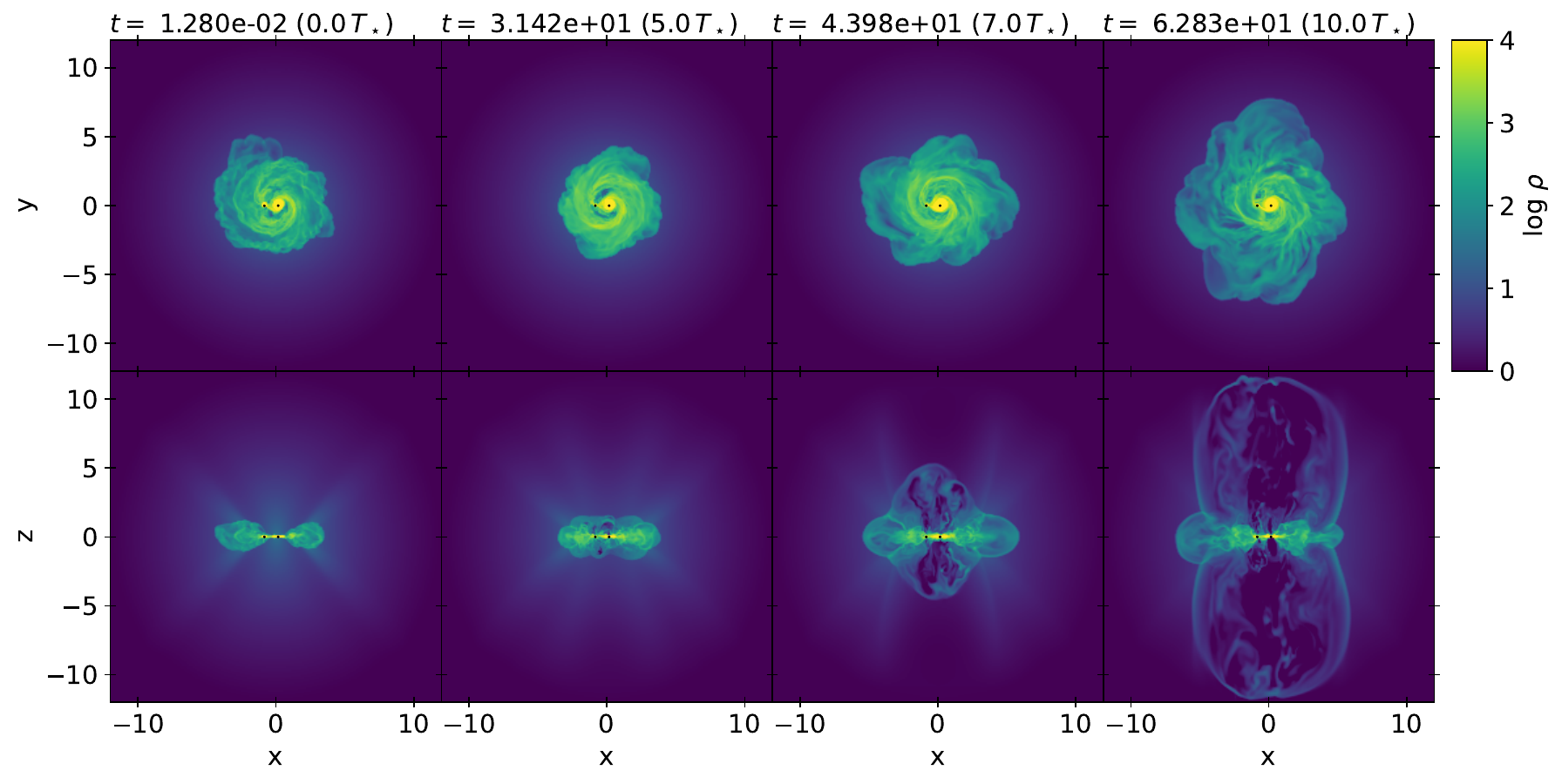}
\caption{Evolution of the fiducial model ($q = 0.2$, $B_{z,0} = 0.1$, with the infalling envelope) at $t = 0$ (the initial condition), $5 T_\star$, $7 T_\star$, and $10 T_\star$ from left to right. The colors show the logarithmic density distributions in the $z=0$ plane (upper panels) and the $y=0$ plane (lower panels). The black circles represent the sink particles, with the right and left particles corresponding to the primary and secondary stars, respectively. The radii of the circles are equal to the sink radius, $r_\mathrm{sink}$. The entire computational domain is shown.
An animation is available in the HTML version.
 \label{f2.pdf}}
\end{figure*}

\begin{figure*}[t]
\plotone{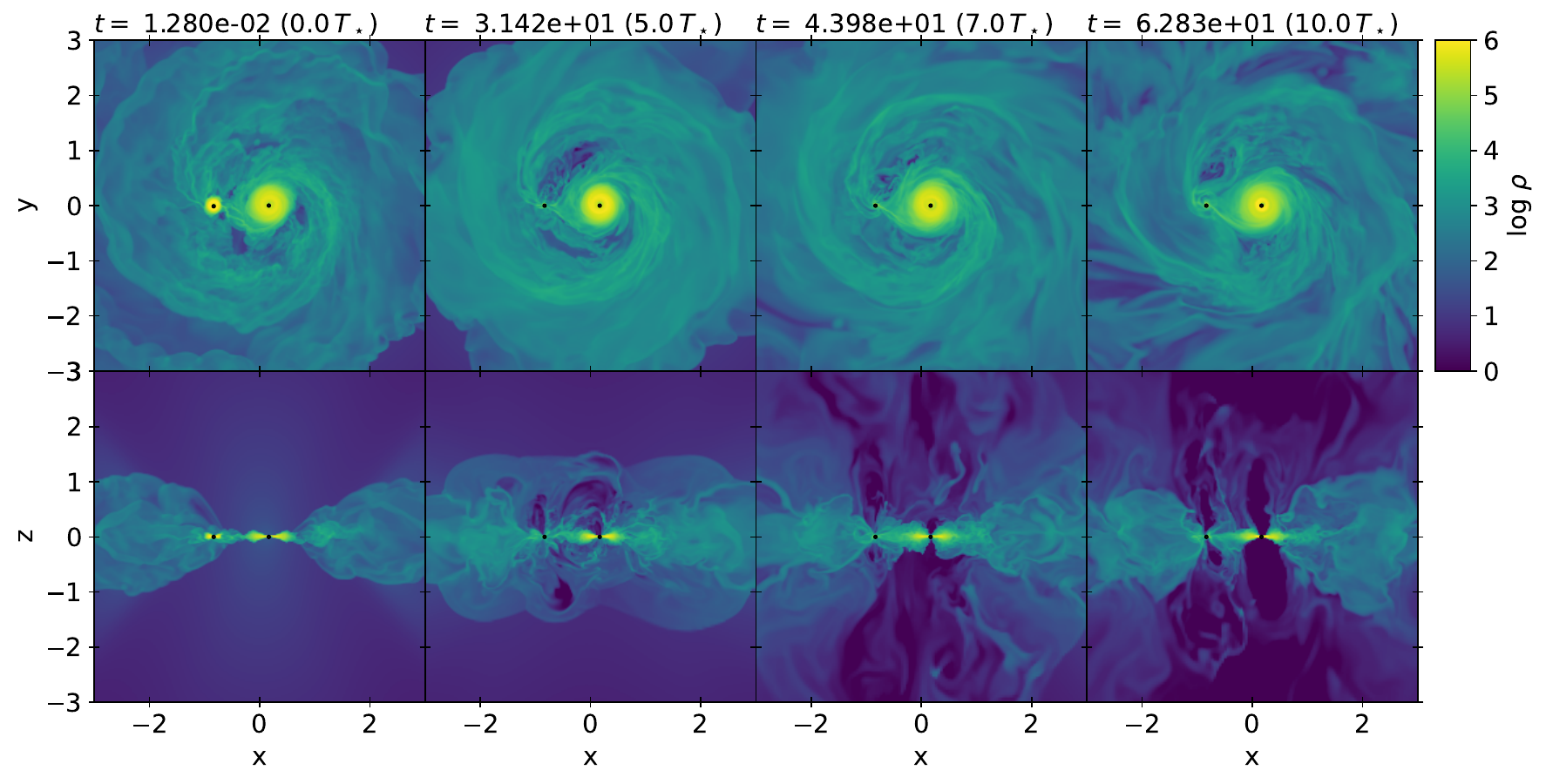}
\caption{Same as Figure~\ref{f2.pdf}, but the magnifications of the central region of $[-3, 3]^2$. 
An animation is available in the HTML version.
 \label{f3.pdf}}
\end{figure*}

The circumstellar and circumbinary disks in the simulations launch outflows that eventually reach the top and bottom surfaces of the cylindrical boundary. At this point, the boundary condition is switched from the gas-injection to the out-going boundary condition for outflows by applying the zero-gradient boundary condition. While the out-going boundary condition allows the outflows to flow out of the computational domain, some waves are reflected by the boundary and return to the computational domain. These reflections have little impact on the evolution of the disks, as the outflows reach the boundaries only in the very last stages (see the lower right panel of \ref{f2.pdf} for the fiducial model).

\section{Results}
\label{sec:results}

\subsection{Overall evolution of the fiducial model}

Evolution of the fiducial model with $B_{z,0} = 0.1$ is shown in Figures~\ref{f2.pdf} and \ref{f3.pdf}. At the initial stage ($t = 0$), each star is surrounded by a circumstellar disk, and the binary stars are surrounded by a circumbinary disk. In the circumbinary disk, two spiral arms are excited because of the gravitational torque of the binary stars. The initial condition is constructed by a pure hydrodynamical calculation. Unlike typical binary accretion models \citep[e.g.,][]{Moody19}, this model has no clear cavity or gap at the initial stage. This is because the gas accreting in the vertical direction has small angular momenta to fill the cavity. At $t = 0$, a uniform magnetic field is imposed, and the MHD calculation starts.

The two spiral arms continue to exist in the circumbinary disk during the MHD calculation. The circumbinary disk gradually extends due to the angular momentum redistribution by the magnetic field. The circumstellar disk around the primary star remains while the disk around the secondary star becomes obscured due to magnetic braking. 
This phenomenon can be attributed to differences in magnetic field fluxes within the circumstellar disks. The secondary star experiences a higher accretion rate compared to the primary star (see Figure~\ref{f21.pdf}). As a result, the secondary circumstellar disk accumulates more magnetic flux than the primary circumstellar disk, leading to more significant angular momentum transport due to stronger magnetic braking in the secondary circumstellar \rev{disk}. 

\begin{figure*}[t]
\plotone{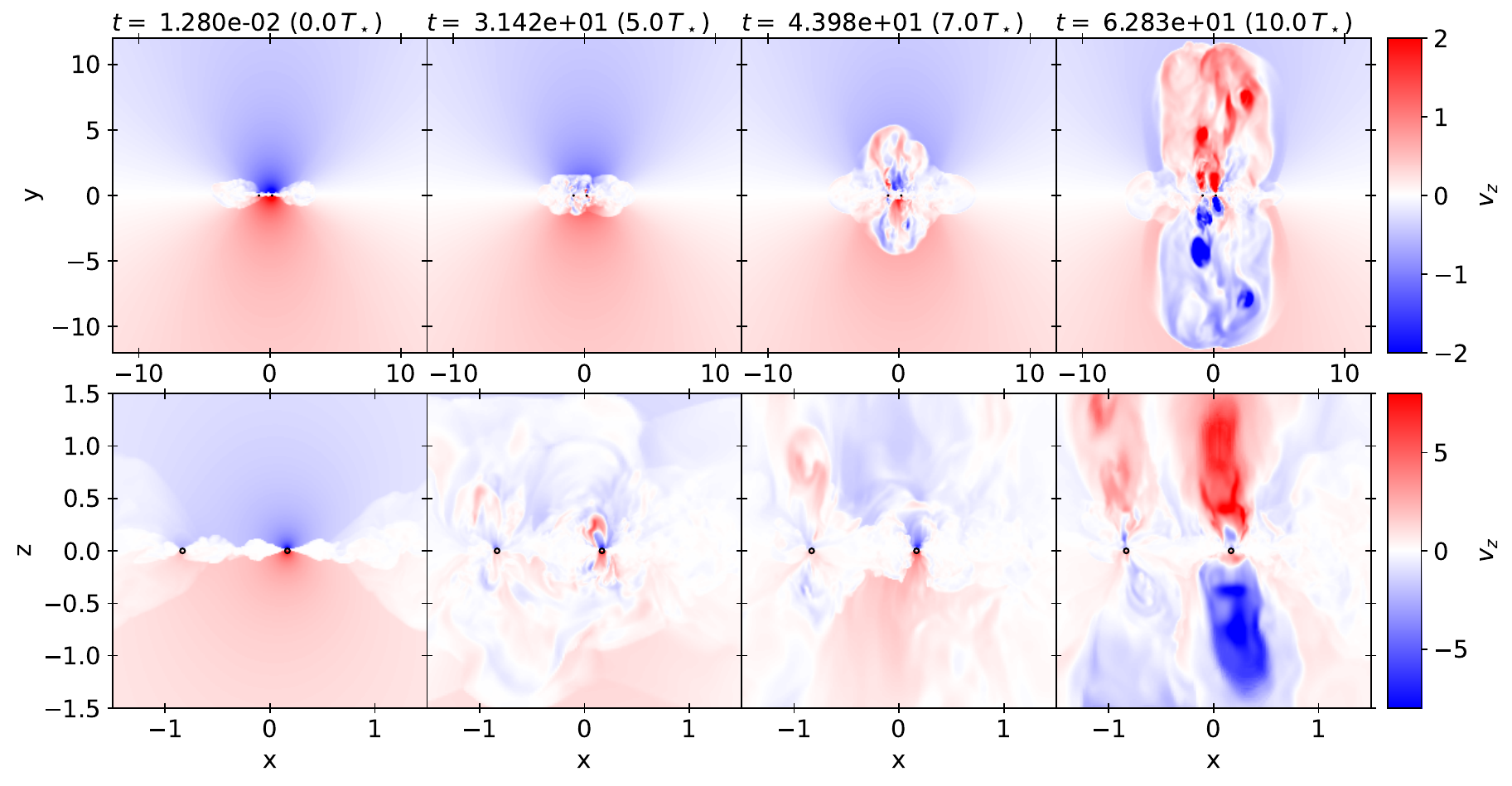}
\caption{Evolution of the outflows for the fiducial model ($q = 0.2$, $B_{z,0} = 0.1$, with the infalling envelope) at $t = 0$ (the initial condition), $5 T_\star$, $7 T_\star$, and $10 T_\star$ from left to right. The colors indicate the distributions of $v_z$ in the $z=0$ plane. The lower panels are the magnifications of the central region of $[-1.5, 1.5]^2$. The black circles show the sink particles; the right and left sink particles correspond to the primary and secondary stars, respectively. The radii of the circles are the sink radius, $r_\mathrm{sink}$.
 \label{f4.pdf}}
\end{figure*}

\begin{figure*}[t]
\epsscale{0.4}
\plotone{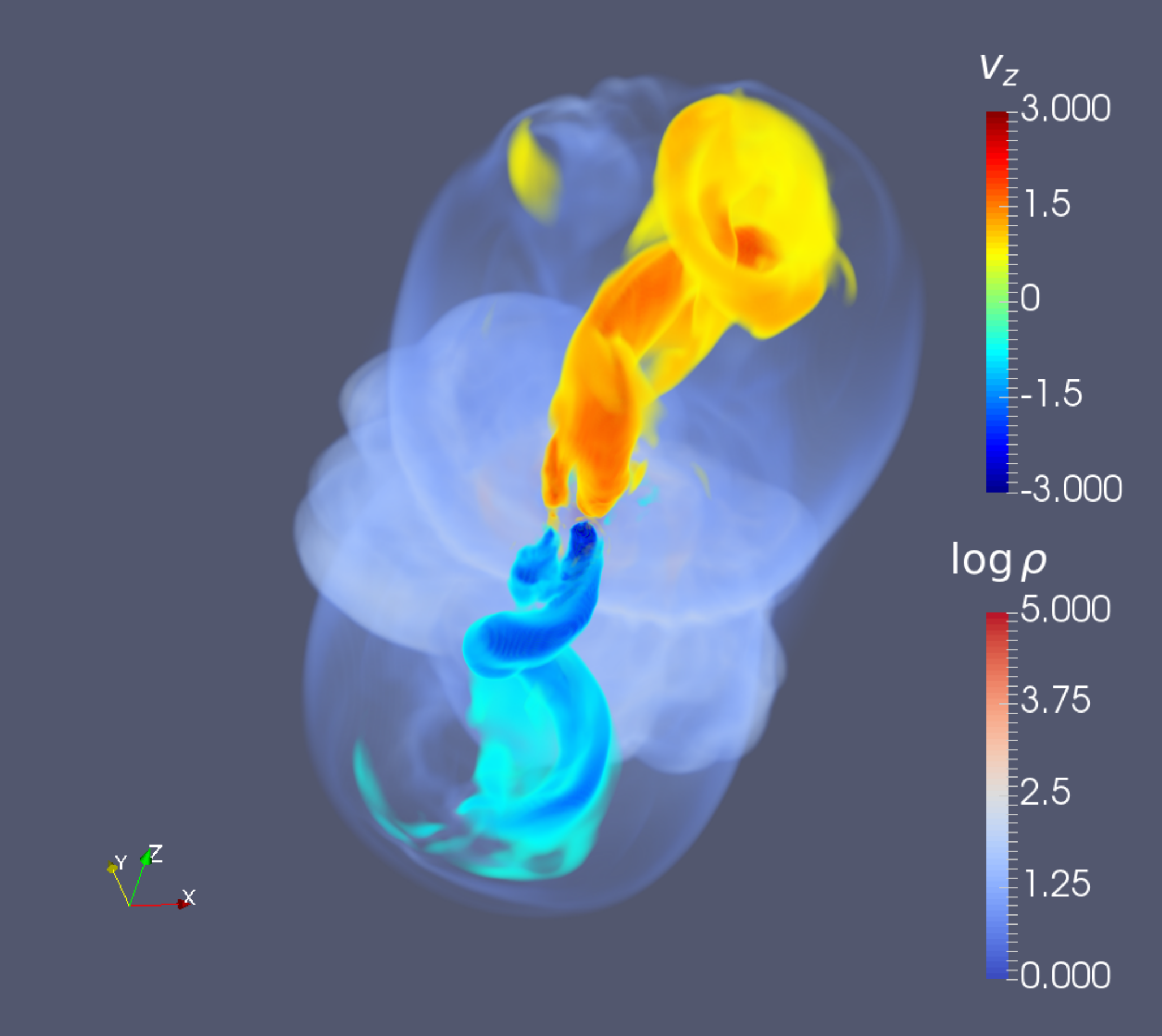}
\plotone{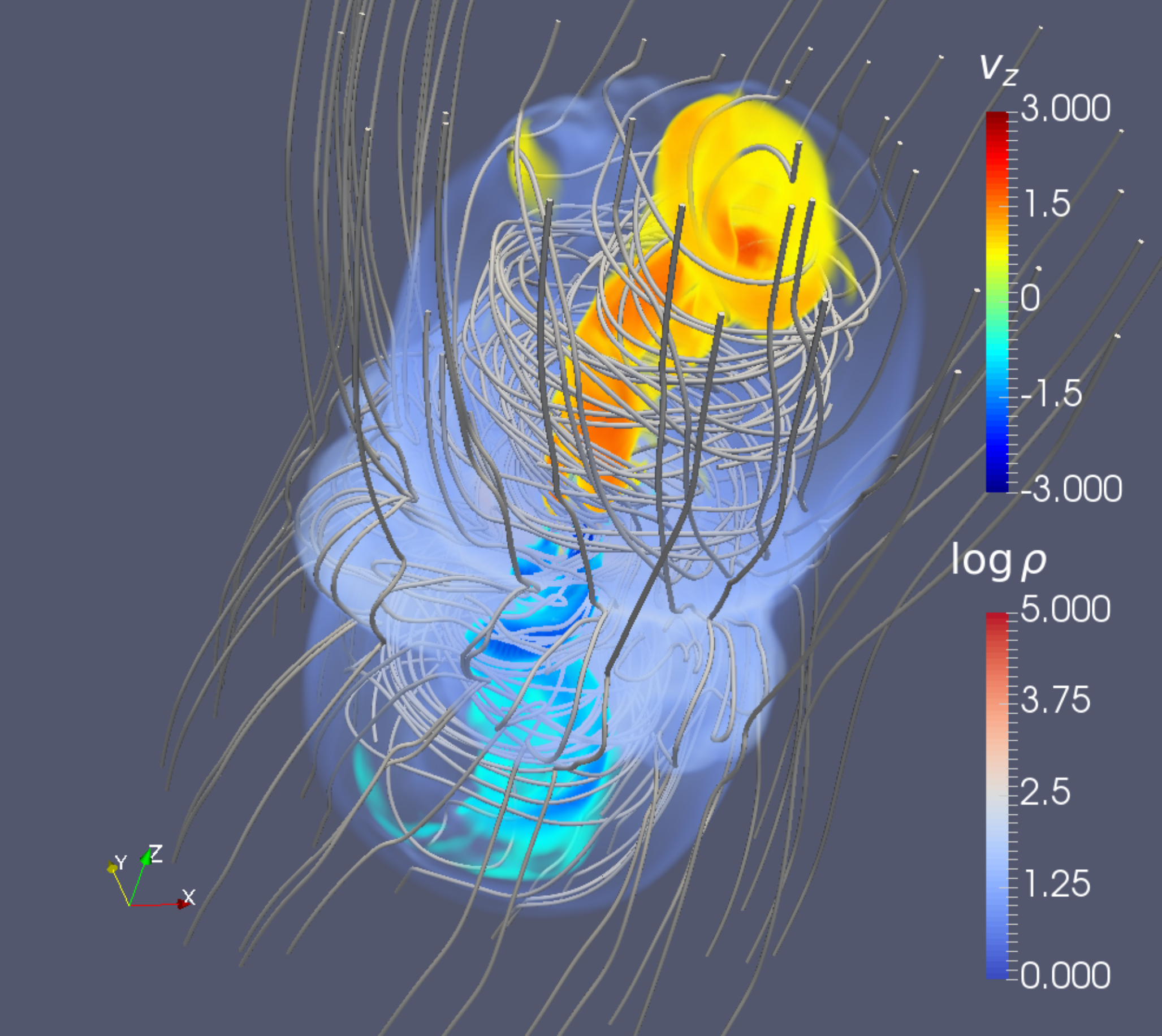}\\
\plotone{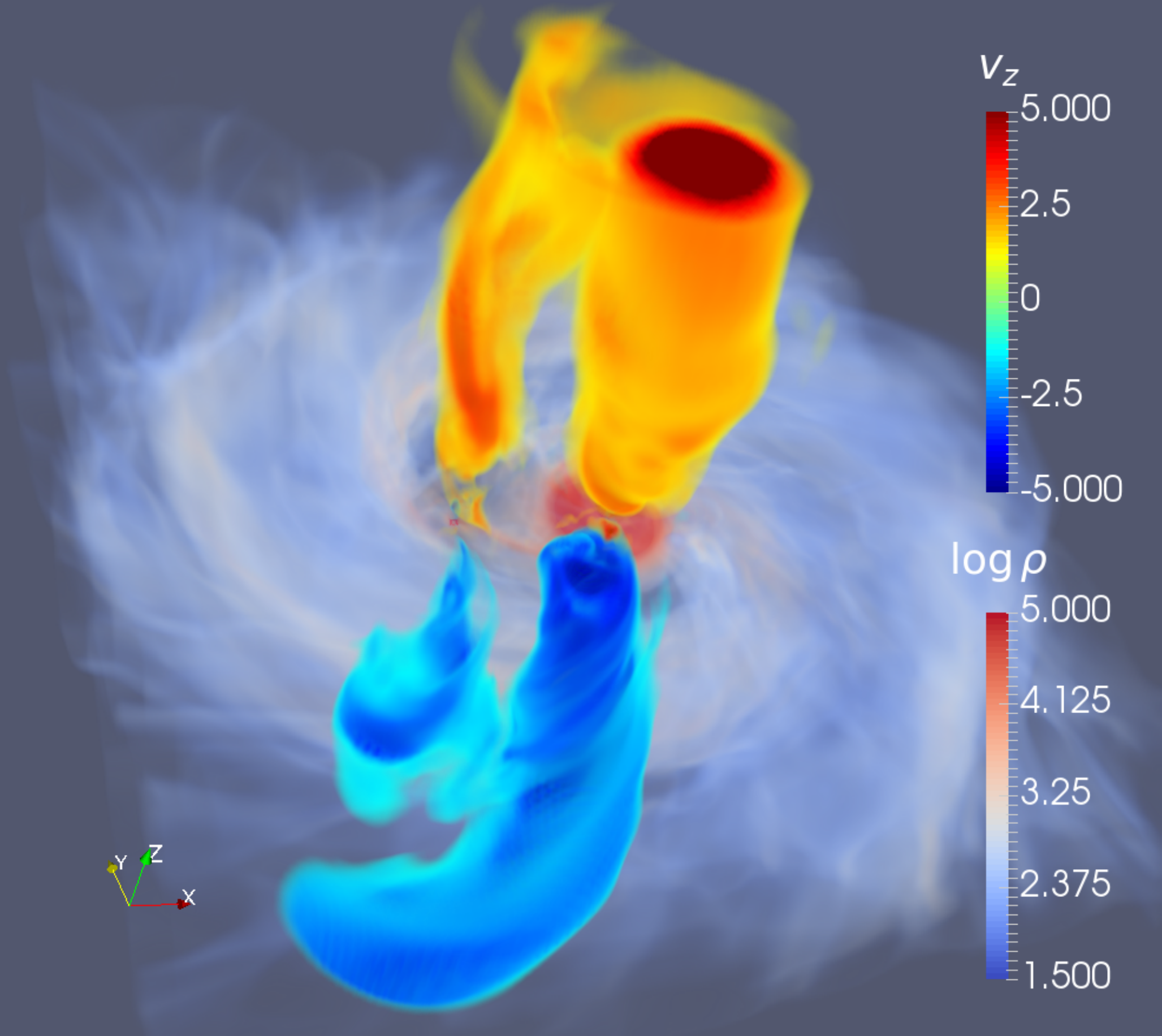}
\plotone{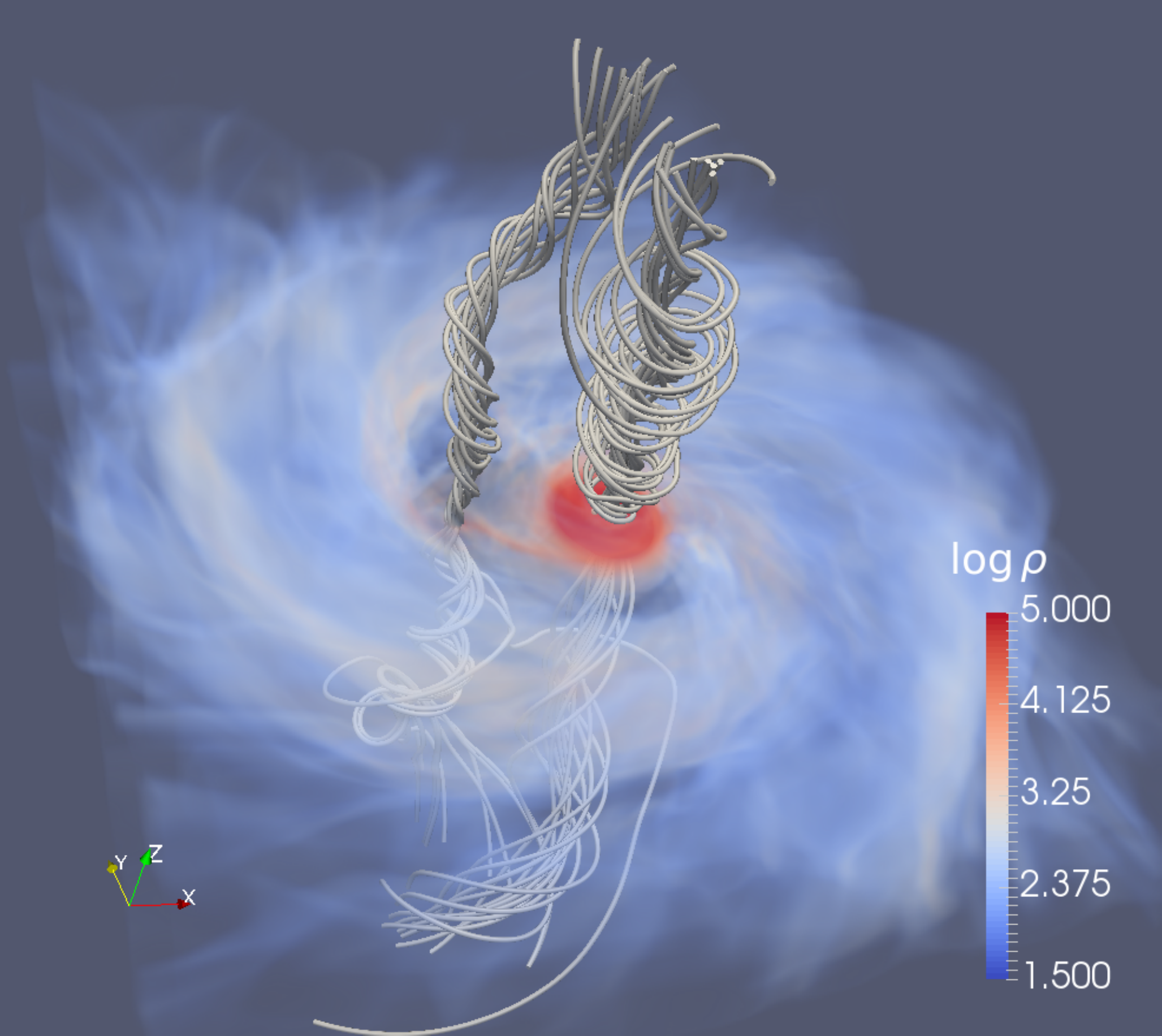}
\caption{3D views of the fiducial model ($q = 0.2$, $B_{z,0} = 0.1$, with the infalling envelope) at $t = 10 T_\star$. The volume rendering displays the gas density distribution on the logarithmic scale. The bipolar structures in red and blue indicate the velocity distributions of the vertical component ($v_z$), revealing that the outflows from the circumstellar disks of the primary and secondary stars propagate in both positive and negative $z$-directions. The upper-right panel displays the same view as the upper-left panel, but with magnetic field lines depicted as gray tubes. The upper panels show the entire computational domain of $[-12, 12]^3$, while the lower panels are magnifications around the binary stars, showing the region of $[-3, 3]^3$.
 \label{vol_field_vel.pdf}}
\end{figure*}

The outflows begin to be launched from the two circumstellar disks at $t \sim 5 T_\star$ (Figure~\ref{f3.pdf}). They propagate in the vertical direction and reach the top and bottom boundary surfaces of the computational domain at $t = 10 T_\star$ (Figure~\ref{f2.pdf}). At the launch points at $t = 10 T_\star$, the outflow from the primary star has a higher velocity ($v_z = \pm (5-8)$) than that from the secondary star ($v_z = \pm (1-3)$) (Figure~\ref{f4.pdf}). 
As showin in the upper left panel of Figure~\ref{vol_field_vel.pdf}, the outflows shown by the red and blue lobes are also twisted due to the orbital motion of the binary stars.
As shown in the lower right panel of Figure~\ref{vol_field_vel.pdf}, the magnetic field lines are tightly twisted along the outflow directions, suggesting that magnetic pressure plays a significant role in accelerating the outflows \citep{Tomisaka02,Machida08,Tomida10}. The higher angular velocity at the inner parts of the circumstellar disks results in outflows being preferentially launched from these regions, while in the sink region, where the gas is decoupled due to high resistivity, outflows are not initiated (see Section~\ref{sec:methods}). Although centrifugal winds, driven by less twisted magnetic field lines, represent another outflow mechanism \citep{Blandford82,Pudritz86}, they are not relevant to this model.

In addition to the fast two outflows stated above, a slow outflow is launched from the circumbinary disk. The slow outflow has velocity of $v_z = \pm (0.2-0.5)$ as shown in Figure~\ref{f4.pdf}. The slow outflow has a cocoon-like shape with highly twisted magnetic fields as shown in the upper right panel of Figure~\ref{vol_field_vel.pdf}. The rotation of the circumbinary disk is responsible for the twisted magnetic field, which implies that the magnetic pressure accelerates the outflow.

\begin{figure*}[t]
\plotone{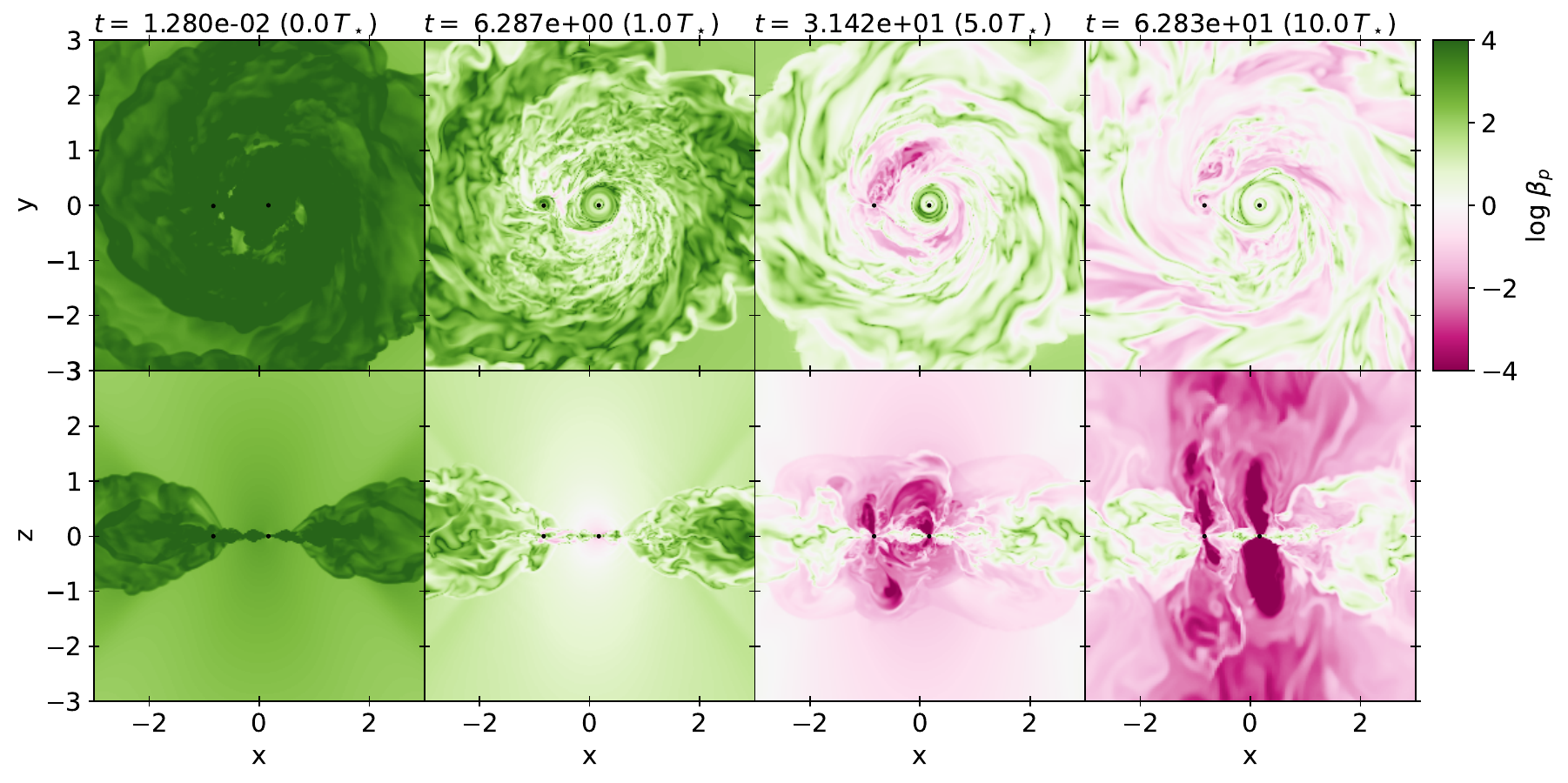}
\caption{Evolution of the plasma $\beta$ for the fiducial model ($q = 0.2$, $B_{z,0} = 0.1$, with the infalling envelope) at $t = 0$ (the initial condition), $T_\star$, $5 T_\star$, and $10 T_\star$ from left to right. The colors show the plasma $\beta$ distributions on the logarithmic scale in the $z=0$ plane (upper panels) and the $y=0$ plane (lower panels). The black circles show the sink particles; the right and left sink particles correspond to primary and secondary stars, respectively. The radii of the circles are the sink radius, $r_\mathrm{sink}$. The central region of $[-3, 3]^2$ is shown. 
 \label{f6.pdf}}
\end{figure*}

Figure~\ref{f6.pdf} shows the evolution of the plasma $\beta$, the ratio of thermal pressure and magnetic pressure denoted by $\beta_p$. For the fiducial model, the infalling gas has $\beta_p \simeq 8 \pi c_s^2 \rho_0 / B_{z,0}^2 = 8 \pi$ due to $c_s = 0.1$ and $B_{z,0} = 0.1$.
At the initial stage ($t = 0$), the circumbinary disk has a typical value of $\beta_p \sim 10^3 - 10^4$ because the density in the disk reaches $(10^2 - 10^3) \rho_0$.
As time passes, $\beta_p$ decreases because the magnetic field strength increases.

The increase in the magnetic field is caused by mainly three processes. The first process is accretion due to the infalling envelope, which brings not only gas but also magnetic flux into the computational domain. This accretion process considerably increases the magnetic flux.
The second process is amplification of the magnetic field by rotation of disks. The rotation of the circumbinary disk and circumstellar disks amplify the toroidal magnetic field, which are associated with the slow and fast outflows, respectively, as shown in Figure ~\ref{vol_field_vel.pdf}. In the fast and slow outflow regions, the toroidal magnetic field dominates over the poloidal fields. 
Observations by \citet{Ching16,Lee18} suggest similar magnetic field structures associated with outflows, where the toroidal magnetic fields are observed wrapping around the outflows.

\begin{figure*}[t]
\epsscale{0.4}
\plotone{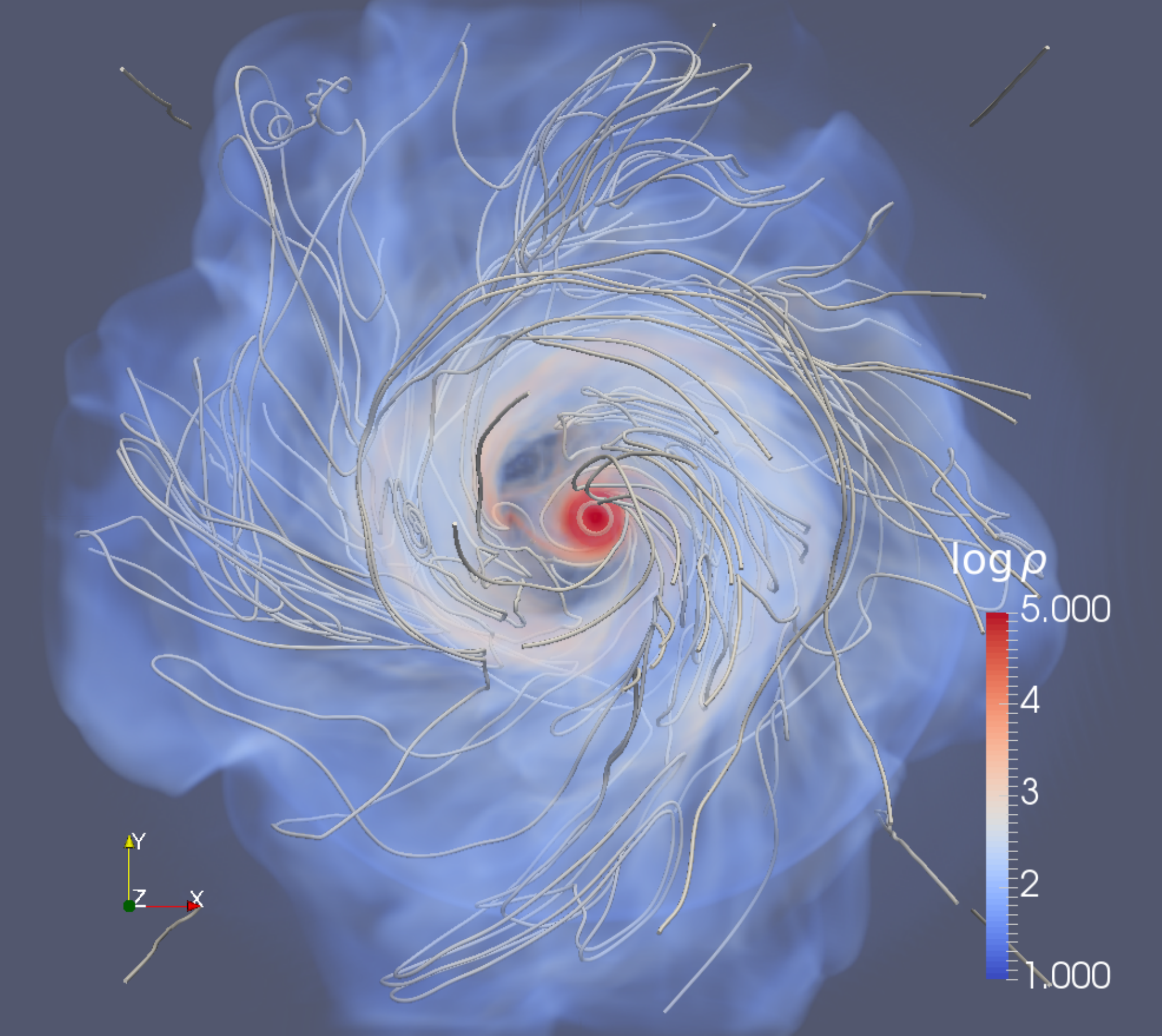}
\plotone{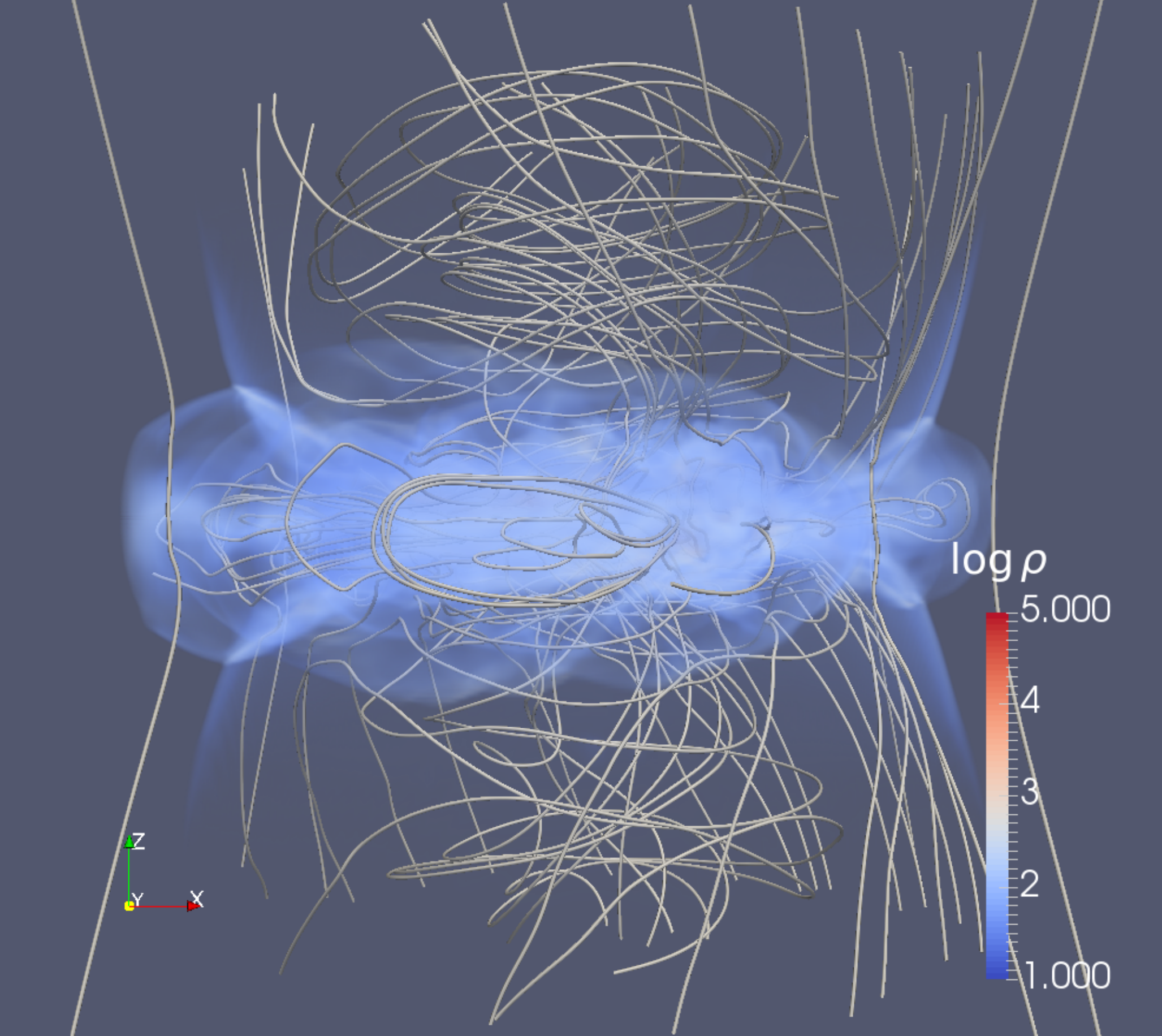}
\caption{
Face-on view (left panel) and edge-on view (right panel) of the fiducial model ($q = 0.2$, $B_{z,0} = 0.1$, with the infalling envelope) at $t = 10 T_\star$. The volume rendering shows the gas density distribution on the logarithmic scale. The gray tubes show the magnetic field lines.  The left panel shows a region of $x, y, z \in [-6, 6]^2\times[-2, 2]$, and the right panel shows a region of $[-6, 6]^3$.
\label{f7a.pdf}}
\end{figure*}

The third process is the amplification of the magnetic field by turbulence in the circumbinary disks. The MRI excites turbulent flows in the circumbinary disk, as shown in the fluctuated density distribution (Figure~\ref{f3.pdf}). Figure~\ref{f7a.pdf} shows magnetic field lines entangled within the circumbinary disk, indicating the development of MRI. Furthermore, the magnetic field lines have an overall spiral structure due to the redistribution of angular momentum. 
The spiral magnetic fields contribute to the spoke-like density structures observed in the face-on views of Figures~\ref{f2.pdf} and \ref{f3.pdf}. These spoke-like density structures are likely caused by channel flows, which are commonly observed in MRI simulations \citep[e.g.,][]{Sano01}. 
Layered accretion also occurs on the surface of the circumbinary disk, dragging the magnetic field lines inward in the cylindrical radial direction, as shown in the upper-right panel of Figure~\ref{vol_field_vel.pdf}. Due to this process of angular momentum redistribution, the gas near the mid-plane slowly moves outward in the cylindrical radial direction, leading to the expansion of the outer part of the disk. Similar processes are seen in MHD simulations of an accretion disk around a single star with vertical magnetic fields \citep{Suzuki14}.

\begin{figure*}[t]
\epsscale{0.5}
\plotone{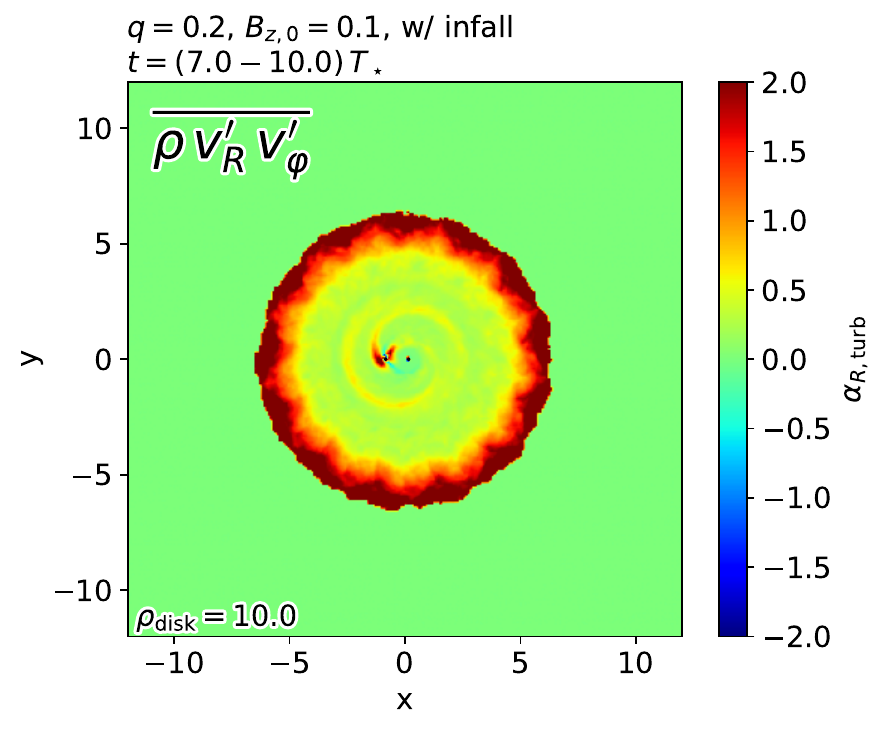}
\plotone{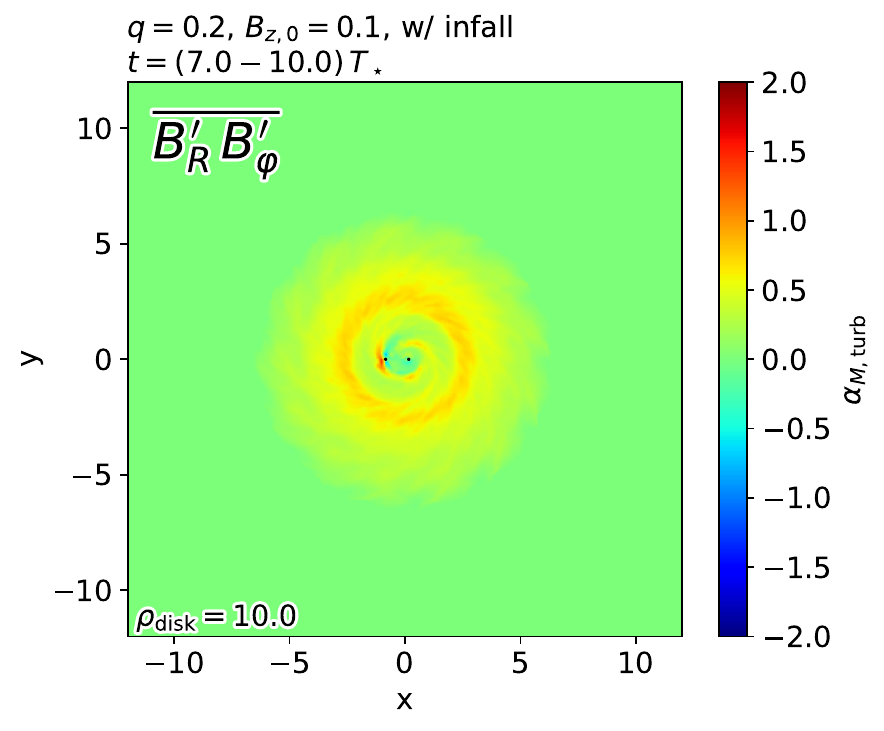}\\
\plotone{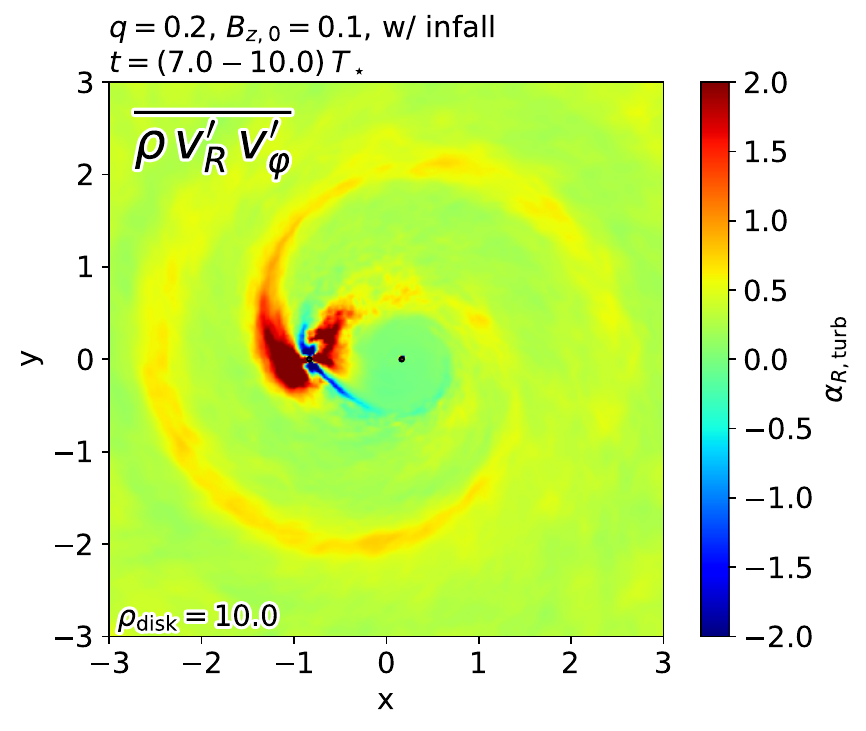}
\plotone{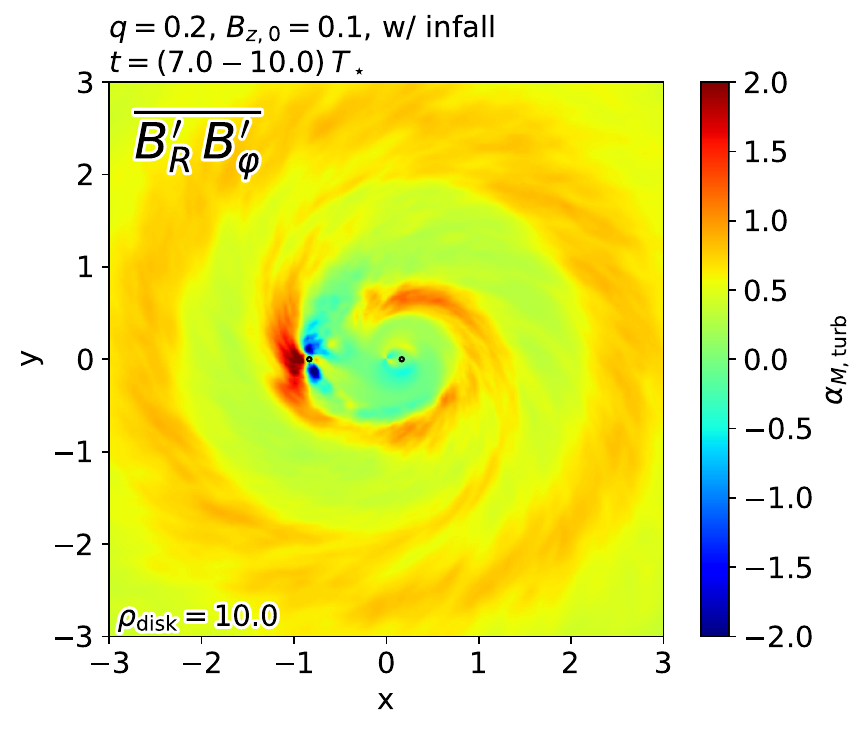}
\caption{
The $\alpha$-parameters of the turbulent components of the Reynolds stress $\alpha_{R, \mathrm{turb}}$ (left panels) and Maxwell stress $\alpha_{M, \mathrm{turb}}$ (right panels) for the fiducial model ($q = 0.2$, $B_{z,0} = 0.1$, with the infalling envelope). They correspond to the $\alpha$-parameters for $\overline{\rho v_R^\prime v_\varphi^\prime}$ and $\overline{B_R^\prime B_\varphi^\prime}$ components, respectively. The lower panels are a magnification of the upper panels. The time average is taken in the period of $t \in [7T_\star, 10T_\star]$. The density of the disk surface is set at $\rho_\mathrm{disk} = 10 \rho_0$. The black circles show the sink particles; the right and left sink particles correspond to the primary and secondary stars, respectively.
\label{f8d.pdf}}
\end{figure*}

\begin{figure*}
\epsscale{0.5}
\plotone{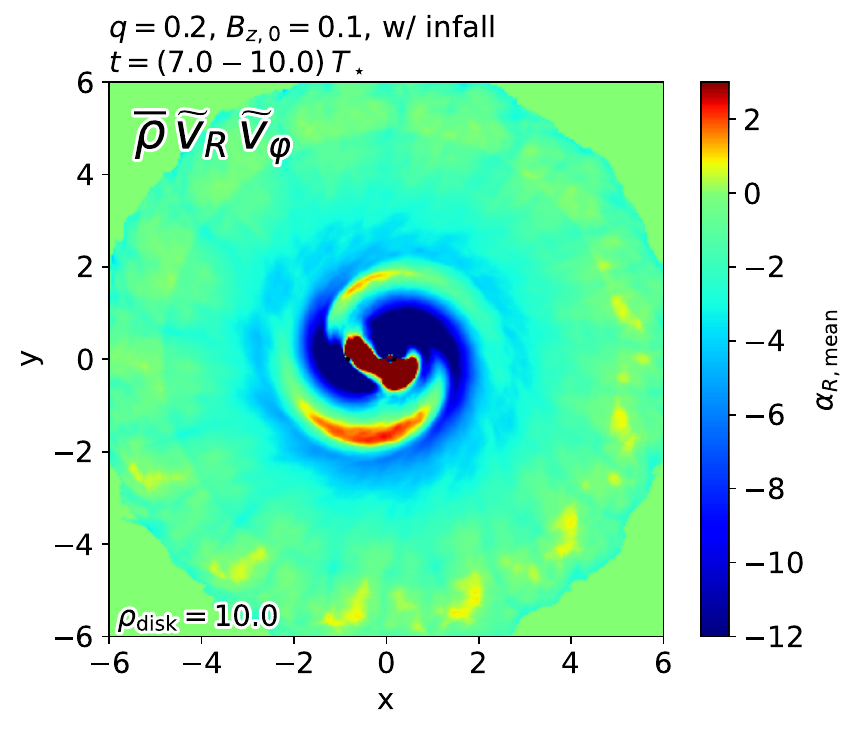}
\plotone{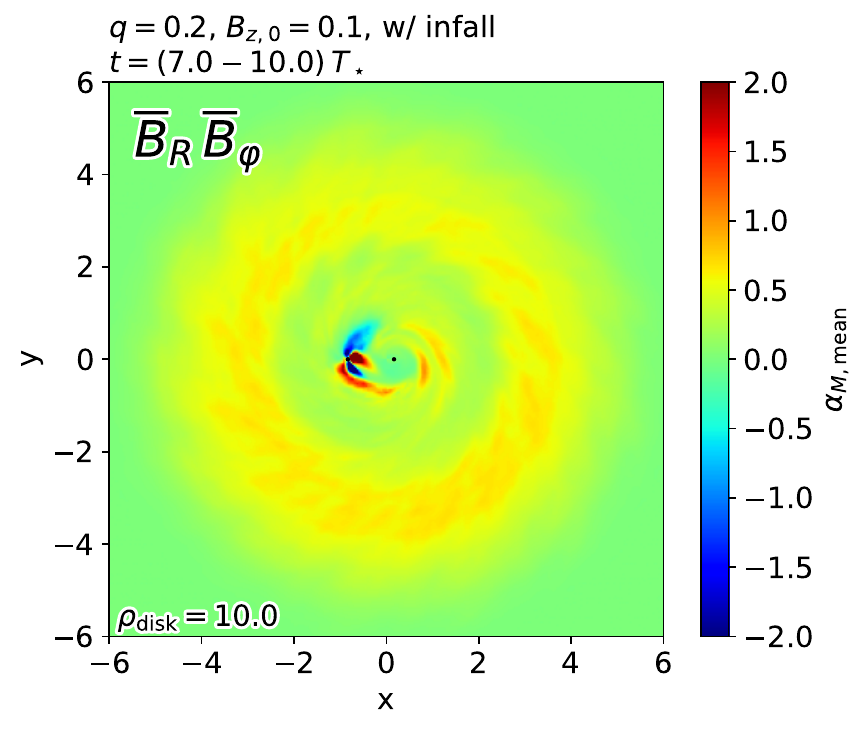}
\caption{
The $\alpha$-parameters of the mean flow/field components of the Reynolds stress $\alpha_{R, \mathrm{mean}}$ ($\overline{\rho} \widetilde{v}_R \widetilde{v}_\varphi$ component) (left panels) and the Maxwell stress $\alpha_{M, \mathrm{mean}}$ ($\overline{B}_R \overline{B}_\varphi$ component) (right panels) for the fiducial model ($q = 0.2$, $B_{z,0} = 0.1$, with the infalling envelope). The time average is taken in the period of $t \in [7T_\star, 10T_\star]$. The color scales are different on the left and right panels.
  \label{f9a.pdf}}
\end{figure*}

\begin{figure}
\epsscale{1.0}
\plotone{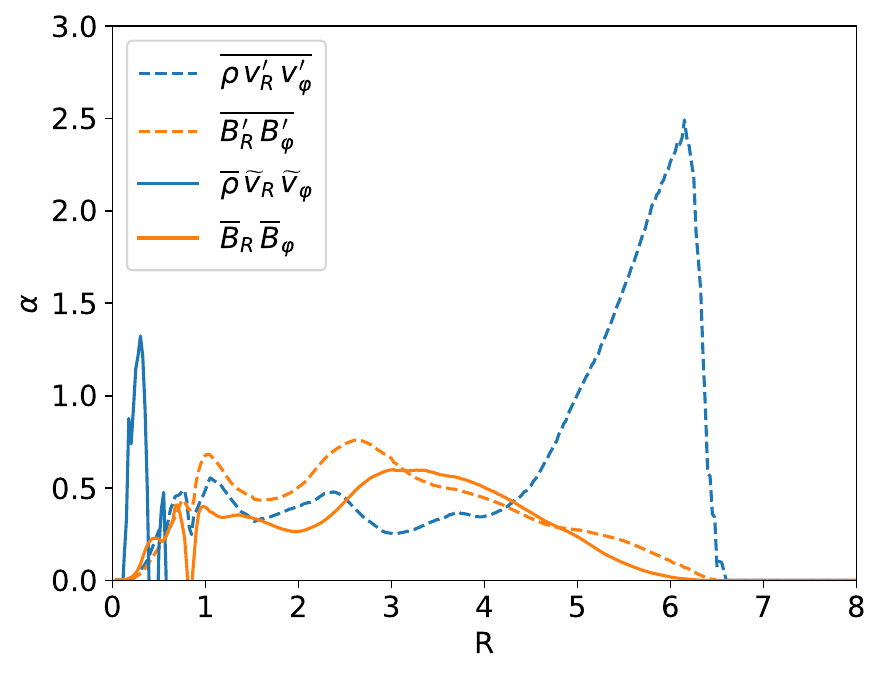}\\
\plotone{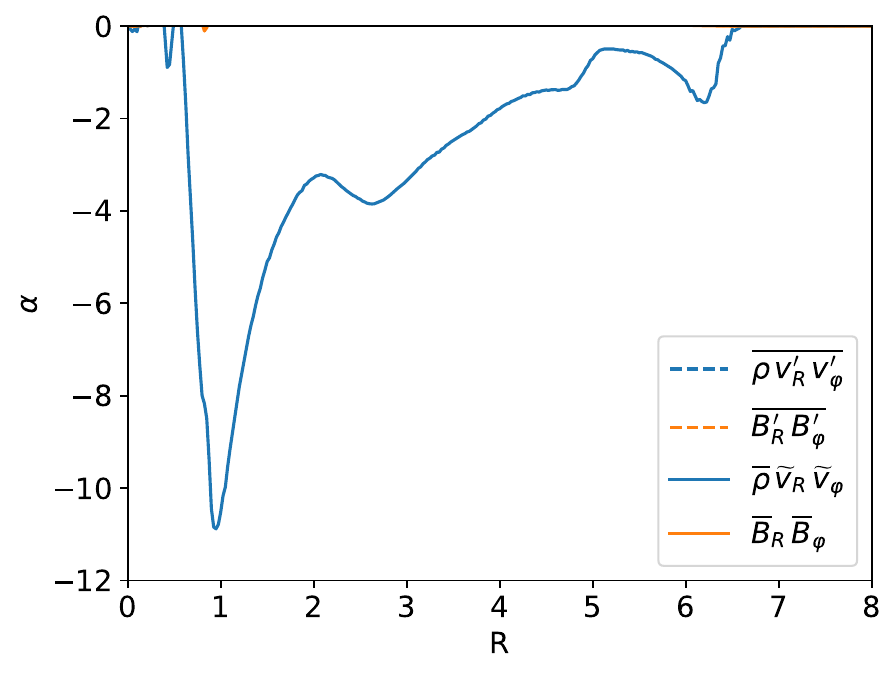}  
\caption{
The distribution of $\alpha$-parameters as a function of radius for the fiducial model ($q = 0.2$, $B_{z,0} = 0.1$, with the infalling envelope). The radial distribution is obtained by averaging the $\alpha$-parameters in the azimuth direction. The time average is taken in the period of $t \in [7T_\star, 10T_\star]$. The upper panel shows the positive values of $\alpha$, while the lower panel shows the negative values.
\label{f10a.pdf}}
\end{figure}

\subsection{Turbulence in the circumbinary disks}

The turbulence is generated in the circumbinary disk. In order to investigate turbulent level, we measure the $\alpha$-parameters for the Reynolds and the Maxwell stresses of the turbulent components, $\alpha_{R,\mathrm{turb}}$ and $\alpha_{M,\mathrm{turb}}$, according to Appendix~\ref{seq:estimate_alpha_parameters}.

\begin{figure*}[t]
\epsscale{0.9}
\plotone{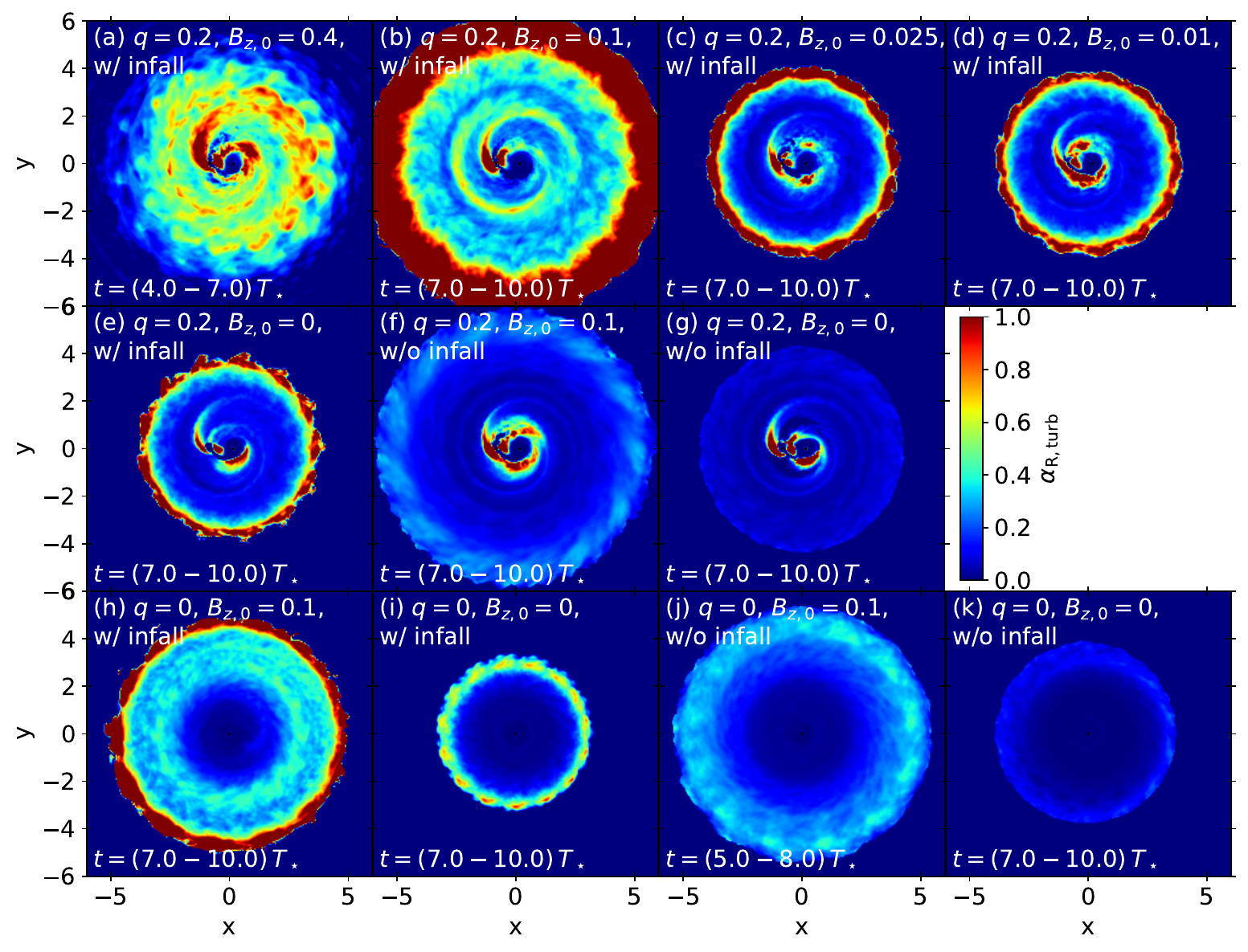}
\caption{The $\alpha$-parameters of the turbulent component the Reynolds stress $\alpha_{R, \mathrm{turb}}$ ($\overline{\rho v_R^\prime v_\varphi^\prime}$ component) for all the models.
\label{f15.pdf}}

\plotone{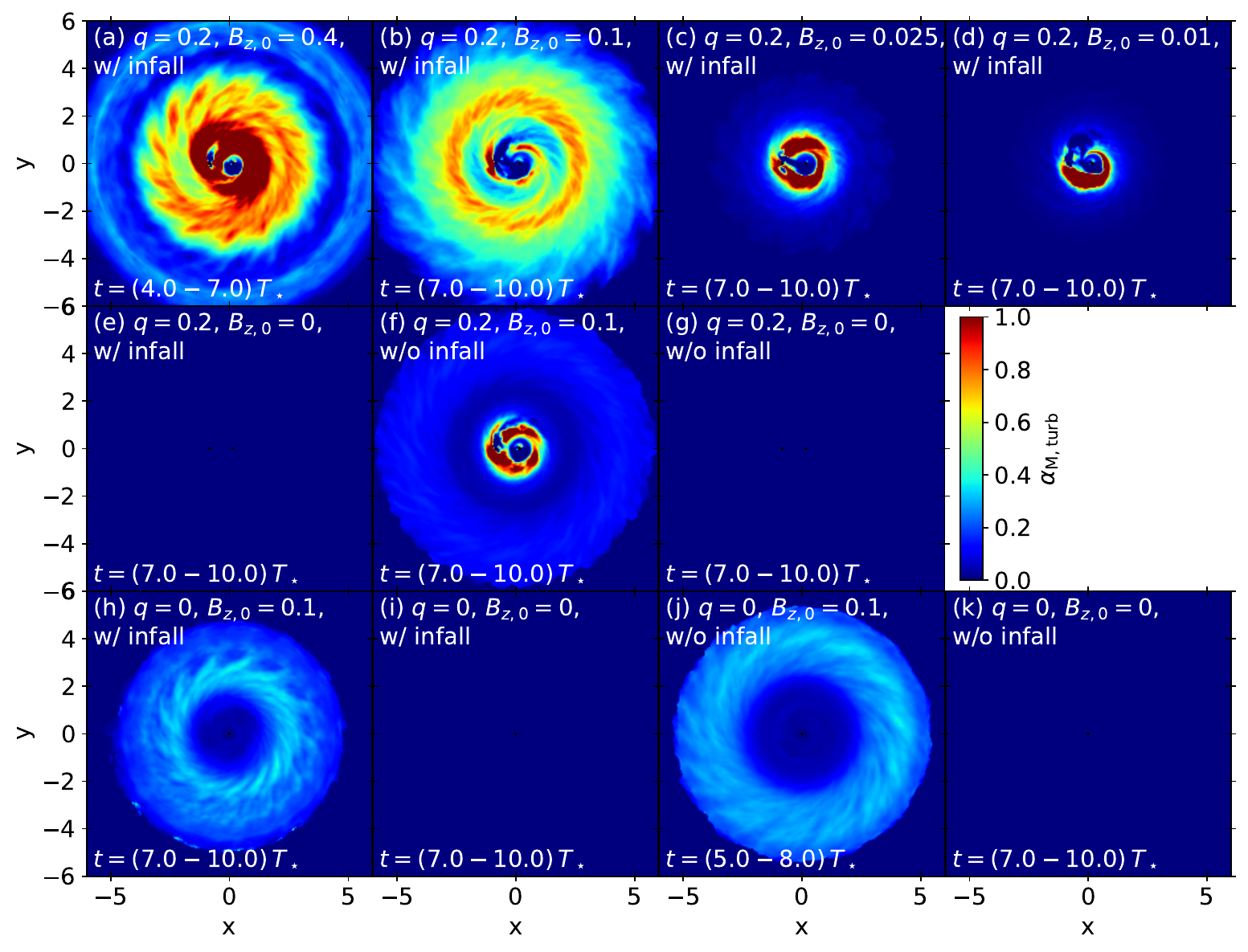}
\caption{The $\alpha$-parameters of the turbulent component the Maxwell stress $\alpha_{R, \mathrm{turb}}$ ($\overline{B_R^\prime B_\varphi^\prime}$ component) for all the models.
\label{f16.pdf}}
\end{figure*}

Figure~\ref{f8d.pdf} shows $\alpha_{R,\mathrm{turb}}$ and $\alpha_{M,\mathrm{turb}}$ for the fiducial model. As shown in lower panels of Figure~\ref{f8d.pdf}, both the $\alpha$-parameters exhibit almost the same level, having large values of $0.5-0.7$ along the spiral arms. This indicates that the spiral arms enhance a turbulent level in the circumbinary disk. Even in the inter-arm regions, the $\alpha$-parameters exhibit a moderate level of $\sim 0.3$, which is larger than that shown in the corresponding single star model (see section~\ref{sec:alpha_parameters_comparison}, and panels (h) of Figures~\ref{f15.pdf} and \ref{f16.pdf}).

The $\alpha$-parameters are measured in the rotating frame in which the binary stars are rest. Even in the rotating frame, the spiral arms are not perfectly steady and fluctuate with time because of inviscid flow. The spiral arm of the secondary star fluctuates more. The high $\alpha$-parameters along the spiral arms is partially attributed to such fluctuation of them. However, they can also disturb gas flows and enhance a turbulent level in the circumbinary disk. 

A high value of $\alpha_{R,\mathrm{turb}}$ is observed at the outer edge of the circumbinary disk (upper left panel of Figure~\ref{f8d.pdf}). This is not due to turbulence, but instead can be attributed to the non-circular shape of the disk. This shape causes temporal changes in density at the outer edge, leading to the high value of $\alpha_{R,\mathrm{turb}}$.

Angular momentum transport in the circumbinary disk is not only due to MRI turbulence but also due to mean flows and mean magnetic fields. Figure~\ref{f9a.pdf} shows the $\alpha$-parameters contributed by the mean flow $\overline{\rho} \widetilde{v}_R \widetilde{v}_\varphi$ and the mean magnetic field $\overline{B}_R \overline{B}_\varphi$, which are denoted by $\alpha_{R, \mathrm{mean}}$ and $\alpha_{M, \mathrm{mean}}$, respectively (see Appendix~\ref{seq:estimate_alpha_parameters}).

The mean flow of $\overline{\rho} \widetilde{v}_R \widetilde{v}_\varphi$ has a large amplitude that follows the shape of the spiral arms, as shown in Figure~\ref{f9a.pdf} (left). Along the spiral arms, the peak value is $\alpha_{R, \mathrm{mean}} \sim 2$, while it has large negative values $\alpha_{R, \mathrm{mean}} \sim -10$ in inter-arm regions. This suggests that the spiral arms efficiently transport angular momentum outward in the spiral arms and inward in the inter-arm regions. The gravitational torque from the binary stars generates the spiral arms and is responsible for the associated angular momentum transport. This transport is also seen in hydrodynamical simulations by \citet{Matsumoto19} and MHD simulations by \citet{Shi12}, where gas exhibits expansion motion along the spiral arms and infall motion in the inter-arm regions. These motions have been observed by ALMA for protobinary systems L1551~NE \citep{Takakuwa14, Takakuwa17} and L1551~IRS~5 \citep{Takakuwa20}.

The angular momentum transport by the mean magnetic field is shown in Figure~\ref{f9a.pdf} (right). The ring-shaped region with a radius of $\sim 3$ exhibits $\alpha_{M, \mathrm{mean}} \sim 0.5$. In this region, the angular momentum is transferred outward by the mean magnetic field of $\overline{B}_R \overline{B}_\varphi$, whose magnetic field configuration can be seen in Figure~\ref{f7a.pdf}. The magnetic field is stretched in the radial direction in the circumbinary disk.

Figure~\ref{f10a.pdf} shows the radial distribution of the $\alpha$-parameters for comparing the various components quantitatively. In the region of $1 \lesssim R \lesssim 4$, the $\alpha$-parameters of the turbulent components ($\overline{\rho v_R^\prime v_\varphi^\prime}$ and $\overline{B_R^\prime B_\varphi^\prime}$) and the mean magnetic field component ($\overline{B}_R \overline{B}_\varphi$) exhibit $\sim 0.5$. The large value of the velocity turbulent component ($\overline{\rho v_R^\prime v_\varphi^\prime}$) at $R \sim 6$ is due to the outer edge of the circumbinary disk, as noted in the upper-left panel of Figure~\ref{f8d.pdf}.

The mean flow component ($\overline{\rho} \widetilde{v}_R \widetilde{v}_\varphi$) exhibits a large negative value in the circumbinary disk, as expected from the left panel of Figure~\ref{f9a.pdf}. It overwhelms the other $\alpha$ components and is attributed to gas accretion through the circumbinary disk to the central circumstellar disks or binary stars.

\subsection{Gas structures: dependence on binary parameters}
\label{sec:density_comparison}

\begin{figure*}
\epsscale{0.9}
\plotone{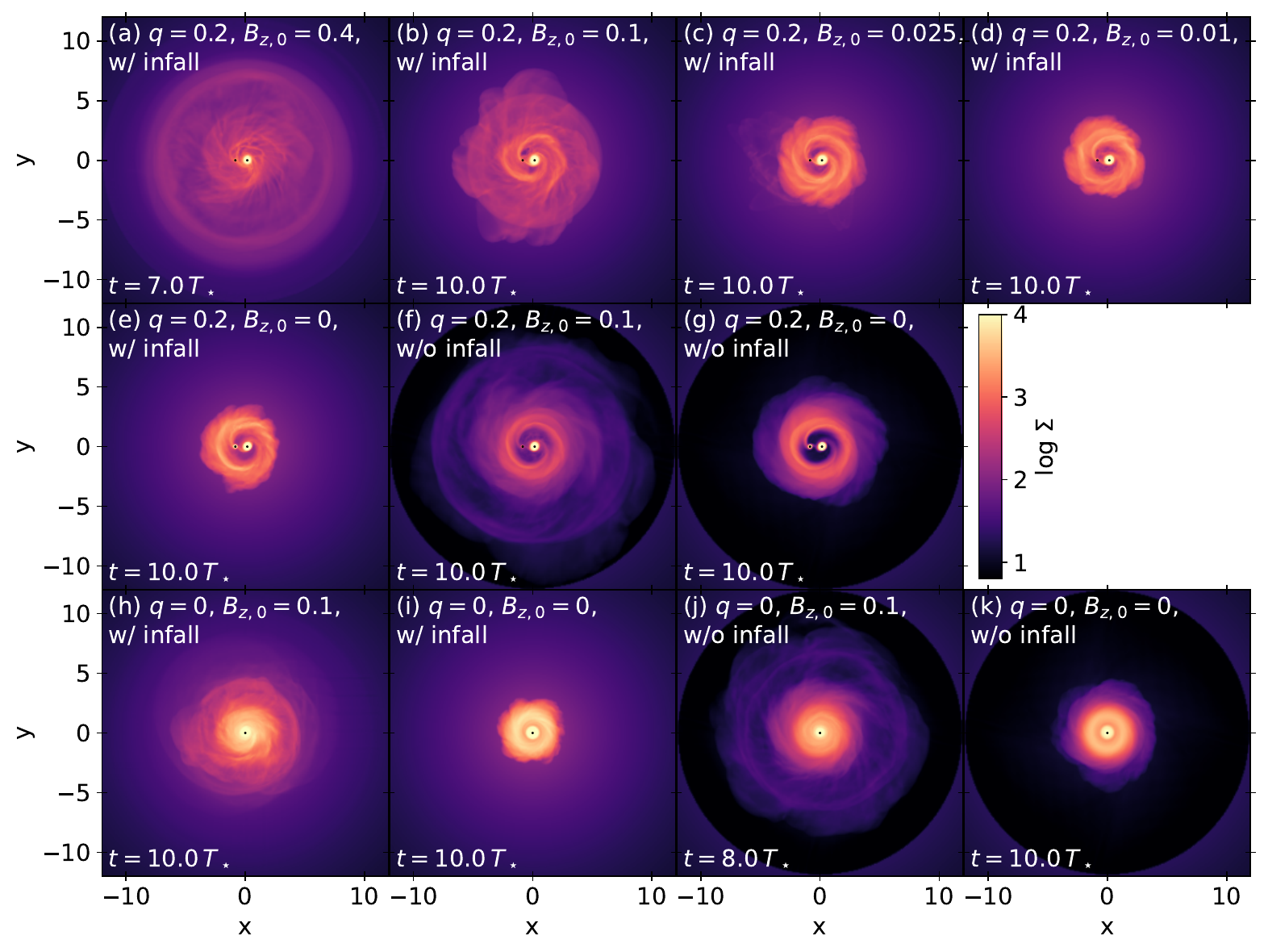}
\caption{
Surface density distributions for all the models at $t \sim 10 T_\star$. The whole computational domains are shown. The model parameters are indicated in each panel, with the terms ``w/ infall'' and ``w/o infall'' referring to the models with and without an infalling envelope, respectively.
\label{f11.pdf}
}

\plotone{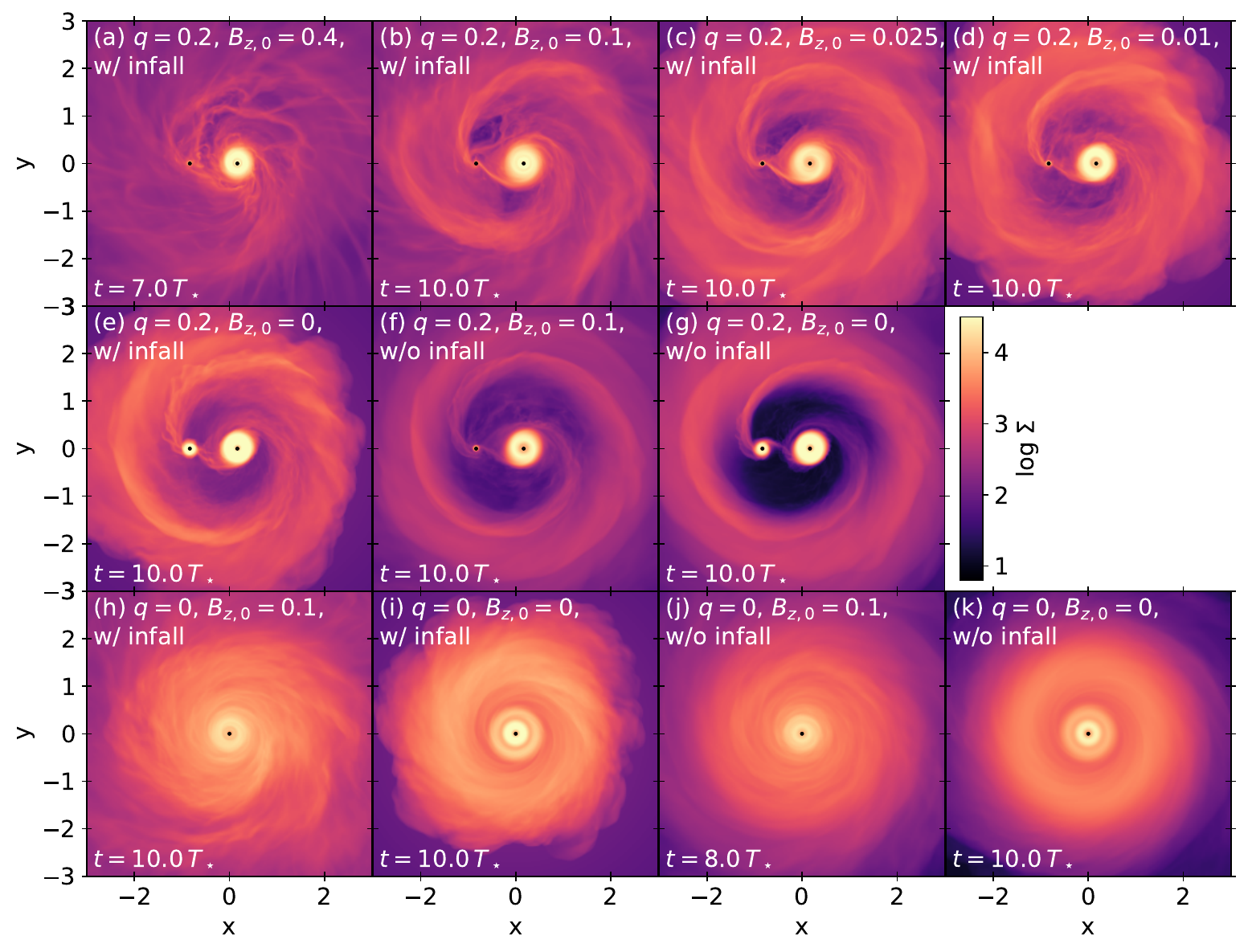}
\caption{Surface density distributions for all the models at $t \sim 10 T_\star$. Each panel shows the magnification of Figure~\ref{f11.pdf}.
\label{f12.pdf}
}
\end{figure*}

\begin{figure*}
\epsscale{0.9}
\plotone{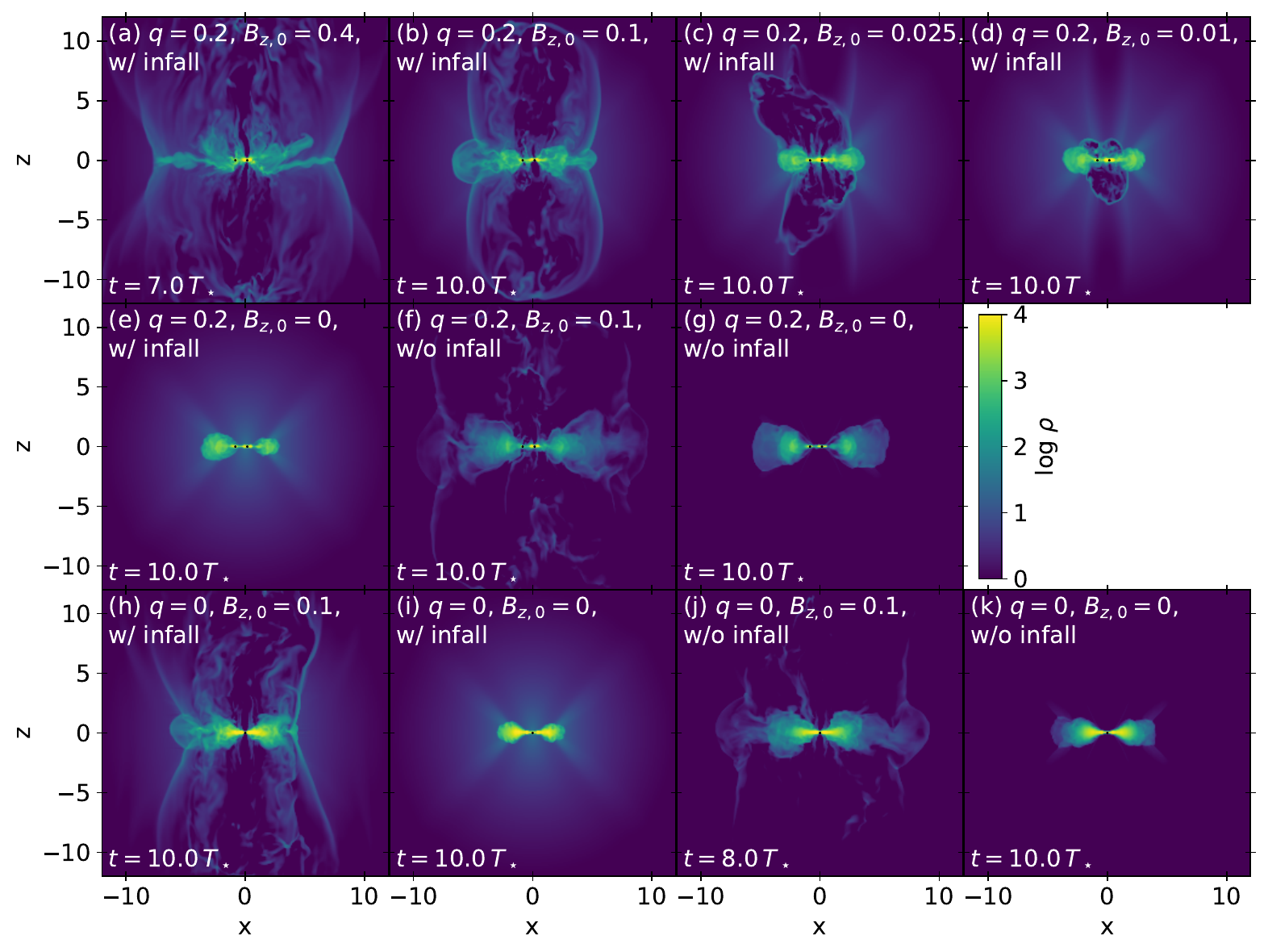}
\caption{Density distributions in the $y=0$ plane (the meridional plane) for all the models at $t \sim 10 T_\star$. The whole computational domains are shown. 
\label{f13.pdf}
}

\plotone{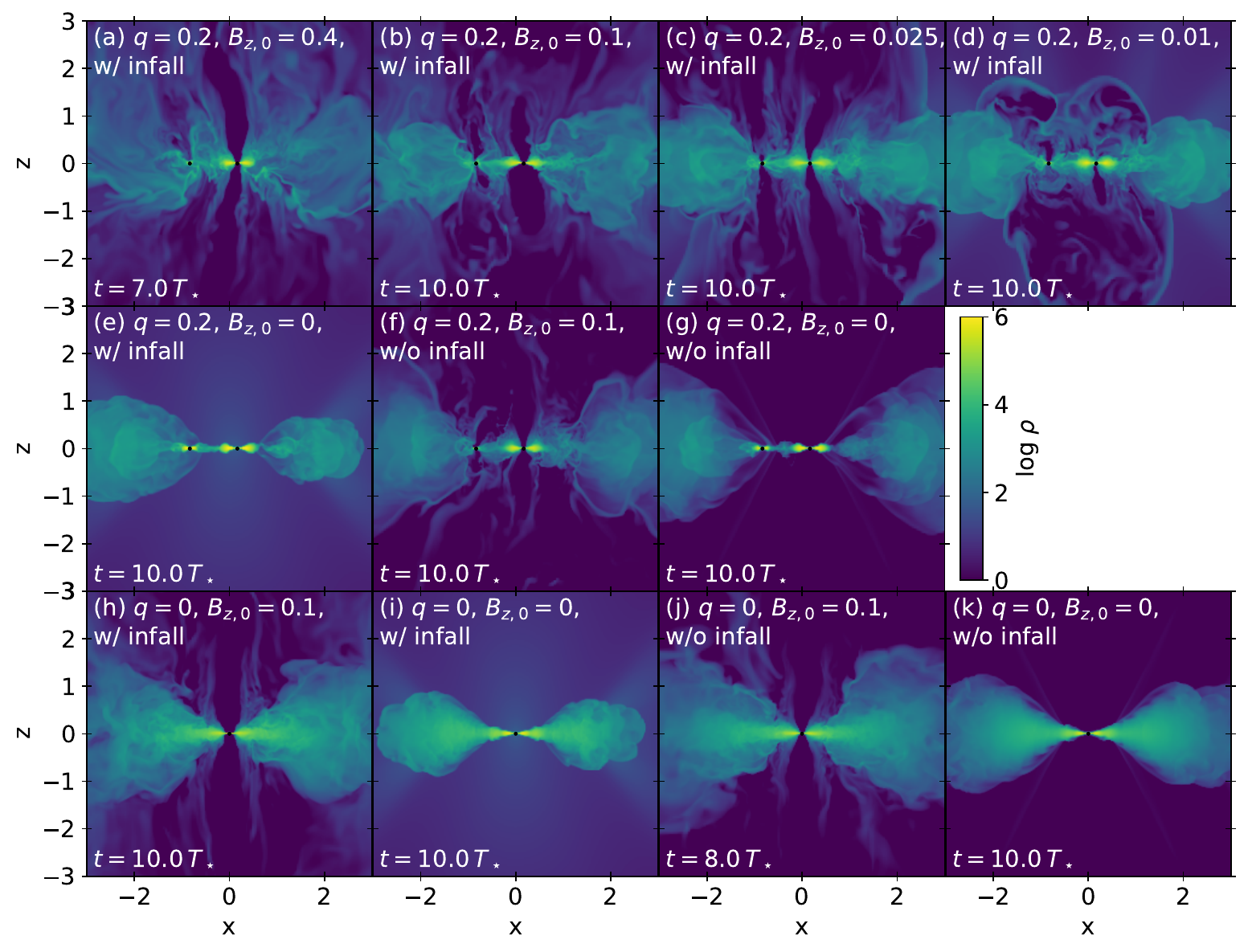}
\caption{Density distributions in the $y=0$ plane (the meridional plane) for all the models at $t \sim 10 T_\star$. Each panel shows the magnification of Figure~\ref{f13.pdf}.
\label{f14.pdf}
}
\end{figure*}

Figures~\ref{f11.pdf} and \ref{f12.pdf} show the surface density distributions in face-on view, comparing the structures of the circumbinary gas among all the models. We compare the models at almost the same stage of $t\sim 10T_\star$, although the weak field models and non-magnetized models are calculated for a longer time.

When comparing Figure~\ref{f11.pdf}a--e, we see that a model with stronger magnetic fields has an extended circumbinary disk, which is attributed to angular momentum redistribution by MRI. Because of MRI, more turbulent density structure is seen in a model with a stronger magnetic field, as shown in Figure~\ref{f12.pdf}a--e. Furthermore, the cavity of the circumbinary disk and spiral arms are obscure for the strong magnetic field models.

A model with a stronger magnetic field also shows stronger outflows. When we observe outflows at the same stage of $t\sim 10T_\star$, a model with a stronger magnetic field launches outflows farther, as shown in panels (a)--(e) of Figures~\ref{f13.pdf} and \ref{f14.pdf}. Note that the very weak magnetic field model (Figure~\ref{f13.pdf}d) exhibits short outflows at $t=10 T_\star$, because the outflow growth rate is slower compared to the fiducial case. Consequently, they extend to the upper and lower boundary surfaces ($z=\pm12$) at a later stage of $t\simeq 20T_\star$.

Panels (f)--(g) of Figures~\ref{f11.pdf} and \ref{f12.pdf} show the binary models without infalling envelopes. Comparing models with the same magnetic field strength (e.g., Figures~\ref{f11.pdf}b and \ref{f11.pdf}f; Figures~\ref{f11.pdf}e and \ref{f11.pdf}g), we observe that the models without an infalling envelope have extended circumbinary disks because there is no ram pressure from the infalling gas.

The models with infalling envelopes exhibit more disturbed density distributions in the circumbinary disks compared to those without, as seen when comparing Figures~\ref{f12.pdf}b and \ref{f12.pdf}f, which have the same initial magnetic field strength of $B_{z,0}=0.1$. This difference is mainly due to two factors. The first is difference in magnetic flux between the models with and without an infalling envelope. The infalling envelope not only brings mass but also magnetic flux through the boundary surfaces into the computational domain. As a result, the model with an infalling envelope has 2.5 times more magnetic flux than the model without at the stages shown in the figures, and therefore exhibits a more developed MRI. The increase in the magnetic flux is disscussed in section~\ref{sec:limitation}.

The second factor is disturbance by the infalling gas onto the circumbinary disk. 
A fraction of the accretion energy is converted into heat at the accretion shock. However, because a barotropic equation of state is assumed here, this heat is radiated away, keeping the gas temperature constant. A certain part of the kinetic energy of the infalling gas is converted to the kinetic energy of the turbulent motion \citep{Klessen2010}. This effect is also reported in previous simulations by \citet{Matsumoto19}. The accretion-driven turbulence is also observed when comparing non-magnetized models in Figures~\ref{f12.pdf}e and \ref{f12.pdf}g.
We also note that the non-magnetized model without an infalling envelope (Figure~\ref{f12.pdf}g) exhibits a cavity with the highest contrast because it has the lowest turbulent level among the binary models examined here, and the absence of accreting gas that would otherwise fill the cavity.

Panels (h)--(k) of Figures~\ref{f11.pdf} and \ref{f12.pdf} show the models of single stars. A comparison between the binary star model and the single star model with the same initial magnetic field (e.g., Figures~\ref{f12.pdf}b and \ref{f12.pdf}h) shows that the binary model has a more turbulent density distribution than the single star model, suggesting that the orbital motion of the binary stars disturbs the circumbinary disk.

\subsection{Alpha parameters: dependence on binary parameters}
\label{sec:alpha_parameters_comparison}

Figures~\ref{f15.pdf} and \ref{f16.pdf} compare the $\alpha$-parameters of turbulent components of Reynolds stress and Maxwell stress, respectively, among the models examined here.

Comparison of panels (a)--(e) in Figures~\ref{f15.pdf} and \ref{f16.pdf} reveals the influence of magnetic fields on the level of turbulence in the circumbinary disk; models with stronger initial magnetic fields have stronger turbulence except for the vicinity of binary stars and the disk edges.

The effect of the infalling envelope in the models with $B_{z,0}=0.1$ can be observed by comparing panels (b) and (f) in Figures~\ref{f15.pdf} and \ref{f16.pdf}. Similarly, panels (e) and (g) show the effect of the envelope in the models with $B_{z,0}=0$. In both cases, the models with infalling envelopes exhibit higher $\alpha$-parameters within the circumbinary disks, which is consistent with the differences in density distributions discussed in section~\ref{sec:density_comparison}.

\begin{figure}[t]
\epsscale{1}
\plotone{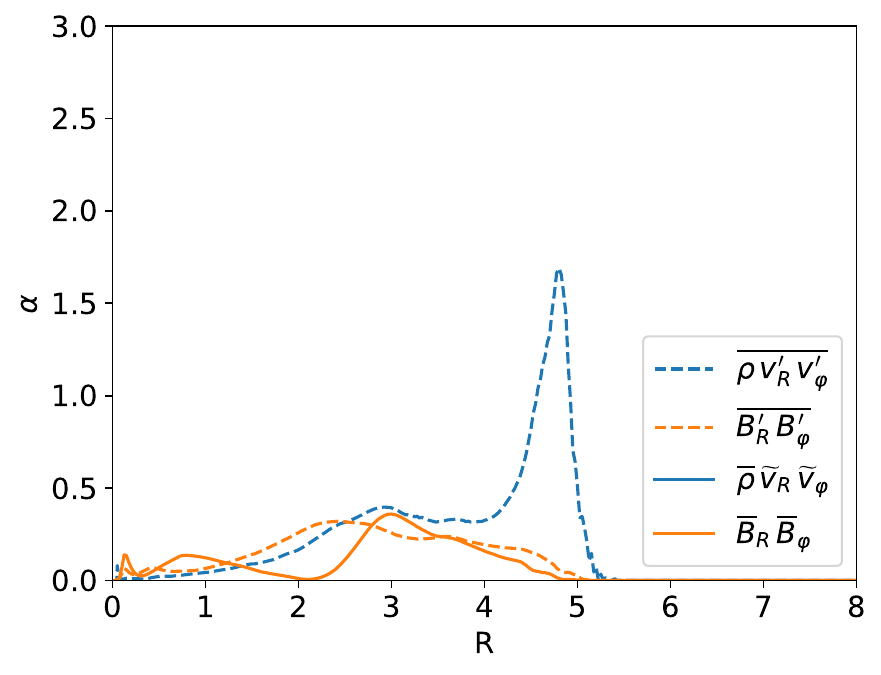}\\
\plotone{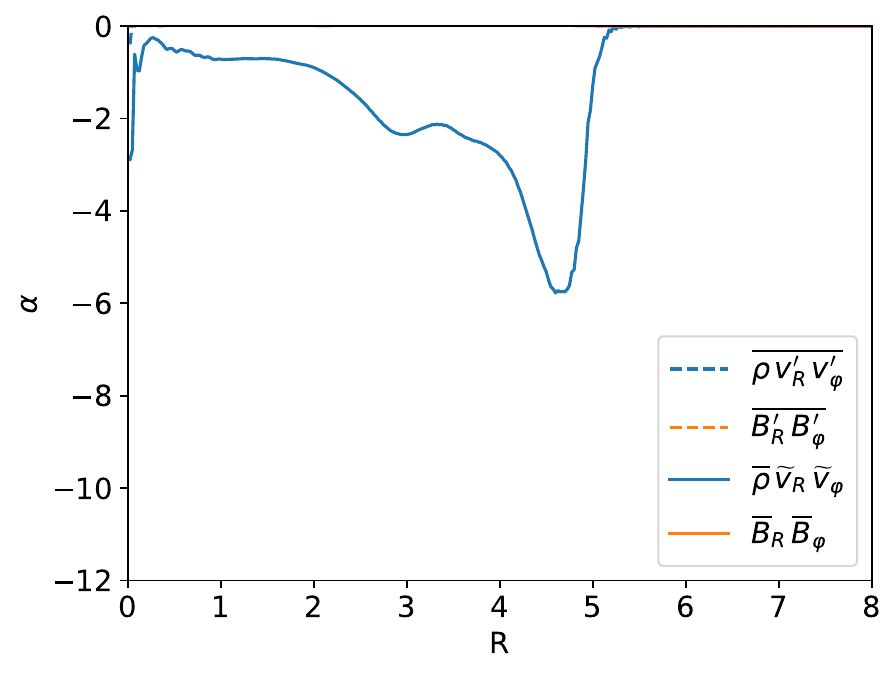}  
\caption{
  Same as Figure~\ref{f10a.pdf} but for the single star model ($q = 0$) with $B_{0,z} = 0.1$ and an infalling envelope.
\label{f17a.pdf}}
\end{figure}

The panels (b) and (h), (e) and (i), (f) and (j), and (g) and (k) demonstrate the impact of binarity on the turbulence levels in disks for various models with/without magnetization and infalling envelopes. These figures indicate that the binary models tend to have a higher level of turbulence than the single star models in their disks. 
For example, the single star model in panel (h) exhibits a very small turbulent level of $\alpha_{R,\mathrm{turb}}\sim 0.01$ and $\alpha_{M,\mathrm{turb}}\sim 0.05$ in the central region ($R \lesssim 1$), and a moderate level of $\alpha_{R,\mathrm{turb}}\sim \alpha_{M,\mathrm{turb}}\sim 0.3$ at $R \sim 3$ (refer to Figure~\ref{f17a.pdf}), but it is lower than that of the corresponding binary model, which is shown in Figure~\ref{f10a.pdf}.

\subsection{Angular momentum transport}
\label{sec:angular_momentum_transport}

In the previous sections, we explored the $\alpha$-parameters, which are indicators of angular momentum transport in the radial or horizontal direction. In this section, we investigate angular momentum fluxes to examine transport in both the horizontal and vertical directions.

\begin{figure}[t]
\epsscale{1}
\plotone{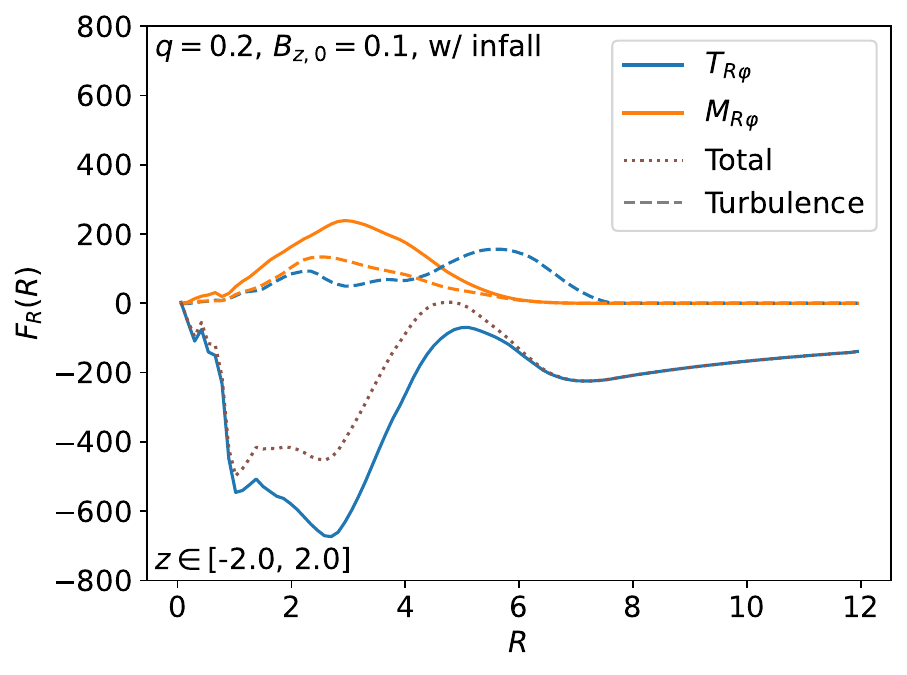}\\
\plotone{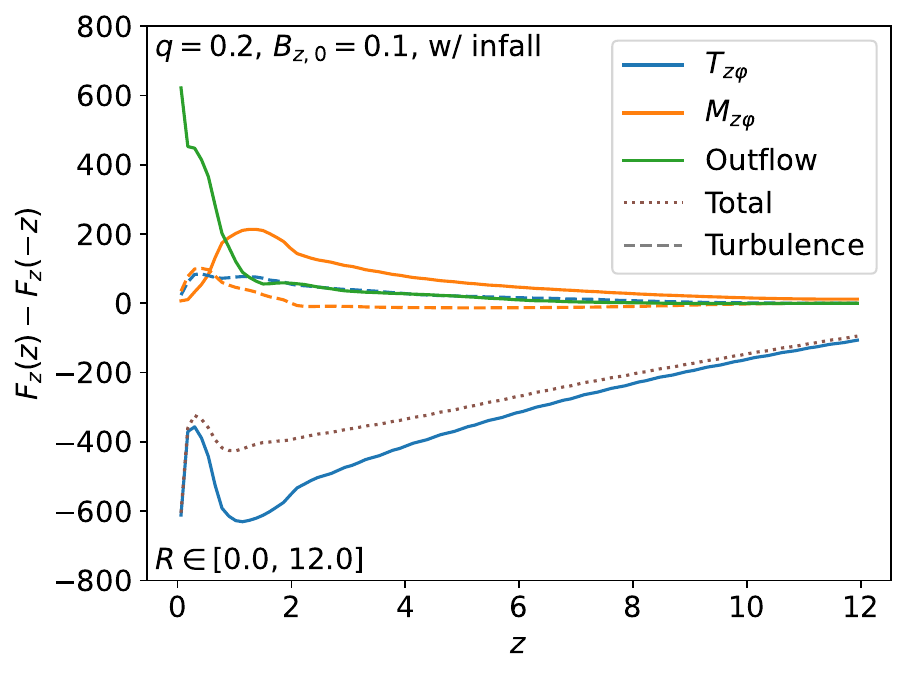}  
\caption{
Angular momentum fluxes for the the fiducial model ($q = 0.2$, $B_{z,0} = 0.1$ with the infalling envelope). 
The upper panel shows the angular momentum flux in the $R$-direction $F_R(R)$ as a function of $R$. The flux is measured through the side of a cylinder with a radius of $R$ and a height of $z \in [-2, 2]$. The lower panel shows the net flux of the angular momentum in the $z$-direction $F_z(z) - F_z(-z)$ as a function of $z$. The net flux is measured through the top and bottom surfaces of a cylinder with a radius of 12 and height of $\pm z$. The time average is taken in the period of $t \in [7T_\star, 10T_\star]$. A positive flux means that angular momentum is extracted from the region of interest (outgoing flux). Each line represents a contribution of each component; e.g., the solid blue line shows a total contribution of the Reynolds stress $T_{R\varphi}$, while the dashed blue line shows the contribution of turbulent component of the Reynolds stress. The methods for evaluating the contributions are described in Appendix~\ref{sec:am_transport}.
\label{f18a.pdf}}
\end{figure}

Figure~\ref{f18a.pdf} displays the angular momentum fluxes in both the $R$ and $z$-directions for the fiducial model. Negative flux values for $F_R$ and $F_z$ represent inflow of angular momentum, while positive values represent outflow. A detailed description of the methods for calculating these fluxes is shown in Appendix~\ref{sec:am_transport}.

We found that, in the circumbinary disk region ($1 \lesssim R \lesssim 5$), the $T_{R\varphi}$ component (blue solid line) is dominant, corresponding to the gas flow associated with spiral arms. The turbulent components of $T_{R\varphi}$ and $M_{R\varphi}$ (dashed lines) exhibit significant positive values, indicating an outgoing transfer of angular momentum due to MRI turbulence. The turbulent component of $M_{R\varphi}$ (orange dashed line) accounts for half of the total positive angular momentum flux of $M_{R\varphi}$ (orange solid line) in this region. The other half is attributable to the coherent component of the magnetic field, resulting from the winding of the magnetic field (see Figure~\ref{f7a.pdf}). The turbulent component of $T_{R\varphi}$ peaks at the outer edge of the circumbinary disk ($R\sim 6$) due to the the non-circular outline of the disk (see Figure~\ref{f2.pdf}).

In the infalling envelope ($R \gtrsim 5$),  the $T_{R\varphi}$ component (blue solid line) dominates the angular momentum flux in the radial direction. Similarly, in the vertical direction, the $T_{z\varphi}$ component (blue solid line) is the dominant factor across the range of the model. Notably, these values are negative, indicating that the infalling envelope transports both mass and angular momentum towards the central region in both the radial and vertical directions.

The angular momentum flux due to the outflows (green line) is shown in the lower panel of Figure~\ref{f18a.pdf}. The outflows are responsible for a large portion of the outward flux of angular momentum in the $z$ direction for $z\lesssim 1$, but it does not exceed the angular momentum inflow due to the infalling envelope. Additionally, we note that the outward angular momentum transport in the upper region of the circumbinary disk ($z \gtrsim 1$) is carried out by magnetic braking ($M_{z\varphi}$ component; orange line) rather than outflows \citep[c.f.,][]{Marchand19,Lee21}.

The angular momentum flux represents the angular momentum flowing into or out of the region of interest. On the other hand, to consider spin-up or down of the region of interest, it is necessary to consider specific angular momentum, which is approximately angular momentum per mass. We estimate the timescale of specific angular momentum change ($\tau_\mathrm{sam}$), following the method provided in Appendix~\ref{sec:change_in_sam}. In the following, the rate of change in specific angular momentum ($\tau_\mathrm{sam}^{-1}$) is shown, of which definition is given by equation~(\ref{eq:1overtausam}). A positive $\tau_\mathrm{sam}^{-1}$ represents increase in the specific angular momentum.

\begin{figure}[t]
\epsscale{1}
  \plotone{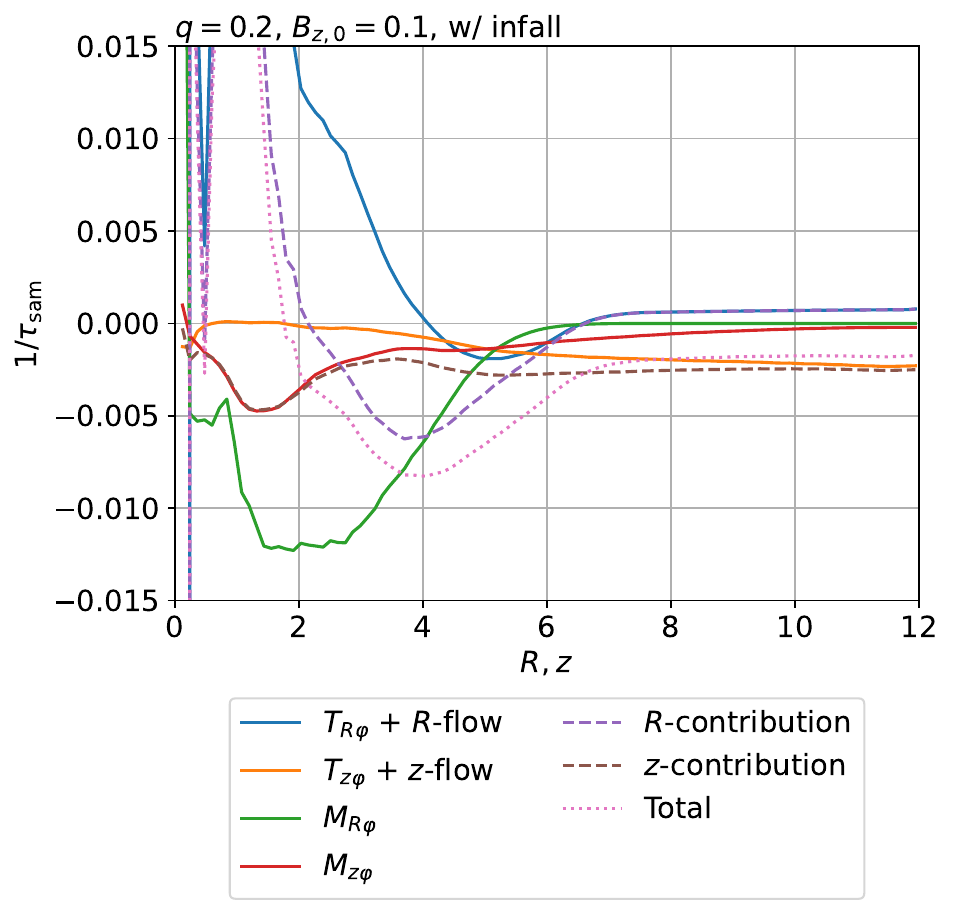}
\caption{
The rate of change in specific angular momentum ($\tau_\mathrm{sam}^{-1}$) inside a cylinder as a function of radius and height of the cylinder for the fiducial model ($q = 0.2$, $B_{z,0} = 0.1$, with the infalling envelope). The each line show the contribution of each component to $\tau_\mathrm{sam}^{-1}$ (see Appendix~\ref{sec:change_in_sam}).
The time average is taken in the period of $t \in [7T_\star, 10T_\star]$.
\label{f19.pdf}}
\end{figure}

Figure~\ref{f19.pdf} shows $\tau_\mathrm{sam}^{-1}$ for the fiducial model.
In the circumbinary disk region of $1 \lesssim R \lesssim 5$, there are a negative contribution due to magnetic field (green line) and a positive contribution due to gas flow (blue line) in $\tau_\mathrm{sam}^{-1}$, indicating that the magnetic field, including MRI turbulence, reduces specific angular momentum, and the gas flow associated with the spiral arms increases the specific angular momentum. As a result of these competing contributions, the specific angular momentum decreases in outer region the circumbinary disk and increases in the inner region the disk (dotted line).

In the outer region of $R, z \gtrsim 7$, a negative $\tau_\mathrm{sam}^{-1}$ in total (dotted line) is exhibited,  which is responsible for the angular momentum transport in the $z$-direction, caused mostly by gas flow due to the outflow and magnetic braking (orange and red lines).

\begin{figure}[t]
\epsscale{1}
\plotone{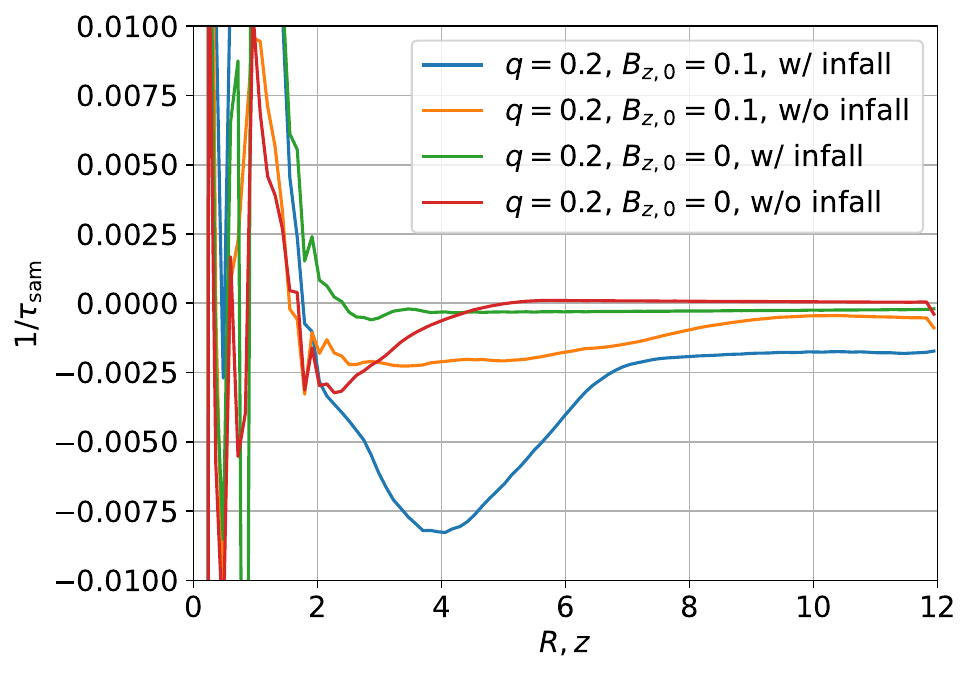}\\
\plotone{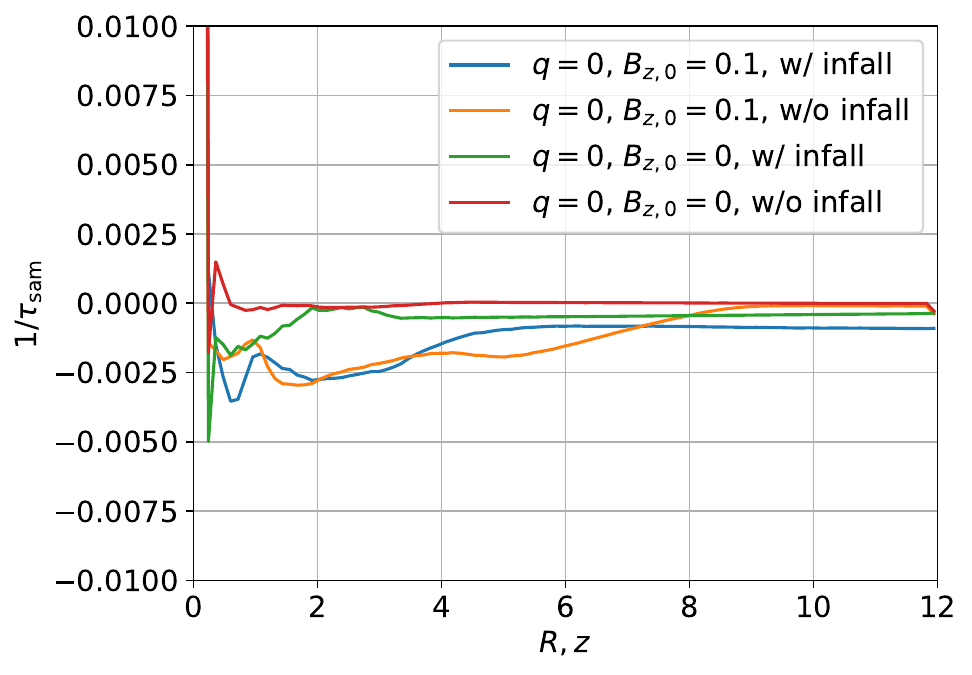}  
\caption{
The rate of change in specific angular momentum ($\tau_\mathrm{sam}^{-1}$) inside a cylinder as a function of radius and height of the cylinder. The upper and lower panels are for binary and single star models, respectively. A negative value means a decrease in the specific angular momentum. The time averages are taken in the periods depicted in Figure~\ref{f15.pdf}. 
\label{f20a.pdf}}
\end{figure}

Figure~\ref{f20a.pdf} shows $\tau_\mathrm{sam}^{-1}$ for representative models. The binary model with magnetic field and infalling envelope (the fiducial model; blue line) shows high negative values of $\tau_\mathrm{sam}^{-1}$ at $R, z \sim 4$. The models with magnetic fields (blue and orange lines) tend to show higher negative values of $\tau_\mathrm{sam}^{-1}$ than those without (green and red lines). This is because the magnetic effect, including MRI, plays an important role in the angular momentum transport. The single star models (lower panel of Figure~\ref{f20a.pdf}) exhibit smaller negative values of $\tau_\mathrm{sam}^{-1}$ than the binary models, indicating that the orbital motion of the binary stars enhances MRI turbulence, as mentioned above.

The high negative value of $\tau_\mathrm{sam}^{-1}$ observed in the magnetic binary models indicates a reduction in specific angular momentum. This suggests that the binary separation could decrease due to the transport of angular momentum facilitated by the magnetic field. A more accurate estimation of the changes in binary separation will be the focus of future investigations.

\subsection{Mass accretion}
The mass accretion onto binary stars has been a topic of debate \citep{Bate97,Ochi05,Hanawa10,Young15,Young15b,Satsuka17,Matsumoto19}, and in general, binary stars accrete gas in a way that increases the mass ratio, leading to the evolution of the system towards a twin binary system. To describe this change in the mass ratio, the $\Gamma$ parameter is defined as 
\begin{equation}
  \Gamma = \frac{\dot{q}/q}{\dot{M}_\mathrm{tot}/M_\mathrm{tot}} = \frac{(1+q)(\dot{M}_2 - q \dot{M}_1)}{q (\dot{M}_1 + \dot{M}_2)},
\end{equation}
where a positive value of $\Gamma$ indicates an increase in the mass ratio \citep{Bate97,Young15,Matsumoto19}.

\begin{figure}[t]
\epsscale{1}
\plotone{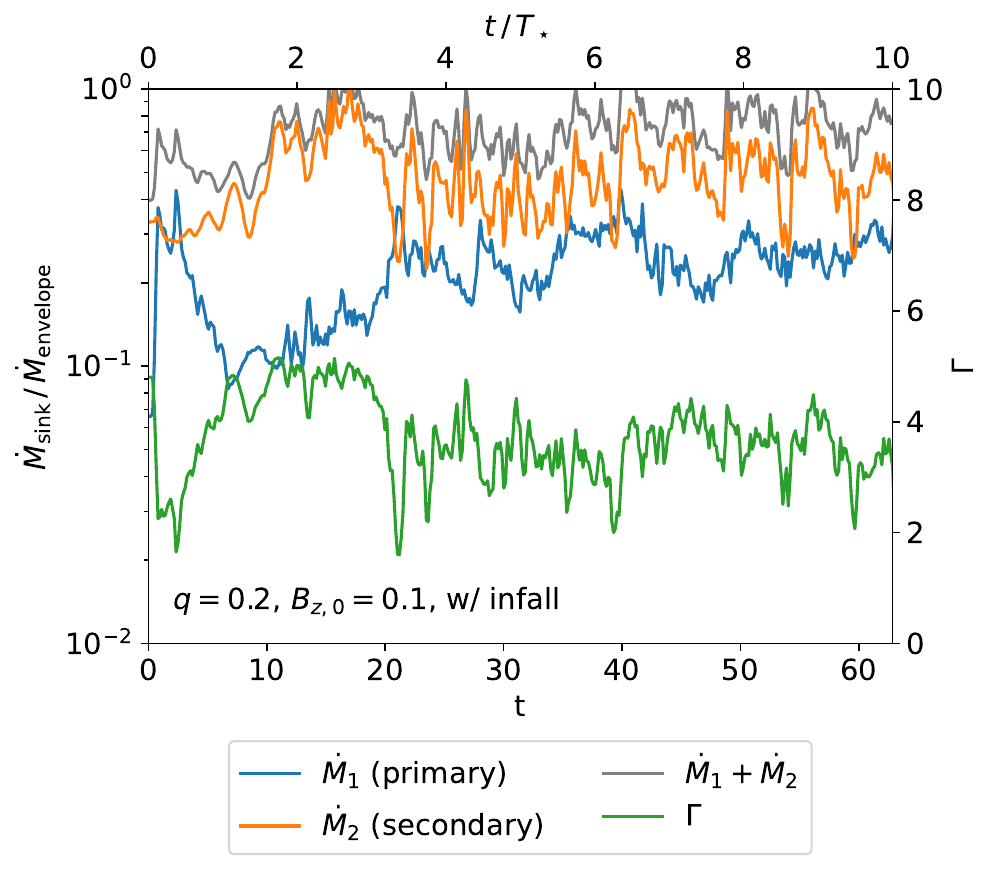}
\caption{
Mass accretion rate of the primary and secondary stars as a function of time for the fiducial model ($q = 0.2$, $B_{z,0} = 0.1$, with the infalling envelope). The mass accretion rates of stars are normalized by the mass accretion rate of the envelope at $t = 0$ (gas injection rate at the boundary surfaces). 
The rate of change in the mass ratio $\Gamma$ is also shown.
\label{f21.pdf}}
\end{figure}

Figure~\ref{f21.pdf} depicts the mass accretion rates onto the primary and secondary stars and the $\Gamma$ parameter as functions of time for the fiducial model. Even in the model with magnetic field, the mass accretion rate onto the secondary star is higher than that onto the primary star in consistent with the previous hydrodynamical simulations \citep{Bate97,Young15,Matsumoto19}. The $\Gamma$ parameter exhibits time variability, but it remains positive throughout the evolution, suggesting that the binary system evolves towards twin binaries. The time-average value of $\Gamma$ is 3.691.

In Figure~\ref{f21.pdf}, the total mass accretion rate for both the primary and secondary stars is also plotted. The total mass accretion rate remains just below unity for most of the time, implying that the total mass of gas in the computational domain is gradually increasing. We observe that the mass of the circumbinary disk also experiences gradual growth over time. This suggests that the disk's evolution is not in a perfect steady state; however, the growth rate is notably slow. Specifically, $\dot{M}_\mathrm{CBD} / M_\mathrm{CBD} \sim (10^{-2} - 10^{-3}) \Omega_\star$ at $t \sim 10T_\star$ for the fiducial model, where $M_\mathrm{CBD}$ denotes the mass of the circumbinary disk. Such a slow growth rate for the circumbinary disk was also noted in the previous hydrodynamical model \citep{Matsumoto19}.

\begin{figure}[t]
\epsscale{1}
\plotone{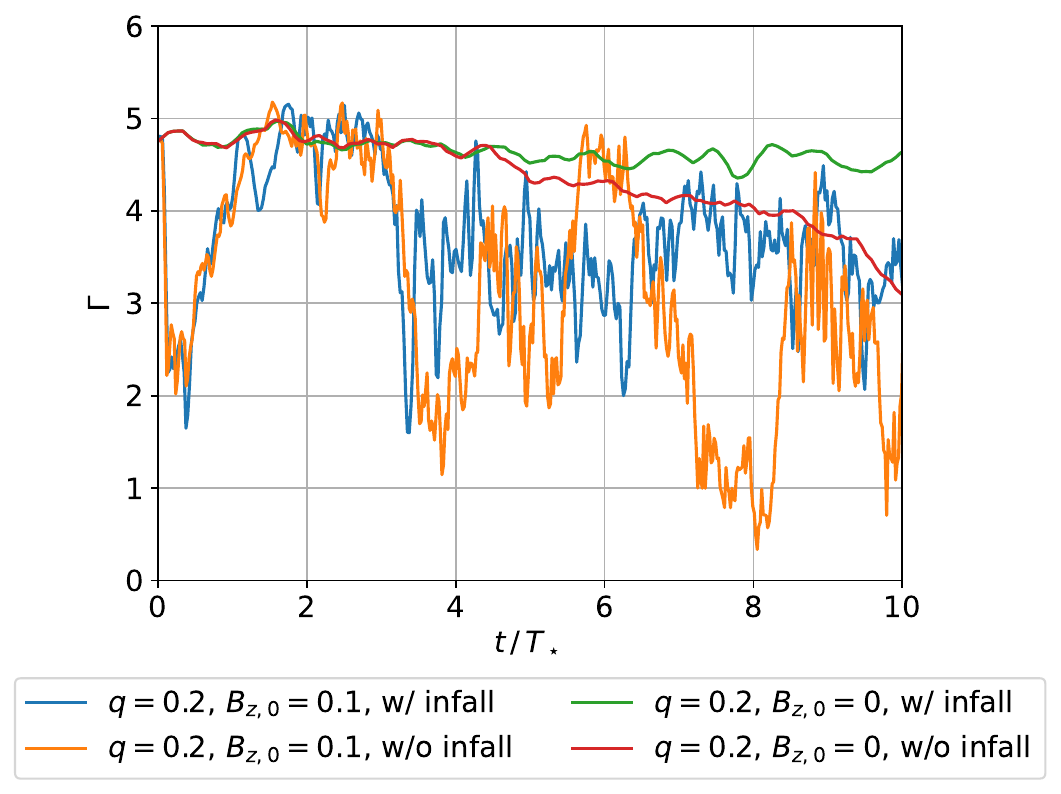}
\caption{
The rate of change in the mass ratio $\Gamma$ as a function of time for the representative binary models.
\label{f22.pdf}}
\end{figure}

Figure~\ref{f22.pdf} shows the $\Gamma$ parameters for four representative models, including models with and without a magnetic field and an infalling envelope. All the models shown have a positive $\Gamma$. The models with a magnetic field exhibit significant variability in $\Gamma$ due to the time variability in the mass accretion rates. In contrast, the models without a magnetic field show low time variability. It is worth noting that the presence of the infalling envelope affects the evolution of the mass ratio since it provides gas that falls onto the primary or secondary stars/circumstellar disks. However, $\Gamma$ remains positive in both cases with and without infalling envelopes.

\section{Discussion}
\label{sec:discussion}

\subsection{Implication for formation of close twin binaries}

The current models offer a possible formation scenario for close binary systems within the framework of the disk fragmentation scenario. 

Based on the analysis of Gaia data, \citet{El-badry19} discovered the existence of a sharp excess of equal-mass binaries (twin excess) in wide ranges of binary star masses and separations \citep[see also][]{Raghavan10,Moe17}. The existence of the twin excess indicates the contribution of mass accretion from circumbinary disks, which has been reproduced by hydrodynamical simulations \citep{Bate97,Young15,Young15b,Matsumoto19}.

Both MHD and hydrodynamical models show the same tendency of accretion rates, where binaries tend to evolve to twin binaries. In contrast to the hydrodynamical models, the MHD models exhibit considerable time-variability in the accretion rates, as well as in the parameter $\Gamma$, which is proportional to $\dot{q}$. Despite the minor changes of accretion rates, the MHD models are also consistent with the twin excess.

\citet{Hwang22} reported that twin binaries with wide separations are likely to have high eccentric orbits, proposing a scenario in which such wide twin binaries could form by scattering from dynamical interactions in their birth environment after accreting gas from a circumbinary disk on a $\sim 100$~au scale. While this scenario explains the formation of wide twin binaries, the formation scenario of close twin binaries has been unknown. As shown in section~\ref{sec:angular_momentum_transport}, the magnetic field transfers angular momentum from the system, and it can make the separation of binaries small and provide a possible formation scenario for close twin binaries.

\subsection{Implications for observations}

\citet{Hara2021} reported the detection of a wiggling structure in an outflow as observed by ALMA. They suggest that the structure is a result of the orbital motion of the primary star, which acts as the driving source. Our simulation, as shown in Figure~\ref{vol_field_vel.pdf}, successfully reproduces the presence of twisted outflows, with the outflow from the primary star being stronger than that from the secondary star. This stronger outflow from the primary star is likely to propagate farther and corresponds to the wiggling structure observed in the ALMA data. Twisted outflows are a natural consequence of binary stars.

The Class I object BHB07-11 is a binary system showing spiral streamers at the central region and also shows an extended flat envelope according to the ALMA observations \citep{Alves17,Alves19}. The outflows are launched at the boundary between the inner dense circumbinary disk and the extended flat envelope. This structure resembles what the magnetized models reproduce. Figure~\ref{vol_field_vel.pdf} (upper right panel) exhibits outflows from the circumbinary disk, and the circumbinary disk is extended from the \rev{launching} radius because of angular momentum redistribution due to the MRI. 
The present models explain not only the density structure but also the velocity structure. \citet{Alves17} reported that the extended envelope exhibits a rotational velocity slightly higher than the Keplerian velocity. This excess in rotational velocity is approximately of the order of the sound speed, as observed in the position-velocity diagram of H$_{12}$CO line emission. Similarly, our fiducial model also shows faster rotation than Keplerian by approximately $\sim c_s$, which is not depicted in a figure, but aligns well with the observed envelope. Consequently, the extended circumbinary disk in our model could correspond to the envelope observed by ALMA

Asymmetry has been observed in multiple circumbinary disks, including L1551~NE \citep{Takakuwa17}, L1551~IRS5 \citep{Takakuwa20}, and HD~142527 \citep{Fukagawa13}. \citet{Matsumoto19} conducted hydrodynamical simulations that replicated the $m=1$ asymmetry in circumbinary disks, which arises due to the gravitational torque of binary stars on the circumbinary disk \citep[see also][]{Shi12,Ragusa17}. This asymmetry also manifests in our current models without a magnetic field (Figures~\ref{f11.pdf}e and \ref{f12.pdf}e). In contrast to the hydrodynamical models, models with a magnetic field exhibit reduced asymmetry in circumbinary disks due to the MRI, because it dilutes the asymmetry. A similar effect has also been observed in simulations of protoplanetary disks containing planets \citep{Zhu14},  where the presence of asymmetry depends on the type of magnetic field model used; a non-ideal MHD model with ambipolar diffusion reproduces asymmetry in the disk, whereas an ideal-MHD model produces strong turbulence and an axisymmetric feature. \citet{Noble21} also reported that a higher magnetization weakens substructures, such as lumps.
Therefore, the magnetic field appears to regulate the degree of asymmetry in circumbinary disks.
We also note that the MRI dilutes substructures of circumbinary disks, e.g., spiral arms and cavities, and the effect of the magnetic field likely controls the density contrast of the substructures.

\subsection{Planet formation}
 
Planets around binary stars are classified into two groups: those orbiting each star and those orbiting the pair of binary stars. The latter is known as circumbinary planets. To date, only 14 circumbinary planets have been discovered in 12 binary systems by Kepler and TESS \citep[e.g.,][]{Doyle11, Welsh12}. The small number of circumbinary planets discovered to date suggests a challenging environment for the formation of planets in circumbinary disks because orbital motion of binary stars generates turbulence in the disk \citep[e.g.,][]{Pierens20}.

Our simulations demonstrate that, in the models without magnetic fields, distinct spiral arms emerge, which produce pressure bumps and vortices that potentially enhance the capture of dust particles \citep{Barge95,Klahr97}. In the magnetic models, MRI induces turbulence, which hinders dust settlement in a circumbinary disk.  Moreover, turbulence disrupts the spiral arm structure, weakening the dust capture mechanism. To confirm the effect of magnetic fields on dust settlement and coagulation, it is necessary to perform MHD simulations that incorporate the advection and growth of dust particles \citep[as has been done in studies of single-star cases;][]{Tsukamoto21}.

\subsection{Limitations of the present models}
\label{sec:limitation}
In the present models, the binary system accretes not only gas but also magnetic flux when considering the infalling envelope. This leads to the so-called magnetic flux problem \citep[e.g.,][]{Shu06,Zhao11}, wherein a protostar would end up with a much higher magnetic flux than what is typically observed in young stars. A potential solution to this problem involves the inclusion of magnetic diffusion processes, such as Ohmic dissipation, ambipolar diffusion, and the Hall effect \citep[e.g.,][]{Wurster18}. The incorporation of these processes weakens the magnetic effects. Therefore, the real process is likely an intermediate situation between our previous hydrodynamical model \citep{Matsumoto19} and the current ideal-MHD model.

Magnetic diffusion is more effective in the circumbinary and circumstellar disks than in the infalling envelope, due to the comparison of timescales of infall and magnetic diffusion \citep{Nakano02}. In particular, the circumstellar disk is likely to be significantly influenced by magnetic diffusion due to its high density. This implies that the magnetic diffusion reduces the magnetic flux threading these disks \citep{Tsukamoto15}, and influences the MRI turbulence and the launching of outflows \citep[e.g.,][]{Bai13,Masson16}. These magnetic diffusion processes will be included in future work. 

  Even though our current work primarily focuses on the circumbinary disk rather than the circumstellar disks, we acknowledge several issues concerning the circumstellar disks. These disks are considerably thinner than the circumbinary disk. The disk thickness is around 0.1 for the primary circumstellar disk and about 0.05 for the secondary circumstellar disk at the most thick parts. The cell width in these regions is $\Delta x = 0.0058$ (at the FMR grid level $\ell =4$), indicating that the thickest regions of the circumstellar disks are resolved with more than 20 cells. 
\rev{
However, in the vicinity of the sink particles, for instance, at a point $r_\mathrm{sink}$ away from the sink particle surface, the disk thickness reduces to approximately $H \sim 0.01$. This thickness is resolved by only a few cells, and it is consistent with the estimated value of $H = 0.004 - 0.007$. This estimate comes from the relation $H/R = c_p (R/GM_1)^{1/2}$, where $c_p = 0.4-0.6$ represents the sound speed as given by the barotropic equation of state assumed here.
}
Therefore, accurately resolving the disk thickness of the circumstellar disks remains a challenging aspect of global 3D simulations.

One might speculate that the numerical resolution near the sink particles in the circumstellar disks is insufficient to fully resolve the MRI. However, \rev{in the case of the secondary circumstellar disk,} as discussed in Appendix~\ref{eq:numerical_resolution}, it remains stable against the MRI during most of the evolutionary stages. 
\rev{For the primary circumstellar disk, we confirm that the very thin area mentioned above is also stable against the MRI because of the thinness and strong magnetic field, and that the outflow is accelerated mainly in a region outside the thin area.} While the numerical resolution may affect the turbulence in the inner regions of the circumstellar disks, it is unlikely to significantly affect the mean magnetic field structures that drive the outflows from the disks. Consequently, the outflows and angular momentum transport associated with them are likely to be reproduced properly.

\rev{
The sink particles are utilized in the present models, indicating that structures finer than the sink radius are not resolved in the simulations. To mitigate the influence of magnetic effects caused by the unresolved structures inside the sink radius, Ohmic dissipation is implemented to decouple the gas from the magnetic field. In actual protostars, there should be unresolved circumstellar disks, which could be around $r_\mathrm{sink} \sim 2 \,\mathrm{au}$ in size, assuming a binary separation of $a_b \sim 100 \,\mathrm{au}$. These small-scale disks could drive high-speed jets, as noted by \citet{Machida08}, but such phenomena are not reproduced in the present models. To incorporate these effects, employing prescriptions for a sink particle is a potential method \citep[e.g.,][]{Cunningham11,Grudic21}.
}

\section{Summary}
\label{sec:summary}

We conducted three-dimensional MHD simulations of accreting binary systems to investigate the impact of magnetic fields on circumbinary materials. We varied the strength of the magnetic fields and compared the results between models with and without magnetic fields, as well as between binary and single-star models. Additionally, we examined the effects of infalling envelopes. Our findings are summarized as follows:

\begin{enumerate}
\item The binary models produce twisted twin outflows with high velocity from the circumstellar disks around the binary stars and a wide outflow with low velocity from the circumbinary disk. The twisted structure is attributed to the orbital motion of the binary stars. The angular momentum redistribution by MRI expands the circumbinary disk. These characteristic structures reproduced in the magnetic field models have recently been observed by ALMA.
\item A circumbinary disk in a binary system is more turbulent
 ($\alpha_{R,\mathrm{turb}}\sim \alpha_{M,\mathrm{turb}} \sim 0.5$ for the fiducial model)
  compared to a circumstellar disk in a single-star system. Turbulence is driven by MRI and is intensified by disturbances from spiral arms, which are caused by the orbital motion of binary stars. We also confirmed that accretion from the infalling envelope contributes to the turbulence in the circumbinary disk.
\item The angular momentum is transferred both in radial and vertical directions. In the radial direction, MRI turbulence and gas flow associated with spiral arms are the main drivers of angular momentum transfer in the circumbinary disk. Specifically, MRI turbulence reduces the specific angular momentum of the circumbinary disk.
In the vertical direction, both outflows and magnetic braking transfer angular momentum, also reducing the specific angular momentum. 
\item Even in the MHD cases, the primary star accretes more gas than the secondary star, as observed in previous hydrodynamic models. This suggests that binary systems tend to evolve into twin binaries. In the MHD models, accretion rates exhibit variability. 
\end{enumerate}

\begin{acknowledgments}
We would like to thank James M. Stone for fruitful discussions and comments.
Numerical computations were carried out on XC50 (ATERUI II) at Center for Computational Astrophysics, National Astronomical Observatory of Japan.
This research was supported in part 
by the Hosei Society of Humanity and Environment, and 
by JSPS KAKENHI Grant Numbers
JP18H05437,
JP17K05394, 
JP23K03464.
\end{acknowledgments}

\appendix
\restartappendixnumbering

\section{Numerical Resolution for MRI}
\label{eq:numerical_resolution}

  The development of the MRI in a numerical simulation hinges on the numerical resolution \citep[e.g.,][]{Sano04}. To probe the numerical resolution requisite for accurately reproducing the MRI, we evaluated its characteristic wavelengths and the MRI quality factors \citep{Noble10,Hawley11,Shiokawa12} for the circumbinary disk (CBD), the primary circumstellar disk (PCSD), and the secondary circumstellar disk (SCSD).

The MRI wavelengths are characterized by the following equations:
\begin{eqnarray}
\lambda_{\mathrm{MRI}, z} &= \frac{2 \pi v_{A,z}}{\Omega},\\
\lambda_{\mathrm{MRI}, \varphi} &= \frac{2 \pi v_{A,\varphi}}{\Omega},
\end{eqnarray}
pertaining to the $z$ and $\varphi$ directions, respectively \citep{Noble10,Hawley11,Shiokawa12}. The term $\lambda_{\mathrm{MRI}, z}$ approximately corresponds to the most unstable wavelength in the initial conditions where the magnetic fields align in the $z$-direction.
The MRI quality factors represent the ratios of the characteristic MRI wavelengths, $\lambda_\mathrm{MRI}$, to the numerical cell sizes. They are defined as $Q_z = \lambda_{\mathrm{MRI, z}}/\Delta x$ and $Q_\varphi = \lambda_{\mathrm{MRI, \varphi}}/\Delta x$ \citep{Noble10,Hawley11,Shiokawa12}. Note that the cell width $\Delta x$ depends on location because of the fixed mesh refinement.

\begin{figure}[t]
\epsscale{1}
\plotone{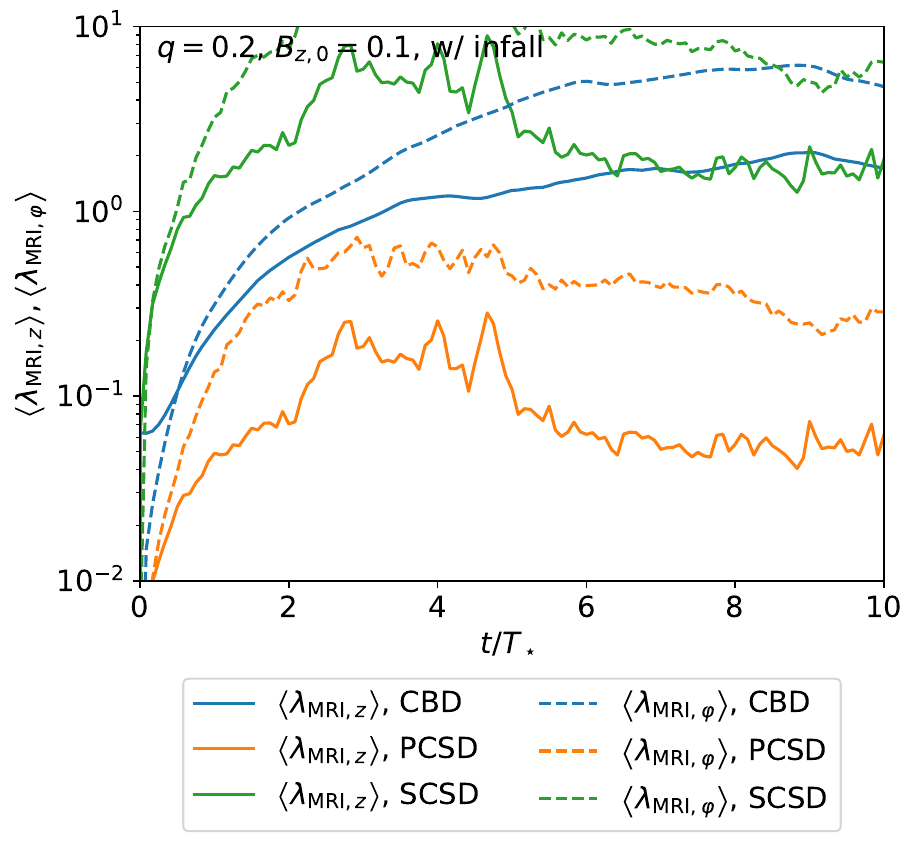}
\caption{
Volume averages of MRI wavelengths for the vertical and azimuthal directions as a function of time for the circumbinary disk (CBD), the primary circumstellar disk (PCSD), and the secondary circumstellar disk (SCSD) in the fiducial model ($q = 0.2$, $B_{z,0} = 0.1$, with the infalling envelope).
  \label{fa1.pdf}}
\end{figure}

\begin{figure*}[t]
\epsscale{0.5}
\plotone{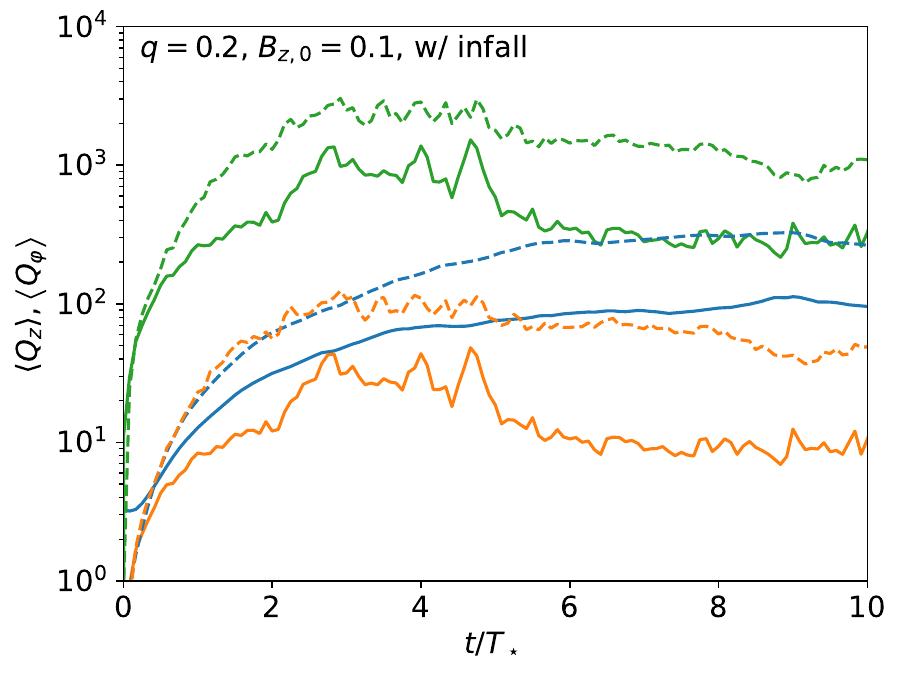}
\plotone{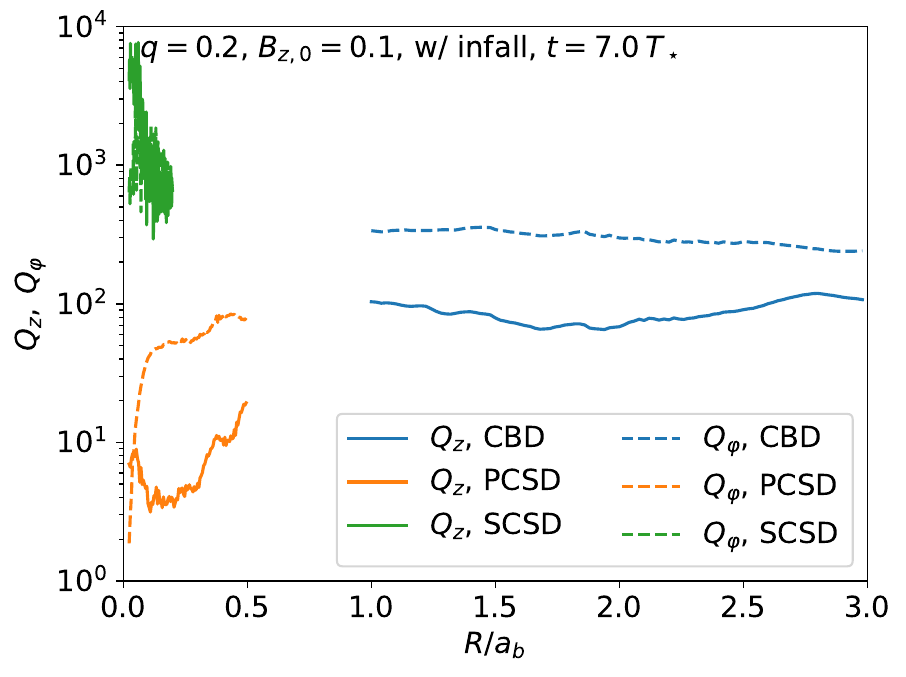}
\caption{
\rev{
The MRI quality factors for the circumbinary disk (CBD), the primary circumstellar disk (PCSD), and the secondary circumstellar disk (SCSD) in the fiducial model ($q = 0.2$, $B_{z,0} = 0.1$, with the infalling envelope).
The left panel shows the volume average of the MRI quality factor for each disk as a function of time. The right panel shows the cylindrical radial distributions of the MRI quality factors at $t = 7\,T_\star$ (corresponding to the stage shown in the third row of Figure~\ref{f2.pdf}). For the PCSD and SCSD, the centers of the radial distributions coincide with the positions of the respective sink particles. The legend in the right panel is common for both the left and right panels.}
  \label{fa2.pdf}}
\end{figure*}

Figures~\ref{fa1.pdf} and ~\ref{fa2.pdf} depict the MRI wavelengths ($\lambda_{\mathrm{MRI}, z}$ and $\lambda_{\mathrm{MRI}, \varphi}$) and the MRI quality factors ($Q_z$ and $Q_\varphi$) throughout the evolution for the CBD, PCSD, and SCSD of the fiducial model. \rev{As regions of the disks, we consider the regions with $\rho > \rho_\mathrm{disk}$ ($=10 \rho_0$), and}
\begin{eqnarray}
 & R \in [1, 3], \;\; z \in [-1, 1], \;\; \text{  for CBD},\\
 &\left[(x-x_1)^2 + (y-y_1)^2\right]^{1/2} \in [r_\mathrm{sink}, 0.5], \nonumber \\ 
 & \quad \quad \quad \quad \quad z \in [-0.1, 0.1], \;\; \text{  for PCSD},\\
 &\left[(x-x_2)^2 + (y-y_2)^2\right]^{1/2} \in [r_\mathrm{sink}, 0.2], \nonumber \\ 
 & \quad \quad \quad \quad \quad z \in [-0.05, 0.05], \;\; \text{  for SCSD},
\end{eqnarray}
where $(x_1, y_1)$ and $(x_2, y_2)$ represent the positions of the primary and secondary stars, respectively, in the $x-y$ plane.
\rev{The volume averages of $\lambda_\mathrm{MRI}$ and $Q$ in these regions are shown in Figure~\ref{fa1.pdf} and in the left panel of Figure~\ref{fa2.pdf}, while the azimuthally and vertically averaged cylindrical radial distribution of $Q$ is presented in the right panel of Figure~\ref{fa2.pdf}.}

For the CBD, at the initial stage, the CBD has $\left<\lambda_{\mathrm{MRI}, z}\right> = 0.06$, which increases to values between $\sim 1$ and $2$. The vertical extent of the CBD lies in the range $z \in [-1, 1]$ (see Figure~\ref{f3.pdf}). Consequently, the CBD undergoes influence from the MRI throughout its evolution.
The MRI quality factor starts with $\left<Q_z\right> = 3.3$ at the onset, but it quickly exceeds 10. Since the required $Q_z$ value to resolve the characteristic wavelength is $Q_z \gtrsim 6$ \citep{Sano04}, the MRI in the CBD is marginally resolved in the early stages of $t \lesssim T_\star$. Following this period, both $\left<Q_z\right>$ and $\left<Q_\varphi\right>$ evolve within the 10-100 range \rev{(left panel of Figure~\ref{fa2.pdf}), and they exhibit spatially constant distributions (right panel of Figure~\ref{fa2.pdf}),} effectively resolving a typical MRI mode \rev{ across the entirety of the CBD.}.

\rev{
The PCSD begins with a value of $\left<\lambda_{\mathrm{MRI}, z}\right> = 0.002$, which subsequently increases to values between $\sim 0.05-0.2$. This is comparable to the disk thickness of $z \in [-0.1, 0.1]$, implying that the PCSD is also influenced by the MRI. Initially, the MRI quality factor is $\left<Q_z\right> = 0.3$. However, by the early phase at $t = 0.8 T_\star$, it exceeds the threshold value of approximately 6. Subsequently, $\left<Q_z\right>$ evolves within the range of $9-40$. The radial distribution of $\left<Q_z\right>$ indicates that it exceeds the threshold value of 6, but drops to around 4 at $R/a_b \sim 0.1 - 0.2$. Consequently, a typical MRI mode is resolved across almost the entire PCSD, although it is marginally resolved at $R/a_b \sim 0.1 - 0.2$.
}

For the SCSD, the initial value of $\left<\lambda_{\mathrm{MRI}, z}\right>$ is 0.06, consistent with the disk thickness of the SCSD $\sim 0.05$. This value exhibits a rapid increase in the very early stages, rising beyond $\sim 1$, which suggests that the SCSD becomes stable against the MRI.  Due to these elongated wavelengths, the MRI quality factors are notably high.

\section{Estimate of alpha parameters}
\label{seq:estimate_alpha_parameters}

The $\alpha$-parameters for the Reynolds and Maxwell stresses are computed using the fluctuating components of the velocity and magnetic field, represented by $\bm{v}^\prime$ and $\bm{B}^\prime$, respectively. The calculation is performed as follows.

All physical variables were transformed from the laboratory frame to the rotating frame with an angular velocity of the binary stars, $\Omega_\star$, using the equation:
\begin{equation}
\bm{U}^\mathrm{rot} = R(\Omega_\star t) \bm{U},
\end{equation}
where $\bm{U} = (\rho, \bm{v}, \bm{B})$ denotes a state vector and $R(\theta)$ denotes the rotation operator with angle $\theta$. This transformation allows the binary stars to be observed as stationary in the rotating frame. In this frame, time averages are performed for all physical variables using the equation:
\begin{equation}
\overline{\bm{U}} = \frac{1}{3T_\star} \int_{7 T_\star}^{10 T_\star} \bm{U}^\mathrm{rot} dt,
\end{equation}
where the period of the time average is set at $t \in [7T_\star, 10T_\star]$ for the fiducial model.

With the time-averaged variables, we can extract the fluctuating parts due to turbulence by subtracting the time-averaged values from the rotating-frame variables:
\begin{eqnarray}
\bm{v}^\prime &=& \bm{v}^\mathrm{rot} - \widetilde{\bm{v}}, \label{eq:vprime}\\
\bm{B}^\prime &=& \bm{B}^\mathrm{rot} - \overline{\bm{B}}, \label{eq:bprime}
\end{eqnarray}
where we use the Favre average $\widetilde{\bm{v}}$, defined as
\begin{equation}
\widetilde{\bm{v}} = \frac{\overline{\rho \bm{v}}}{\overline{\rho}},
\end{equation}
because we consider compressible flows in our simulations.

The time-averaged stress components related to angular momentum transport, $v_R^\prime v_\varphi^\prime$ (the Reynolds stress) and $B_R^\prime B_\varphi^\prime$ (the Maxwell stress), are computed using the fluctuating parts defined by Equations~(\ref{eq:vprime})--(\ref{eq:bprime}), as follows:
\begin{eqnarray}
\alpha_{R,\mathrm{turb}} &=& \frac{\int \overline{\rho v_R^\prime v_\varphi^\prime} dz}{\int \overline{p}dz}, \label{eq:alpha_R}\\
\alpha_{M,\mathrm{turb}} &=& \frac{\int - \overline{B_R^\prime B_\varphi^\prime} dz}{4\pi \int \overline{p}dz}, \label{eq:alpha_M}
\end{eqnarray}
where integration along the $z$-direction is performed for a region with density larger than $\rho_\mathrm{disk}$:
\begin{equation}
\int dz = \int_{\rho \ge \rho_\mathrm{disk}} dz.
\end{equation}
We set $\rho_\mathrm{disk} = 10 \rho_0$, which traces the surfaces of the circumbinary disk. 

Equations (\ref{eq:alpha_R}) and (\ref{eq:alpha_M}) are $\alpha$ parameters contributed by turbulence.
We also define the $\alpha$ parameters of mean flow and mean magnetic field as follows,
\begin{eqnarray}
\alpha_{R,\mathrm{mean}} &=& \frac{\int \overline{\rho} \widetilde{v}_R \widetilde{v}_\varphi dz}{\int \overline{p}dz}, \label{eq:alpha_R_mean}\\
\alpha_{M,\mathrm{mean}} &=& \frac{\int - \overline{B}_R \overline{B}_\varphi dz}{4\pi \int \overline{p}dz}, \label{eq:alpha_M_mean}
\end{eqnarray}
Because of $\overline{\rho v_R v_\varphi} = \overline{\rho} \widetilde{v}_R \widetilde{v}_\varphi + \overline{\rho v_R^\prime v_\varphi^\prime}$ and
$\overline{B_R B_\varphi} = \overline{B}_R \overline{B}_\varphi + \overline{B_R^\prime B_\varphi^\prime}$,
we also define $\alpha$-parameters for total stress
\begin{eqnarray}
\alpha_{R,\mathrm{total}} &=& \alpha_{R,\mathrm{mean}} + \alpha_{R,\mathrm{turb}}, \label{eq:alpha_R_total}\\
\alpha_{M,\mathrm{total}} &=& \alpha_{M,\mathrm{mean}} + \alpha_{M,\mathrm{turb}}. \label{eq:alpha_M_total}
\end{eqnarray}

\section{Angular momentum transport through an enclosed cylinder}
\label{sec:am_transport}
We consider a cylindrical region with a radius of $R_\mathrm{max}$ and a range of $z \in [z_\mathrm{min}, z_\mathrm{max}]$.
The angular momentum transport in the $R$ and $z$ directions are expressed by
\begin{eqnarray}
  \frac{\partial}{\partial t} (\rho j) &+& \frac{1}{R}\frac{\partial}{\partial R} \left[R^2 (T_{R\varphi}+M_{R\varphi})\right] \nonumber \\
  &+& \frac{\partial}{\partial z} \left[R (T_{z\varphi}+M_{z\varphi})\right] = R \rho g_\varphi,
  \label{eq:am_conservation}
\end{eqnarray}
where 
$j = R v_\varphi$ is a specific angular momentum, and
\begin{eqnarray}
  T_{R \varphi} &=& \rho v_R v_\varphi,\\
  T_{z \varphi} &=& \rho v_z v_\varphi,\\
  M_{R \varphi} &=& -B_R B_\varphi/4\pi,\\
  M_{z \varphi} &=& -B_z B_\varphi/4\pi.
\end{eqnarray}

The volume integration of $\int dV = \int_0^{R_\mathrm{max}} R dR \int_0^{2 \pi} d\varphi \int_{z_\mathrm{min}}^{z_\mathrm{max}} dz$ is applied to Equation~(\ref{eq:am_conservation}), and an integral form is obtained, given by
\begin{eqnarray}
  \lefteqn{ \frac{\partial}{\partial t} \int \rho j dV } \nonumber\\
  &+& \int_0^{2 \pi} d\varphi \int_{z_\mathrm{min}}^{z_\mathrm{max}} dz R^2\left[T_{R\varphi}+M_{R\varphi}\right]_0^{R_\mathrm{max}} \nonumber\\
  &+& \int_0^{2 \pi} d\varphi \int_0^{R_\mathrm{max}} R^2 dR \left[T_{z\varphi}+M_{z\varphi}\right]_{z_\mathrm{min}}^{z_\mathrm{max}} \nonumber\\
  &=& \int_0^{2 \pi} d\varphi \int_0^{R_\mathrm{max}} R^2 dR \int_{z_\mathrm{min}}^{z_\mathrm{max}} dz \rho g_\varphi.
  \label{eq:am_conservation2}
\end{eqnarray}
The first term in the left-hand side of Equation~(\ref{eq:am_conservation2}) denotes the rate of change of the angular momentum of the region of interest.

We define the fluxes of angular momenta in the $R$ and $z$ directions as functions of $R$ and $z$, respectively, from the second and third terms in the left-hand side of Equation~(\ref{eq:am_conservation2}). These are given by
\begin{eqnarray}
  F_R (R) &=& \int_0^{2 \pi} d\varphi \int_{z_\mathrm{min}}^{z_\mathrm{max}} dz R^2 \left(T_{R\varphi}+M_{R\varphi}\right),
  \label{eq:appendix_F_R}\\
F_z (z) &=& \int_0^{2 \pi} d\varphi \int_0^{R_\mathrm{max}} R^2 dR \left(T_{z\varphi}+M_{z\varphi}\right),
\label{eq:appendix_F_z}
\end{eqnarray}
where $F_R(R)$ is the angular momentum flux through the side of an enclosed cylinder with a radius of $R$ and a range of $[z_\mathrm{min}, z_\mathrm{max}]$, and $F_z(z)$ is the angular momentum flux through the base of an enclosed cylinder with a radius of $R_\mathrm{max}$ and a height of $z$.

We also define a gravitational torque component as a function of $R$ from the right-hand side of Equation~(\ref{eq:am_conservation2}). This is given by 
\begin{equation}
S_g(R) = \int_0^{2 \pi} d\varphi \int_0^R R^\prime {}^2 dR^\prime \int_{z_\mathrm{min}}^{z_\mathrm{max}} dz \rho g_\varphi.
\label{eq:appendix_F_g}
\end{equation}
where $S_g(R)$ is the gravitational torque acting on the region of interest of the cylinder with the radius $R$ and the height, $[z_\mathrm{min}, z_\mathrm{max}]$, although it is not an angular momentum flux.

With Equations~(\ref{eq:am_conservation2})---(\ref{eq:appendix_F_g}), change in the angular momentum is expressed by
\begin{equation}
\frac{\partial J}{\partial t} + F_R(R_\mathrm{max}) + F_z(z_\mathrm{max}) - F_z(z_\mathrm{min}) = S_g(R_\mathrm{max}),
\label{eq:dJ/dt}
\end{equation}
where $J = \int \rho j dV$.

Figure~\ref{f18a.pdf} displays several contributions to the fluxes separately. The line of $T_{R\varphi}$ represents the contribution of the $T_{R\varphi}$ term in equation~(\ref{eq:appendix_F_R}), i.e.,
\begin{equation}
F_R(R) = \int_0^{2 \pi} d\varphi \int_{-2}^{2} dz R^2 T_{R\varphi}. \;\;\;\text{(for $T_{R\varphi}$ component)}
\end{equation}
The contributions of $M_{R\varphi}$, $T_{z\varphi}$, and $M_{z\varphi}$ are calculated in the same manner. The turbulent contributions, which are shown by the dashed lines in Figure~\ref{f18a.pdf}, is calculated, taking into account only the fluctuating components of the stresses, e.g.,
\begin{eqnarray}
& F_R(R) =\int_0^{2 \pi}  d\varphi \int_{-2}^{2} dz R^2 \overline{\rho v_R^\prime v_\varphi^\prime} . \quad \quad \quad \quad \quad \quad  \nonumber\\ 
& \quad \quad \quad \quad \text{(for turbulent contribution of $T_{R\varphi}$)}
\end{eqnarray}
The outflow contribution (green line in the lower panel of Figure~\ref{f18a.pdf}) is estimated by considering the $T_{z\varphi}$ term in Equation~(\ref{eq:appendix_F_z}), but taking into account the cells with $\widetilde{v}_z > 0$ in the $z > 0$ region and $\widetilde{v}_z < 0$ in the $z < 0$ region when evaluating the integration.
In the scale of the circumbinary disk, contribution of $S_g(R)$ is much less than $F_r(R)$, and we only show $F_r(R)$ in Figure~\ref{f18a.pdf} (upper panel) for simplicity.

\section{Change in mean specific angular momentum}
\label{sec:change_in_sam}
Change in the mean specific angular momentum in the region of interest is give by
\begin{equation}
\frac{\partial}{\partial t}\left(\frac{J}{M}\right) = \frac{J}{M}\left(\frac{1}{J}\frac{\partial J}{\partial t} - \frac{1}{M}\frac{\partial M}{\partial t} \right).
\label{eq:dJ/Md/t}
\end{equation}
We consider the angular momentum and mass inside a cylinder with the radius $R_\mathrm{max}$ and the height $[z_\mathrm{min}, z_\mathrm{max}]$, and they are given by,
\begin{eqnarray}
J &=& \int \rho j dV \nonumber \\
  &=& \int_0^{R_\mathrm{max}} R dR\int_0^{2 \pi} d\varphi \int_{z_\mathrm{min}}^{z_\mathrm{max}} dz \rho j,
\label{eq:J_integ}\\
M &=& \int \rho dV \nonumber \\
&=& \int_0^{R_\mathrm{max}} R dR\int_0^{2 \pi} d\varphi \int_{z_\mathrm{min}}^{z_\mathrm{max}} dz \rho,
\label{eq:M_integ}
\end{eqnarray}
where $j=Rv_\varphi$, and the volume integration $\int dV$ is performed over the region of interest.
The time derivative of mass $\partial M/\partial t$ is given by Gauss' law as follows,
\begin{eqnarray}
\frac{\partial M}{\partial t} &=& -\int \rho \bm{v} \cdot d\bm{S} \nonumber \\
  &=& -\int_0^{2 \pi} d\varphi \int_{z_\mathrm{min}}^{z_\mathrm{max}} dz \left[ R \rho v_R \right]_0^{R_\mathrm{max}}\nonumber\\
&-&\int_0^{R_\mathrm{max}} dR\int_0^{2 \pi} d\varphi \left[ R \rho v_z \right]_{z_\mathrm{min}}^{z_\mathrm{max}}
\label{eq:dM/dt}
\end{eqnarray}
Using Equation~(\ref{eq:dJ/dt}) without contribution of $S_g$, and Equations (\ref{eq:J_integ})---(\ref{eq:dM/dt}), Equation~(\ref{eq:dJ/Md/t}) can be evaluated.

When the timescale of the specific angular momentum is defined as $\tau_\mathrm{sam}$, the rate of change in the specific angular momentum is given by
\begin{equation}
\frac{1}{\tau_\mathrm{sam}} = \frac{M}{J}
\frac{\partial}{\partial t}\left(\frac{J}{M}\right) = \frac{1}{J}\frac{\partial J}{\partial t} - \frac{1}{M}\frac{\partial M}{\partial t}.
\label{eq:1overtausam}
\end{equation}
When applying equation~(\ref{eq:1overtausam}) to the models in Figures~\ref{f19.pdf} and \ref{f20a.pdf}, both the radius and height of the cylinder were set to $R$, and $\tau_\mathrm{sam}^{-1}$ was calculated as a function of $R$.
In Figure~\ref{f19.pdf}, $\tau_\mathrm{sam}^{-1}$ are shown for several contribution separately. The line of ``$T_\mathrm{R\varphi} + R$-flow'' means the rate of change in specific angular momentum due to $T_\mathrm{R\varphi}$ and radial flow. It is calculated based on Equation~(\ref{eq:1overtausam}), 
\begin{equation}
\frac{1}{\tau_\mathrm{sam}} = \frac{1}{J}\left.\frac{\partial J}{\partial t}\right|_{T_\mathrm{R\varphi}} - \frac{1}{M}\left.\frac{\partial M}{\partial t}\right|_{R-\mathrm{flow}}
\end{equation}
The term $\left. \partial J/\partial t \right|_{T_\mathrm{R\varphi}}$ is evaluated based on Equations~(\ref{eq:appendix_F_R}) and (\ref{eq:dJ/dt}),
\begin{equation}
\left.\frac{\partial J}{\partial t}\right|_{T_\mathrm{R\varphi}} 
= -\int_0^{2 \pi} d\varphi \int_{-R}^{R} dz R^2 T_{R\varphi}.
\end{equation}
The term $\left. \partial M/\partial t \right|_{R-\mathrm{flow}}$ is evaluated based on Equations~(\ref{eq:dM/dt}),
\begin{equation}
\left. \frac{\partial M}{\partial t} \right|_{R-\mathrm{flow}} = -\int_0^{2 \pi} d\varphi \int_{-R}^{R} dz \left[ R \rho v_R \right]_0^{R}.
\end{equation}
Other contributions were calculated in a similar method.


\software{SFUMATO \citep{Matsumoto07}, Paraview}

\bibliography{ms}{}

\begin{thebibliography}{}
\expandafter\ifx\csname natexlab\endcsname\relax\def\natexlab#1{#1}\fi
\providecommand{\url}[1]{\href{#1}{#1}}
\providecommand{\dodoi}[1]{doi:~\href{http://doi.org/#1}{\nolinkurl{#1}}}
\providecommand{\doeprint}[1]{\href{http://ascl.net/#1}{\nolinkurl{http://ascl.net/#1}}}
\providecommand{\doarXiv}[1]{\href{https://arxiv.org/abs/#1}{\nolinkurl{https://arxiv.org/abs/#1}}}

\bibitem[{{Alves} {et~al.}(2019){Alves}, {Caselli}, {Girart}, {Segura-Cox},
  {Franco}, {Schmiedeke}, \& {Zhao}}]{Alves19}
{Alves}, F.~O., {Caselli}, P., {Girart}, J.~M., {et~al.} 2019, Science, 366,
  90, \dodoi{10.1126/science.aaw3491}

\bibitem[{{Alves} {et~al.}(2017){Alves}, {Girart}, {Caselli}, {Franco}, {Zhao},
  {Vlemmings}, {Evans}, \& {Ricci}}]{Alves17}
{Alves}, F.~O., {Girart}, J.~M., {Caselli}, P., {et~al.} 2017, \aap, 603, L3,
  \dodoi{10.1051/0004-6361/201731077}

\bibitem[{{Avara} {et~al.}(2023){Avara}, {Krolik}, {Campanelli}, {Noble},
  {Bowen}, \& {Ryu}}]{Avara23}
{Avara}, M.~J., {Krolik}, J.~H., {Campanelli}, M., {et~al.} 2023, arXiv
  e-prints, arXiv:2305.18538, \dodoi{10.48550/arXiv.2305.18538}

\bibitem[{{Bai} \& {Stone}(2013)}]{Bai13}
{Bai}, X.-N., \& {Stone}, J.~M. 2013, \apj, 769, 76,
  \dodoi{10.1088/0004-637X/769/1/76}

\bibitem[{{Balbus} \& {Hawley}(1991)}]{Balbus91}
{Balbus}, S.~A., \& {Hawley}, J.~F. 1991, \apj, 376, 214,
  \dodoi{10.1086/170270}

\bibitem[{{Barge} \& {Sommeria}(1995)}]{Barge95}
{Barge}, P., \& {Sommeria}, J. 1995, \aap, 295, L1,
  \dodoi{10.48550/arXiv.astro-ph/9501050}

\bibitem[{{Bate}(2000)}]{Bate00}
{Bate}, M.~R. 2000, \mnras, 314, 33, \dodoi{10.1046/j.1365-8711.2000.03333.x}

\bibitem[{{Bate} \& {Bonnell}(1997)}]{Bate97}
{Bate}, M.~R., \& {Bonnell}, I.~A. 1997, \mnras, 285, 33,
  \dodoi{10.1093/mnras/285.1.33}

\bibitem[{{Blandford} \& {Payne}(1982)}]{Blandford82}
{Blandford}, R.~D., \& {Payne}, D.~G. 1982, \mnras, 199, 883,
  \dodoi{10.1093/mnras/199.4.883}

\bibitem[{{Bowen} {et~al.}(2018){Bowen}, {Mewes}, {Campanelli}, {Noble},
  {Krolik}, \& {Zilh{\~a}o}}]{Bowen18}
{Bowen}, D.~B., {Mewes}, V., {Campanelli}, M., {et~al.} 2018, \apjl, 853, L17,
  \dodoi{10.3847/2041-8213/aaa756}

\bibitem[{{Ching} {et~al.}(2016){Ching}, {Lai}, {Zhang}, {Yang}, {Girart}, \&
  {Rao}}]{Ching16}
{Ching}, T.-C., {Lai}, S.-P., {Zhang}, Q., {et~al.} 2016, \apj, 819, 159,
  \dodoi{10.3847/0004-637X/819/2/159}

\bibitem[{{Cunningham} {et~al.}(2011){Cunningham}, {Klein}, {Krumholz}, \&
  {McKee}}]{Cunningham11}
{Cunningham}, A.~J., {Klein}, R.~I., {Krumholz}, M.~R., \& {McKee}, C.~F. 2011,
  \apj, 740, 107, \dodoi{10.1088/0004-637X/740/2/107}

\bibitem[{{Dittmann} \& {Ryan}(2022)}]{Dittmann22}
{Dittmann}, A.~J., \& {Ryan}, G. 2022, \mnras, 513, 6158,
  \dodoi{10.1093/mnras/stac935}

\bibitem[{{Doyle} {et~al.}(2011){Doyle}, {Carter}, {Fabrycky}, {Slawson},
  {Howell}, {Winn}, {Orosz}, {P{\v{r}}sa}, {Welsh}, {Quinn}, {Latham},
  {Torres}, {Buchhave}, {Marcy}, {Fortney}, {Shporer}, {Ford}, {Lissauer},
  {Ragozzine}, {Rucker}, {Batalha}, {Jenkins}, {Borucki}, {Koch}, {Middour},
  {Hall}, {McCauliff}, {Fanelli}, {Quintana}, {Holman}, {Caldwell}, {Still},
  {Stefanik}, {Brown}, {Esquerdo}, {Tang}, {Furesz}, {Geary}, {Berlind},
  {Calkins}, {Short}, {Steffen}, {Sasselov}, {Dunham}, {Cochran}, {Boss},
  {Haas}, {Buzasi}, \& {Fischer}}]{Doyle11}
{Doyle}, L.~R., {Carter}, J.~A., {Fabrycky}, D.~C., {et~al.} 2011, Science,
  333, 1602, \dodoi{10.1126/science.1210923}

\bibitem[{{Duch{\^e}ne} \& {Kraus}(2013)}]{Duchene13}
{Duch{\^e}ne}, G., \& {Kraus}, A. 2013, \araa, 51, 269,
  \dodoi{10.1146/annurev-astro-081710-102602}

\bibitem[{{Dutrey} {et~al.}(2014){Dutrey}, {di Folco}, {Guilloteau}, {Boehler},
  {Bary}, {Beck}, {Beust}, {Chapillon}, {Gueth}, {Hur{\'e}}, {Pierens},
  {Pi{\'e}tu}, {Simon}, \& {Tang}}]{Dutrey14}
{Dutrey}, A., {di Folco}, E., {Guilloteau}, S., {et~al.} 2014, \nat, 514, 600,
  \dodoi{10.1038/nature13822}

\bibitem[{{El-Badry} {et~al.}(2019){El-Badry}, {Rix}, {Tian}, {Duch{\^e}ne}, \&
  {Moe}}]{El-badry19}
{El-Badry}, K., {Rix}, H.-W., {Tian}, H., {Duch{\^e}ne}, G., \& {Moe}, M. 2019,
  \mnras, 489, 5822, \dodoi{10.1093/mnras/stz2480}

\bibitem[{{Fukagawa} {et~al.}(2013){Fukagawa}, {Tsukagoshi}, {Momose}, {Saigo},
  {Ohashi}, {Kitamura}, {Inutsuka}, {Muto}, {Nomura}, {Takeuchi}, {Kobayashi},
  {Hanawa}, {Akiyama}, {Honda}, {Fujiwara}, {Kataoka}, {Takahashi}, \&
  {Shibai}}]{Fukagawa13}
{Fukagawa}, M., {Tsukagoshi}, T., {Momose}, M., {et~al.} 2013, \pasj, 65, L14,
  \dodoi{10.1093/pasj/65.6.L14}

\bibitem[{{Gerrard} {et~al.}(2019){Gerrard}, {Federrath}, \&
  {Kuruwita}}]{Gerrard19}
{Gerrard}, I.~A., {Federrath}, C., \& {Kuruwita}, R. 2019, \mnras, 485, 5532,
  \dodoi{10.1093/mnras/stz784}

\bibitem[{{Gombosi} {et~al.}(2002){Gombosi}, {T{\'o}th}, {De Zeeuw}, {Hansen},
  {Kabin}, \& {Powell}}]{Gombosi02}
{Gombosi}, T.~I., {T{\'o}th}, G., {De Zeeuw}, D.~L., {et~al.} 2002, Journal of
  Computational Physics, 177, 176, \dodoi{10.1006/jcph.2002.7009}

\bibitem[{{Grudi{\'c}} {et~al.}(2021){Grudi{\'c}}, {Guszejnov}, {Hopkins},
  {Offner}, \& {Faucher-Gigu{\`e}re}}]{Grudic21}
{Grudi{\'c}}, M.~Y., {Guszejnov}, D., {Hopkins}, P.~F., {Offner}, S. S.~R., \&
  {Faucher-Gigu{\`e}re}, C.-A. 2021, \mnras, 506, 2199,
  \dodoi{10.1093/mnras/stab1347}

\bibitem[{{Hanawa} {et~al.}(2010){Hanawa}, {Ochi}, \& {Ando}}]{Hanawa10}
{Hanawa}, T., {Ochi}, Y., \& {Ando}, K. 2010, \apj, 708, 485,
  \dodoi{10.1088/0004-637X/708/1/485}

\bibitem[{{Hara} {et~al.}(2021){Hara}, {Kawabe}, {Nakamura}, {Hirano},
  {Takakuwa}, {Shimajiri}, {Kamazaki}, {Di Francesco}, {Machida}, {Tamura},
  {Saigo}, {Matsumoto}, \& {Tomida}}]{Hara2021}
{Hara}, C., {Kawabe}, R., {Nakamura}, F., {et~al.} 2021, \apj, 912, 34,
  \dodoi{10.3847/1538-4357/abb810}

\bibitem[{{Hawley} {et~al.}(2011){Hawley}, {Guan}, \& {Krolik}}]{Hawley11}
{Hawley}, J.~F., {Guan}, X., \& {Krolik}, J.~H. 2011, \apj, 738, 84,
  \dodoi{10.1088/0004-637X/738/1/84}

\bibitem[{{Hayashi} {et~al.}(1993){Hayashi}, {Ohashi}, \& {Miyama}}]{Hayashi93}
{Hayashi}, M., {Ohashi}, N., \& {Miyama}, S.~M. 1993, \apjl, 418, L71,
  \dodoi{10.1086/187119}

\bibitem[{{Hioki} {et~al.}(2007){Hioki}, {Itoh}, {Oasa}, {Fukagawa}, {Kudo},
  {Mayama}, {Funayama}, {Hayashi}, {Hayashi}, {Pyo}, {Ishii}, {Nishikawa}, \&
  {Tamura}}]{Hioki07}
{Hioki}, T., {Itoh}, Y., {Oasa}, Y., {et~al.} 2007, \aj, 134, 880,
  \dodoi{10.1086/519737}

\bibitem[{{Hwang} {et~al.}(2022){Hwang}, {El-Badry}, {Rix}, {Hamilton}, {Ting},
  \& {Zakamska}}]{Hwang22}
{Hwang}, H.-C., {El-Badry}, K., {Rix}, H.-W., {et~al.} 2022, \apjl, 933, L32,
  \dodoi{10.3847/2041-8213/ac7c70}

\bibitem[{{Klahr} \& {Henning}(1997)}]{Klahr97}
{Klahr}, H.~H., \& {Henning}, T. 1997, \icarus, 128, 213,
  \dodoi{10.1006/icar.1997.5720}

\bibitem[{{Klessen} \& {Hennebelle}(2010)}]{Klessen2010}
{Klessen}, R.~S., \& {Hennebelle}, P. 2010, \aap, 520, A17,
  \dodoi{10.1051/0004-6361/200913780}

\bibitem[{{Kratter} {et~al.}(2010){Kratter}, {Matzner}, {Krumholz}, \&
  {Klein}}]{Kratter10}
{Kratter}, K.~M., {Matzner}, C.~D., {Krumholz}, M.~R., \& {Klein}, R.~I. 2010,
  \apj, 708, 1585, \dodoi{10.1088/0004-637X/708/2/1585}

\bibitem[{{Lai} \& {Mu{\~n}oz}(2022)}]{Lai22}
{Lai}, D., \& {Mu{\~n}oz}, D.~J. 2022, arXiv e-prints, arXiv:2211.00028,
  \dodoi{10.48550/arXiv.2211.00028}

\bibitem[{{Lee} {et~al.}(2018){Lee}, {Hwang}, {Ching}, {Hirano}, {Lai}, {Rao},
  \& {Ho}}]{Lee18}
{Lee}, C.-F., {Hwang}, H.-C., {Ching}, T.-C., {et~al.} 2018, Nature
  Communications, 9, 4636, \dodoi{10.1038/s41467-018-07143-8}

\bibitem[{{Lee} {et~al.}(2021){Lee}, {Charnoz}, \& {Hennebelle}}]{Lee21}
{Lee}, Y.-N., {Charnoz}, S., \& {Hennebelle}, P. 2021, \aap, 648, A101,
  \dodoi{10.1051/0004-6361/202038105}

\bibitem[{{Lopez Armengol} {et~al.}(2021){Lopez Armengol}, {Combi},
  {Campanelli}, {Noble}, {Krolik}, {Bowen}, {Avara}, {Mewes}, \&
  {Nakano}}]{Lopez-Armengol21}
{Lopez Armengol}, F.~G., {Combi}, L., {Campanelli}, M., {et~al.} 2021, \apj,
  913, 16, \dodoi{10.3847/1538-4357/abf0af}

\bibitem[{{Machida} {et~al.}(2008){Machida}, {Inutsuka}, \&
  {Matsumoto}}]{Machida08}
{Machida}, M.~N., {Inutsuka}, S.-i., \& {Matsumoto}, T. 2008, \apj, 676, 1088,
  \dodoi{10.1086/528364}

\bibitem[{{Machida} {et~al.}(2011){Machida}, {Inutsuka}, \&
  {Matsumoto}}]{Machida11}
{Machida}, M.~N., {Inutsuka}, S.-I., \& {Matsumoto}, T. 2011, \pasj, 63, 555,
  \dodoi{10.1093/pasj/63.3.555}

\bibitem[{{Marchand} {et~al.}(2020){Marchand}, {Tomida}, {Tanaka},
  {Commer{\c{c}}on}, \& {Chabrier}}]{Marchand19}
{Marchand}, P., {Tomida}, K., {Tanaka}, K. E.~I., {Commer{\c{c}}on}, B., \&
  {Chabrier}, G. 2020, \apj, 900, 180, \dodoi{10.3847/1538-4357/abad99}

\bibitem[{{Masson} {et~al.}(2016){Masson}, {Chabrier}, {Hennebelle}, {Vaytet},
  \& {Commer{\c{c}}on}}]{Masson16}
{Masson}, J., {Chabrier}, G., {Hennebelle}, P., {Vaytet}, N., \&
  {Commer{\c{c}}on}, B. 2016, \aap, 587, A32,
  \dodoi{10.1051/0004-6361/201526371}

\bibitem[{{Masunaga} {et~al.}(1998){Masunaga}, {Miyama}, \&
  {Inutsuka}}]{Masunaga98}
{Masunaga}, H., {Miyama}, S.~M., \& {Inutsuka}, S.-i. 1998, \apj, 495, 346,
  \dodoi{10.1086/305281}

\bibitem[{{Matsumoto}(2007)}]{Matsumoto07}
{Matsumoto}, T. 2007, \pasj, 59, 905, \dodoi{10.1093/pasj/59.5.905}

\bibitem[{{Matsumoto} {et~al.}(2015){Matsumoto}, {Dobashi}, \&
  {Shimoikura}}]{Matsumoto15}
{Matsumoto}, T., {Dobashi}, K., \& {Shimoikura}, T. 2015, \apj, 801, 77,
  \dodoi{10.1088/0004-637X/801/2/77}

\bibitem[{{Matsumoto} \& {Hanawa}(2003)}]{Matsumoto03}
{Matsumoto}, T., \& {Hanawa}, T. 2003, \apj, 595, 913, \dodoi{10.1086/377367}

\bibitem[{{Matsumoto} {et~al.}(2017){Matsumoto}, {Machida}, \&
  {Inutsuka}}]{Matsumoto17}
{Matsumoto}, T., {Machida}, M.~N., \& {Inutsuka}, S.-i. 2017, \apj, 839, 69,
  \dodoi{10.3847/1538-4357/aa6a1c}

\bibitem[{{Matsumoto} {et~al.}(2019{\natexlab{a}}){Matsumoto}, {Miyoshi}, \&
  {Takasao}}]{Matsumoto19HLLD}
{Matsumoto}, T., {Miyoshi}, T., \& {Takasao}, S. 2019{\natexlab{a}}, \apj, 874,
  37, \dodoi{10.3847/1538-4357/ab05cb}

\bibitem[{{Matsumoto} {et~al.}(2019{\natexlab{b}}){Matsumoto}, {Saigo}, \&
  {Takakuwa}}]{Matsumoto19}
{Matsumoto}, T., {Saigo}, K., \& {Takakuwa}, S. 2019{\natexlab{b}}, \apj, 871,
  36, \dodoi{10.3847/1538-4357/aaf6ab}

\bibitem[{{Miyoshi} \& {Kusano}(2005)}]{Miyoshi05}
{Miyoshi}, T., \& {Kusano}, K. 2005, Journal of Computational Physics, 208,
  315, \dodoi{10.1016/j.jcp.2005.02.017}

\bibitem[{{Moe} \& {Di Stefano}(2017)}]{Moe17}
{Moe}, M., \& {Di Stefano}, R. 2017, \apjs, 230, 15,
  \dodoi{10.3847/1538-4365/aa6fb6}

\bibitem[{{Moody} {et~al.}(2019){Moody}, {Shi}, \& {Stone}}]{Moody19}
{Moody}, M. S.~L., {Shi}, J.-M., \& {Stone}, J.~M. 2019, \apj, 875, 66,
  \dodoi{10.3847/1538-4357/ab09ee}

\bibitem[{{Nakano} {et~al.}(2002){Nakano}, {Nishi}, \& {Umebayashi}}]{Nakano02}
{Nakano}, T., {Nishi}, R., \& {Umebayashi}, T. 2002, \apj, 573, 199,
  \dodoi{10.1086/340587}

\bibitem[{{Noble} {et~al.}(2021){Noble}, {Krolik}, {Campanelli}, {Zlochower},
  {Mundim}, {Nakano}, \& {Zilh{\~a}o}}]{Noble21}
{Noble}, S.~C., {Krolik}, J.~H., {Campanelli}, M., {et~al.} 2021, \apj, 922,
  175, \dodoi{10.3847/1538-4357/ac2229}

\bibitem[{{Noble} {et~al.}(2010){Noble}, {Krolik}, \& {Hawley}}]{Noble10}
{Noble}, S.~C., {Krolik}, J.~H., \& {Hawley}, J.~F. 2010, \apj, 711, 959,
  \dodoi{10.1088/0004-637X/711/2/959}

\bibitem[{{Noble} {et~al.}(2012){Noble}, {Mundim}, {Nakano}, {Krolik},
  {Campanelli}, {Zlochower}, \& {Yunes}}]{Noble12}
{Noble}, S.~C., {Mundim}, B.~C., {Nakano}, H., {et~al.} 2012, \apj, 755, 51,
  \dodoi{10.1088/0004-637X/755/1/51}

\bibitem[{{Ochi} {et~al.}(2005){Ochi}, {Sugimoto}, \& {Hanawa}}]{Ochi05}
{Ochi}, Y., {Sugimoto}, K., \& {Hanawa}, T. 2005, \apj, 623, 922,
  \dodoi{10.1086/428601}

\bibitem[{{Offner} {et~al.}(2022){Offner}, {Moe}, {Kratter}, {Sadavoy},
  {Jensen}, \& {Tobin}}]{Offner22}
{Offner}, S. S.~R., {Moe}, M., {Kratter}, K.~M., {et~al.} 2022, arXiv e-prints,
  arXiv:2203.10066, \dodoi{10.48550/arXiv.2203.10066}

\bibitem[{{Onishi} {et~al.}(2002){Onishi}, {Mizuno}, {Kawamura}, {Tachihara},
  \& {Fukui}}]{Onishi02}
{Onishi}, T., {Mizuno}, A., {Kawamura}, A., {Tachihara}, K., \& {Fukui}, Y.
  2002, \apj, 575, 950, \dodoi{10.1086/341347}

\bibitem[{{Pierens} {et~al.}(2020){Pierens}, {McNally}, \&
  {Nelson}}]{Pierens20}
{Pierens}, A., {McNally}, C.~P., \& {Nelson}, R.~P. 2020, \mnras, 496, 2849,
  \dodoi{10.1093/mnras/staa1550}

\bibitem[{{Pudritz} \& {Norman}(1986)}]{Pudritz86}
{Pudritz}, R.~E., \& {Norman}, C.~A. 1986, \apj, 301, 571,
  \dodoi{10.1086/163924}

\bibitem[{{Raghavan} {et~al.}(2010){Raghavan}, {McAlister}, {Henry}, {Latham},
  {Marcy}, {Mason}, {Gies}, {White}, \& {ten Brummelaar}}]{Raghavan10}
{Raghavan}, D., {McAlister}, H.~A., {Henry}, T.~J., {et~al.} 2010, \apjs, 190,
  1, \dodoi{10.1088/0067-0049/190/1/1}

\bibitem[{{Ragusa} {et~al.}(2017){Ragusa}, {Dipierro}, {Lodato}, {Laibe}, \&
  {Price}}]{Ragusa17}
{Ragusa}, E., {Dipierro}, G., {Lodato}, G., {Laibe}, G., \& {Price}, D.~J.
  2017, \mnras, 464, 1449, \dodoi{10.1093/mnras/stw2456}

\bibitem[{{Reipurth} {et~al.}(2014){Reipurth}, {Clarke}, {Boss}, {Goodwin},
  {Rodr{\'\i}guez}, {Stassun}, {Tokovinin}, \& {Zinnecker}}]{Reipurth14}
{Reipurth}, B., {Clarke}, C.~J., {Boss}, A.~P., {et~al.} 2014, in Protostars
  and Planets VI, ed. H.~{Beuther}, R.~S. {Klessen}, C.~P. {Dullemond}, \&
  T.~{Henning}, 267--290, \dodoi{10.2458/azu_uapress_9780816531240-ch012}

\bibitem[{{Sano} \& {Inutsuka}(2001)}]{Sano01}
{Sano}, T., \& {Inutsuka}, S.-i. 2001, \apjl, 561, L179, \dodoi{10.1086/324763}

\bibitem[{{Sano} {et~al.}(2004){Sano}, {Inutsuka}, {Turner}, \&
  {Stone}}]{Sano04}
{Sano}, T., {Inutsuka}, S.-i., {Turner}, N.~J., \& {Stone}, J.~M. 2004, \apj,
  605, 321, \dodoi{10.1086/382184}

\bibitem[{{Satsuka} {et~al.}(2017){Satsuka}, {Tsuribe}, {Tanaka}, \&
  {Nagamine}}]{Satsuka17}
{Satsuka}, T., {Tsuribe}, T., {Tanaka}, S., \& {Nagamine}, K. 2017, \mnras,
  465, 986, \dodoi{10.1093/mnras/stw2709}

\bibitem[{{Shakura} \& {Sunyaev}(1973)}]{Shakura73}
{Shakura}, N.~I., \& {Sunyaev}, R.~A. 1973, \aap, 24, 337

\bibitem[{{Shi} \& {Krolik}(2015)}]{Shi15}
{Shi}, J.-M., \& {Krolik}, J.~H. 2015, \apj, 807, 131,
  \dodoi{10.1088/0004-637X/807/2/131}

\bibitem[{{Shi} {et~al.}(2012){Shi}, {Krolik}, {Lubow}, \& {Hawley}}]{Shi12}
{Shi}, J.-M., {Krolik}, J.~H., {Lubow}, S.~H., \& {Hawley}, J.~F. 2012, \apj,
  749, 118, \dodoi{10.1088/0004-637X/749/2/118}

\bibitem[{{Shiokawa} {et~al.}(2012){Shiokawa}, {Dolence}, {Gammie}, \&
  {Noble}}]{Shiokawa12}
{Shiokawa}, H., {Dolence}, J.~C., {Gammie}, C.~F., \& {Noble}, S.~C. 2012,
  \apj, 744, 187, \dodoi{10.1088/0004-637X/744/2/187}

\bibitem[{{Shu} {et~al.}(2006){Shu}, {Galli}, {Lizano}, \& {Cai}}]{Shu06}
{Shu}, F.~H., {Galli}, D., {Lizano}, S., \& {Cai}, M. 2006, \apj, 647, 382,
  \dodoi{10.1086/505258}

\bibitem[{{Suzuki} \& {Inutsuka}(2014)}]{Suzuki14}
{Suzuki}, T.~K., \& {Inutsuka}, S.-i. 2014, \apj, 784, 121,
  \dodoi{10.1088/0004-637X/784/2/121}

\bibitem[{{Takakuwa} {et~al.}(2017){Takakuwa}, {Saigo}, {Matsumoto}, {Saito},
  {Lim}, {Hanawa}, {Yen}, \& {Ho}}]{Takakuwa17}
{Takakuwa}, S., {Saigo}, K., {Matsumoto}, T., {et~al.} 2017, \apj, 837, 86,
  \dodoi{10.3847/1538-4357/aa6116}

\bibitem[{{Takakuwa} {et~al.}(2014){Takakuwa}, {Saito}, {Saigo}, {Matsumoto},
  {Lim}, {Hanawa}, \& {Ho}}]{Takakuwa14}
{Takakuwa}, S., {Saito}, M., {Saigo}, K., {et~al.} 2014, \apj, 796, 1,
  \dodoi{10.1088/0004-637X/796/1/1}

\bibitem[{{Takakuwa} {et~al.}(2020){Takakuwa}, {Saigo}, {Matsumoto}, {Saito},
  {Lim}, {Yen}, {Ohashi}, {Ho}, \& {Looney}}]{Takakuwa20}
{Takakuwa}, S., {Saigo}, K., {Matsumoto}, T., {et~al.} 2020, \apj, 898, 10,
  \dodoi{10.3847/1538-4357/ab9b7c}

\bibitem[{{Tokuda} {et~al.}(2016){Tokuda}, {Onishi}, {Matsumoto}, {Saigo},
  {Kawamura}, {Fukui}, {Inutsuka}, {Machida}, {Tomida}, {Tachihara}, \&
  {Andr{\'e}}}]{Tokuda16}
{Tokuda}, K., {Onishi}, T., {Matsumoto}, T., {et~al.} 2016, \apj, 826, 26,
  \dodoi{10.3847/0004-637X/826/1/26}

\bibitem[{{Tokuda} {et~al.}(2020){Tokuda}, {Fujishiro}, {Tachihara},
  {Takashima}, {Fukui}, {Zahorecz}, {Saigo}, {Matsumoto}, {Tomida}, {Machida},
  {Inutsuka}, {Andr{\'e}}, {Kawamura}, \& {Onishi}}]{Tokuda20}
{Tokuda}, K., {Fujishiro}, K., {Tachihara}, K., {et~al.} 2020, \apj, 899, 10,
  \dodoi{10.3847/1538-4357/ab9ca7}

\bibitem[{{Tomida} {et~al.}(2010){Tomida}, {Tomisaka}, {Matsumoto}, {Ohsuga},
  {Machida}, \& {Saigo}}]{Tomida10}
{Tomida}, K., {Tomisaka}, K., {Matsumoto}, T., {et~al.} 2010, \apjl, 714, L58,
  \dodoi{10.1088/2041-8205/714/1/L58}

\bibitem[{{Tomisaka}(2002)}]{Tomisaka02}
{Tomisaka}, K. 2002, \apj, 575, 306, \dodoi{10.1086/341133}

\bibitem[{{Tsukamoto} {et~al.}(2015){Tsukamoto}, {Iwasaki}, {Okuzumi},
  {Machida}, \& {Inutsuka}}]{Tsukamoto15}
{Tsukamoto}, Y., {Iwasaki}, K., {Okuzumi}, S., {Machida}, M.~N., \& {Inutsuka},
  S. 2015, \mnras, 452, 278, \dodoi{10.1093/mnras/stv1290}

\bibitem[{{Tsukamoto} {et~al.}(2021){Tsukamoto}, {Machida}, \&
  {Inutsuka}}]{Tsukamoto21}
{Tsukamoto}, Y., {Machida}, M.~N., \& {Inutsuka}, S.-i. 2021, \apjl, 920, L35,
  \dodoi{10.3847/2041-8213/ac2b2f}

\bibitem[{{Tsukamoto} {et~al.}(2018){Tsukamoto}, {Okuzumi}, {Iwasaki},
  {Machida}, \& {Inutsuka}}]{Tsukamoto18}
{Tsukamoto}, Y., {Okuzumi}, S., {Iwasaki}, K., {Machida}, M.~N., \& {Inutsuka},
  S. 2018, \apj, 868, 22, \dodoi{10.3847/1538-4357/aae4dc}

\bibitem[{{Tsukamoto} {et~al.}(2022){Tsukamoto}, {Maury}, {Commer{\c{c}}on},
  {Alves}, {Cox}, {Sakai}, {Ray}, {Zhao}, \& {Machida}}]{Tsukamoto22}
{Tsukamoto}, Y., {Maury}, A., {Commer{\c{c}}on}, B., {et~al.} 2022, arXiv
  e-prints, arXiv:2209.13765, \dodoi{10.48550/arXiv.2209.13765}

\bibitem[{{Welsh} {et~al.}(2012){Welsh}, {Orosz}, {Carter}, {Fabrycky}, {Ford},
  {Lissauer}, {Pr{\v{s}}a}, {Quinn}, {Ragozzine}, {Short}, {Torres}, {Winn},
  {Doyle}, {Barclay}, {Batalha}, {Bloemen}, {Brugamyer}, {Buchhave},
  {Caldwell}, {Caldwell}, {Christiansen}, {Ciardi}, {Cochran}, {Endl},
  {Fortney}, {Gautier}, {Gilliland}, {Haas}, {Hall}, {Holman}, {Howard},
  {Howell}, {Isaacson}, {Jenkins}, {Klaus}, {Latham}, {Li}, {Marcy}, {Mazeh},
  {Quintana}, {Robertson}, {Shporer}, {Steffen}, {Windmiller}, {Koch}, \&
  {Borucki}}]{Welsh12}
{Welsh}, W.~F., {Orosz}, J.~A., {Carter}, J.~A., {et~al.} 2012, \nat, 481, 475,
  \dodoi{10.1038/nature10768}

\bibitem[{{Wurster} {et~al.}(2018){Wurster}, {Bate}, \& {Price}}]{Wurster18}
{Wurster}, J., {Bate}, M.~R., \& {Price}, D.~J. 2018, \mnras, 481, 2450,
  \dodoi{10.1093/mnras/sty2438}

\bibitem[{{Young} {et~al.}(2015){Young}, {Baird}, \& {Clarke}}]{Young15}
{Young}, M.~D., {Baird}, J.~T., \& {Clarke}, C.~J. 2015, \mnras, 447, 2907,
  \dodoi{10.1093/mnras/stu2656}

\bibitem[{{Young} \& {Clarke}(2015)}]{Young15b}
{Young}, M.~D., \& {Clarke}, C.~J. 2015, \mnras, 452, 3085,
  \dodoi{10.1093/mnras/stv1512}

\bibitem[{{Zhao} {et~al.}(2011){Zhao}, {Li}, {Nakamura}, {Krasnopolsky}, \&
  {Shang}}]{Zhao11}
{Zhao}, B., {Li}, Z.-Y., {Nakamura}, F., {Krasnopolsky}, R., \& {Shang}, H.
  2011, \apj, 742, 10, \dodoi{10.1088/0004-637X/742/1/10}

\bibitem[{{Zhu} \& {Stone}(2014)}]{Zhu14}
{Zhu}, Z., \& {Stone}, J.~M. 2014, \apj, 795, 53,
  \dodoi{10.1088/0004-637X/795/1/53}

\end{thebibliography}
\bibliographystyle{aasjournal}

\end{document}